\documentclass[prb,floatfix,twocolumn,showpacs,amsmath,amssymb]{revtex4}
\usepackage{graphicx}
\usepackage{dcolumn}
\usepackage{bm}
\usepackage{color}
\usepackage{ulem}

\begin{document}
\title{
Three-quarter Dirac points, Landau levels and magnetization 
in $\alpha$-(BEDT-TTF)$_2$I$_3$ 
}

\author
{Keita Kishigi}
\affiliation{Faculty of Education, Kumamoto University, Kurokami 2-40-1, 
Kumamoto, 860-8555, Japan}

\author{Yasumasa Hasegawa}
\affiliation{Department of Material Science, 
Graduate School of Material Science, 
University of Hyogo, Hyogo, 678-1297, Japan}

\date{\today}

\begin{abstract}
The energies as a function of the magnetic field ($H$) and the pressure are studied theoretically
in the tight-binding model for the two-dimensional organic conductor,
 $\alpha$-(BEDT-TTF)$_2$I$_3$, in which massless Dirac fermions are realized. 
The effects of the uniaxial pressure ($P$) are studied by using the pressure-dependent hopping parameters. 
The system is semi-metallic 
with the same area of an electron pocket and a hole pocket at $P < 3.0$~kbar, 
where the energies $(\varepsilon_{\rm D}^0$) at the Dirac points locate below the Fermi energy $(\varepsilon_{\rm F}^0$) when $H=0$. 
We find that at $P=2.3$~kbar the Dirac cones are critically tilted. In that case a new type of band crossing occurs at \textit{``three-quarter''-Dirac points}, i.e., the dispersion is quadratic in one direction and linear in the other three directions. 
We obtain new magnetic-field-dependences of the Landau levels $(\varepsilon_n)$; 
$\varepsilon_n-\varepsilon^0_{\textrm{D}} \propto (n H)^{4/5}$ 
at $P=2.3$~kbar ({\it ``three-quarter''-Dirac points}) and 
$|\varepsilon_n-\varepsilon_{\rm F}^0| \propto (n H)^2$ at $P=3.0$~kbar (the critical pressure for the semi-metallic state). 
We also study the magnetization as a function of the inverse magnetic field. 
We obtain two types of quantum oscillations. 
One is the usual de Haas van Alphen (dHvA) oscillation, and the other is 
the unusual dHvA-like oscillation which is seen even in the system without the Fermi surface. 
\end{abstract}

\date{\today}

\maketitle

\section{Introduction}
\begin{figure}[b]
\begin{flushleft} \hspace{0.0cm}(a) \end{flushleft}\vspace{-0.0cm}
\includegraphics[width=0.3\textwidth]{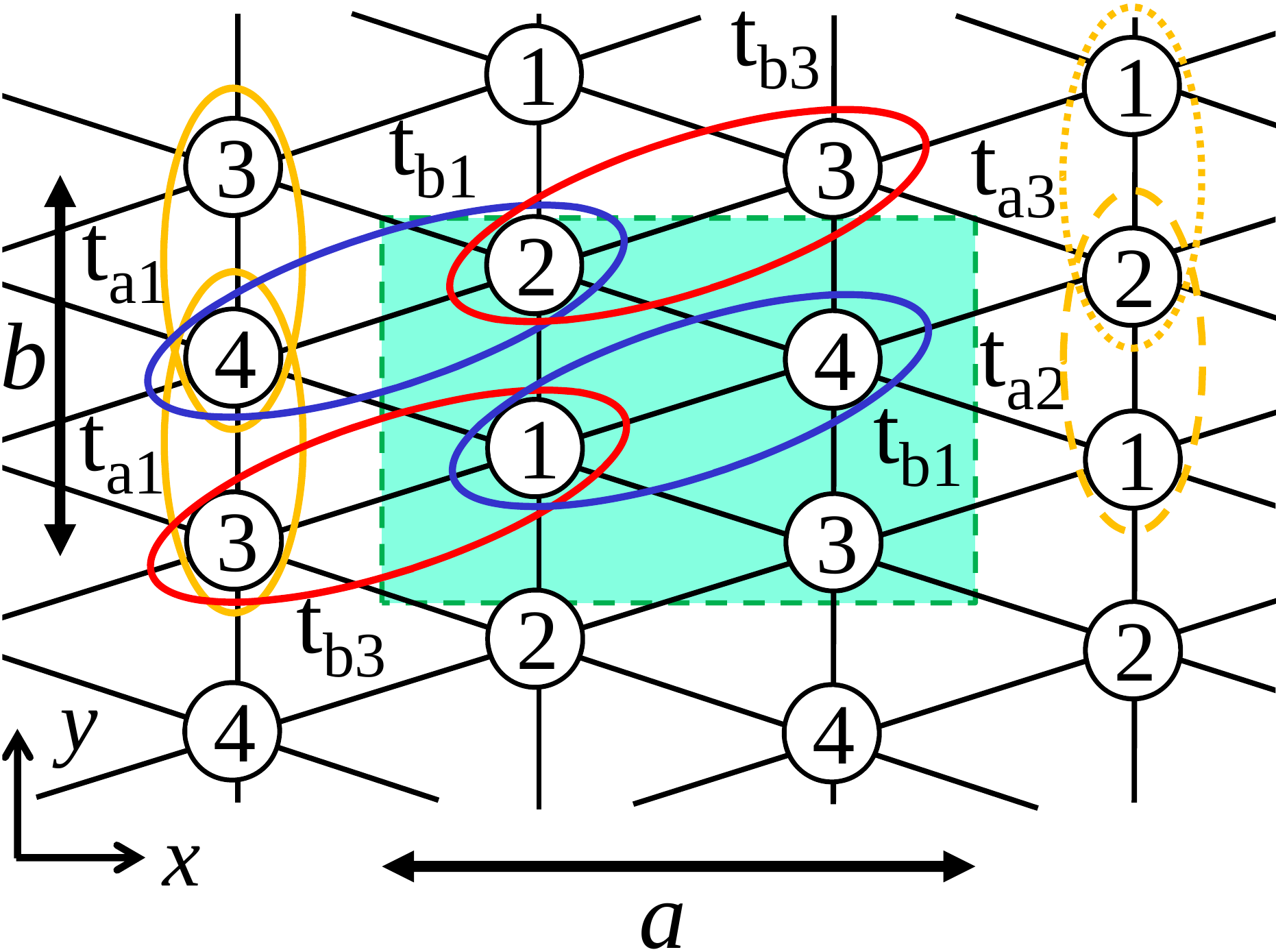}
\begin{flushleft} \hspace{0.0cm}(b) \end{flushleft}\vspace{0.1cm}
\includegraphics[width=0.3\textwidth]{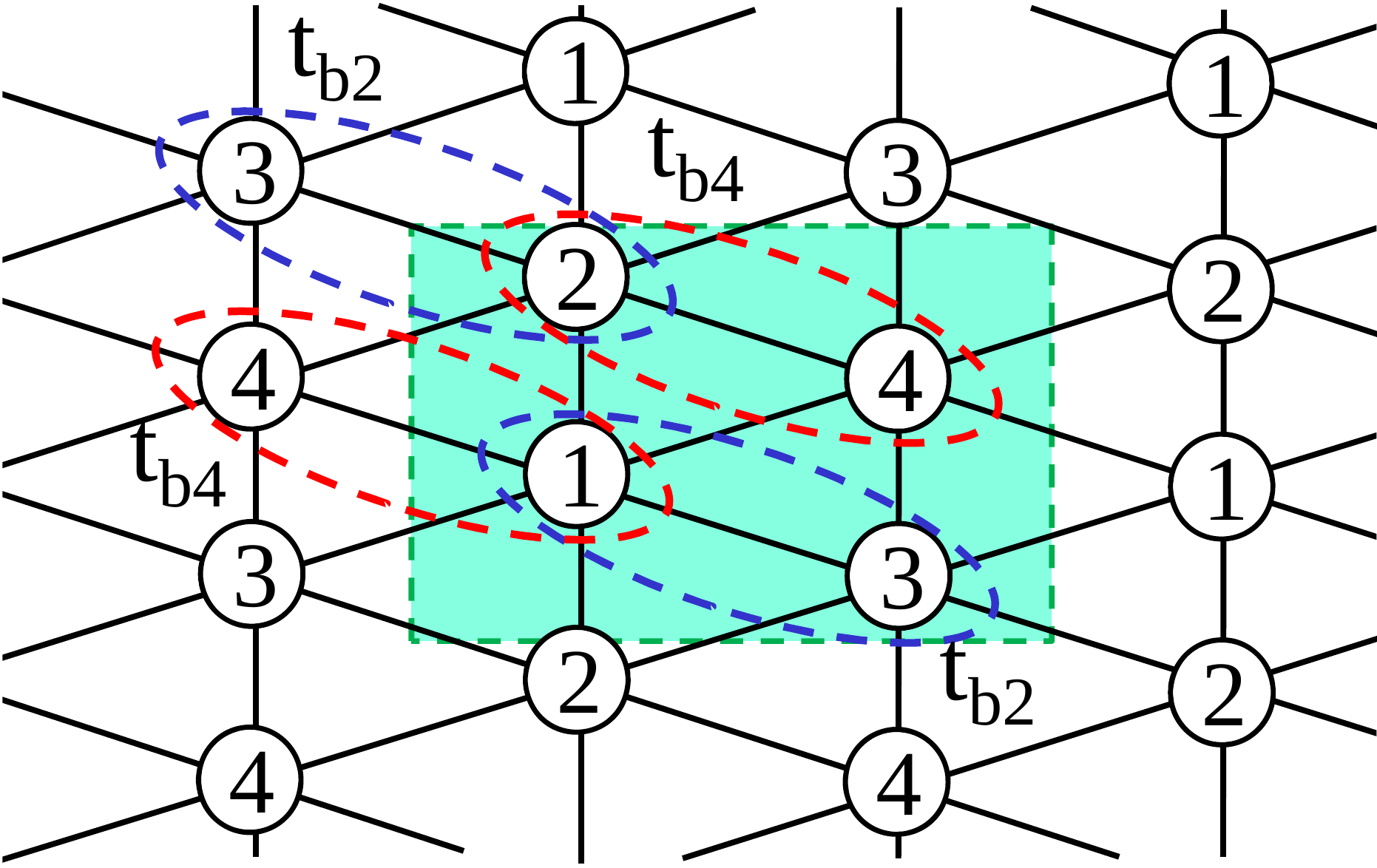}
\caption{
(Color online) The schematic figure of the tight-binding model
 for $\alpha$-(BEDT-TTF)$_2$I$_3$.  
The unit cell is the rectangle in (a) and (b). 
The transfer integrals
 ($t_{\mathrm{a1}}$,  $t_{\mathrm{a2}}$, $t_{\mathrm{a3}}$, $t_{\mathrm{b1}}$, 
$t_{\mathrm{b2}}$, $t_{\mathrm{b3}}$ and $t_{\mathrm{b4}}$) are shown as ovals.
}
\label{fig1}
\end{figure}

$\alpha$-(BEDT-TTF)$_2$I$_3$ is the two-dimensional organic conductor\cite{review,review_2}, which has attracted interest recently due to the realization of massless Dirac fermions\cite{Katayama2006,Kobayashi2007, Kajita2014,Hirata2011,Konoike2012,Osada2008}. 
There are four BEDT-TTF molecules in the unit cell, as shown in Fig. \ref{fig1}, and four 
energy bands are constructed by the highest occupied molecular orbits (HOMO) of BEDT-TTF
 molecules. 
 The electron bands are 3/4-filled, since one electron is
removed from two BEDT-TTF molecules.
 Therefore, the system is semi-metallic when the third and the fourth bands overlap, and
 it is an insulator when there is a gap between two bands. 
 
Katayama, Kobayashi and Suzumura\cite{Katayama2006}
 have theoretically shown the realization of massless Dirac fermions in $\alpha$-(BEDT-TTF)$_2$I$_3$, where the third and the fourth bands touch at two Dirac points. 
Two bands near the Fermi energy can be approximately described by the tilted Weyl equation\cite{Kobayashi2007}. 
The existence of massless Dirac fermions in $\alpha$-(BEDT-TTF)$_2$I$_3$ have been confirmed experimentally\cite{Kajita2014,Hirata2011,Konoike2012,Osada2008}.

The energy dispersion of massless Dirac fermions near the Dirac points is linear, which is called a Dirac cone. Recently, by considering the anisotropy of the nearest-neighbor hoppings on a honeycomb lattice\cite{Hasegawa2006, Dietl2008} it has been found that the dispersion is quadratic in two directions and linear in the two other directions when two Dirac points marge at a time-reversal-invariant point. That special point was named as a semi-Dirac point in VO$_2$/TiO$_2$ nanostructures\cite{Banerjee2009}. 
The semi-Dirac point has been also shown to exist in $\alpha$-(BEDT-TTF)$_2$I$_3$ at high pressure theoretically \cite{Montambaux2009_prb,Suzumura2013}. 

%

When the magnetic field ($H$) is applied to two-dimensional systems, 
the energies are quantized. 
In many papers the effects of the magnetic field have been studied semiclassically\cite{Onsager} which is explained in Appendix \ref{Appendix0}. However, a treatment by a quantum mechanical manner is possible for simple cases. 
For example, the energies are given by 
\begin{equation}
  \varepsilon_{n}^{\textrm{(massive)}} \propto \left( n+\frac{1}{2} \right)  H, \ \ \ n=0, 1, 2
\label{eq2Dfree}
\end{equation}
for two-dimensional massive free electrons\cite{shoenberg} and 
\begin{equation}
\varepsilon_{n} ^{\textrm{(Dirac)}}\propto {\rm sgn}(n)\sqrt{|n| H},  \ \ \ n=0, \pm 1, \pm 2, \cdots \label{eq_1x}
\end{equation} 
for massless Dirac fermions (graphene\cite{Novo2005,McC1956} and 
$\alpha$-(BEDT-TTF)$_2$I$_3$\cite{Georbig2008,Morinari2009}, where the linearization of the energy dispersion has been done). Moreover, on the honeycomb lattice with the semi-Dirac point, 
Dietl, Piechon and Montambaux\cite{Dietl2008} have found new magnetic-field-dependences which are given by
\begin{equation}
 \varepsilon_n^{\textrm{(semi-Dirac)}}\propto\textrm{sgn}(n) g(n)
 \left(\left| n+\frac{1}{2} \right| H \right)^{\frac{2}{3}}, \ n=0, \pm1, \pm2, \cdots 
\label{eqmergedDirac}
\end{equation}
where $g(0) \simeq 0.808$, $g(\pm 1) \simeq 0.994$ and $g(n) \simeq 1$ for $|n| \geq 2$.

In this study, we show the existence of a new type of band crossing that we baptize {\it ``three-quarter''-Dirac points} because the dispersion relation is quadratic in one direction and linear in the other three directions. 
Furthermore, we study the magnetic-field-dependences of the energy in various cases of semi-metallic state, critically tilted Dirac cones, massless Dirac fermions 
and massive Dirac fermions.

In the tight-binding electrons, rich structures such as the broadening of the Landau levels (Harper broadening\cite{Harper}) and recursive gap structures are seen on the square lattice\cite{Hof,H1989,H1990} and on the honeycomb 
lattice\cite{Rammal,HK2006,KH2014}. These characteristic graphs are called the Hofstadter butterfly diagrams. Recently, we have studied the de Haas van Alphen (dHvA) oscillation\cite{shoenberg} in the tight-binding model 
for (TMTSF)$_2$NO$_3$ where electron and hole pockets 
coexist\cite{pouget,fisdw_no3,kang_2009}.
In that system the dHvA oscillation 
has been usually studied in the phenomenological theory of magnetic breakdown\cite{fortin2008,fortin2009} and the Lifshitz and Kosevich (LK) formula \cite{Pippard62,Falicov66}.  
The dHvA oscillation and the LK formula\cite{LK,Igor2004PRL,Igor2011,Sharapov} are explained in Appendix \ref{AppendixC}. 
We have shown that the magnetic-field-dependence of the amplitude of the dHvA oscillation at zero temperature is different from that of the LK formula due to the Harper broadening\cite{KH2016}. 
We have also obtained the dHvA-like oscillation on the honeycomb lattice even if the system is an insulator\cite{KH2014}. 
We investigate the oscillation of the magnetization in 
the Hofstadter butterfly diagrams for $\alpha$-(BEDT-TTF)$_2$I$_3$ in this paper.

In $\alpha$-(BEDT-TTF)$_2$I$_3$, the metal-insulator transition is 
observed at $T=135$ K, which is thought to be caused by the 
charge ordering\cite{KF1995,Seo2000,takano2001,Woj2003}. 
The metal-insulator transition is suppressed by pressure.
Tajima {\it et al.} have observed from the conductivity that the charge 
ordering disappears at an uniaxial pressure, $P\gtrsim 10$ kbar\cite{Tajima2002}.  
In the hydrostatic pressure,  
the charge ordering has not been observed above 17 kbar from the magneto 
conductivity\cite{Tajima2013} 
and above  11$-$12 kbar from the optical investigations\cite{Beyer} and 
conductivity\cite{Dong}. 
In this paper we do not study the interaction between electrons, 
so we do not concern the metal-insulator transition caused by the charge ordering.


\begin{figure}[bt]
\begin{center}
\begin{flushleft} \hspace{0.5cm}(a) \end{flushleft}\vspace{-0.0cm}
\includegraphics[width=0.46\textwidth]{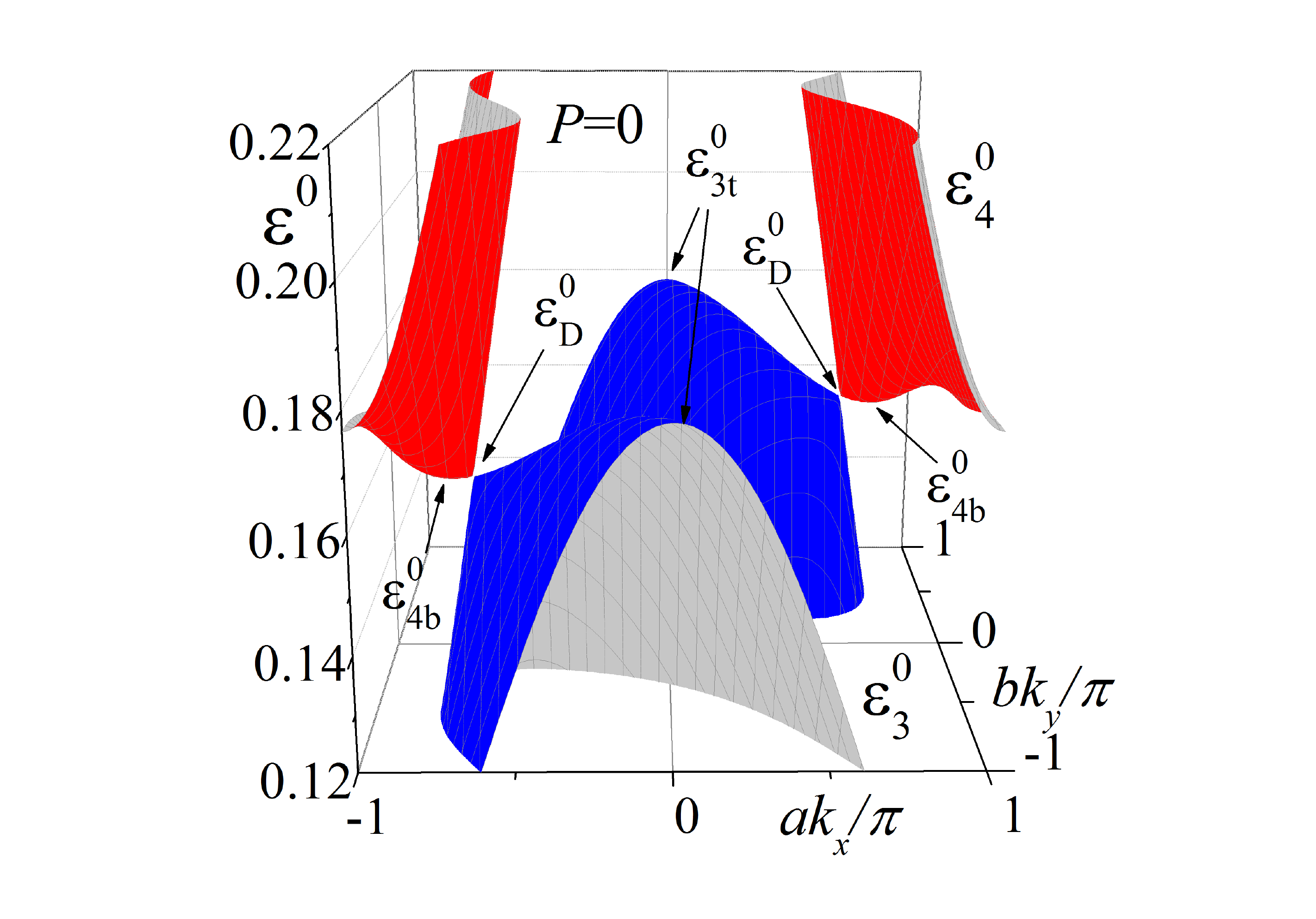}\vspace{-0.0cm}
\begin{flushleft} \hspace{0.5cm}(b) \end{flushleft}\vspace{-0.0cm}
\includegraphics[width=0.46\textwidth]{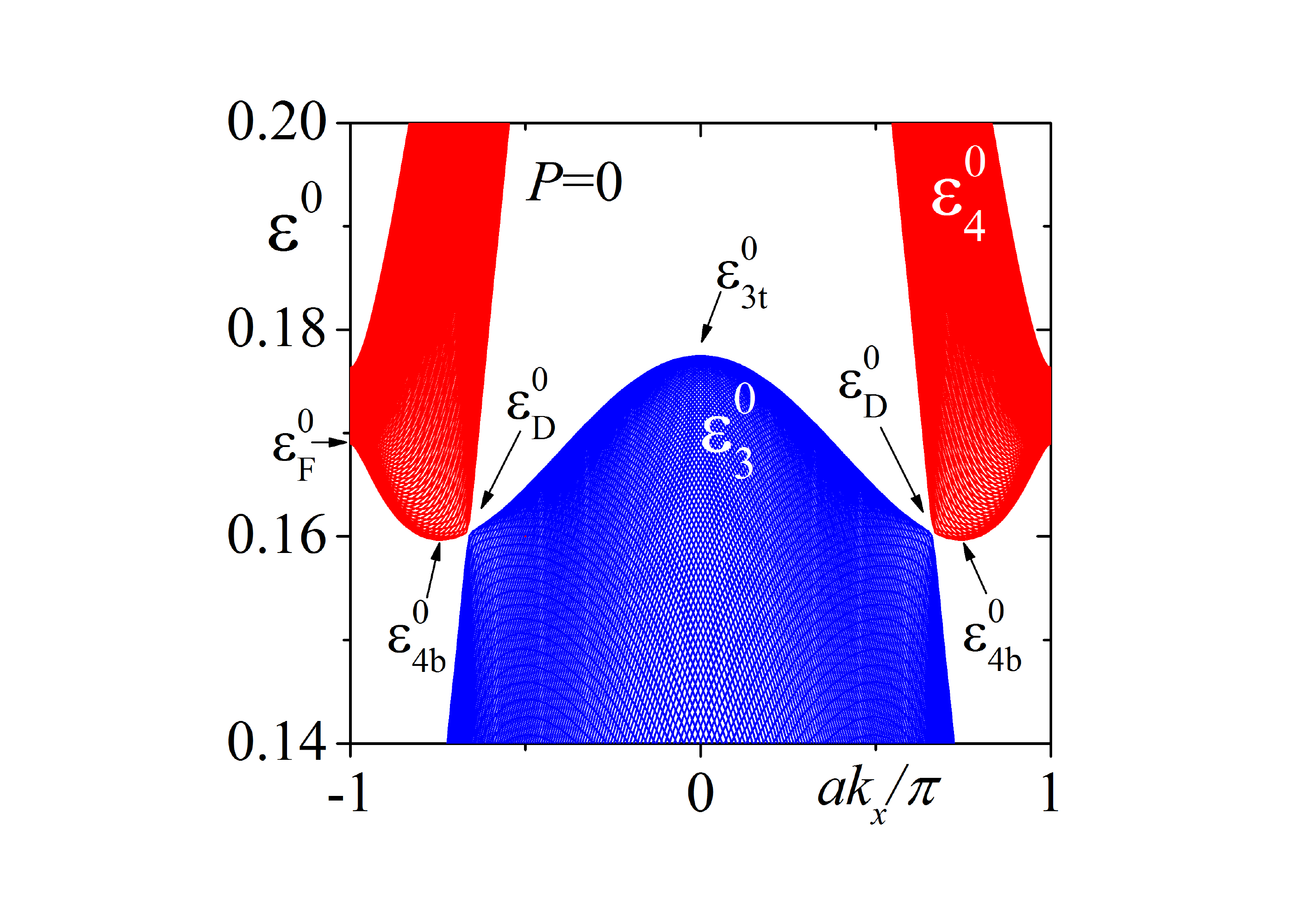}\vspace{-0.0cm}
\begin{flushleft} \hspace{0.5cm}(c) \end{flushleft}\vspace{-0.0cm}
\includegraphics[width=0.46\textwidth]{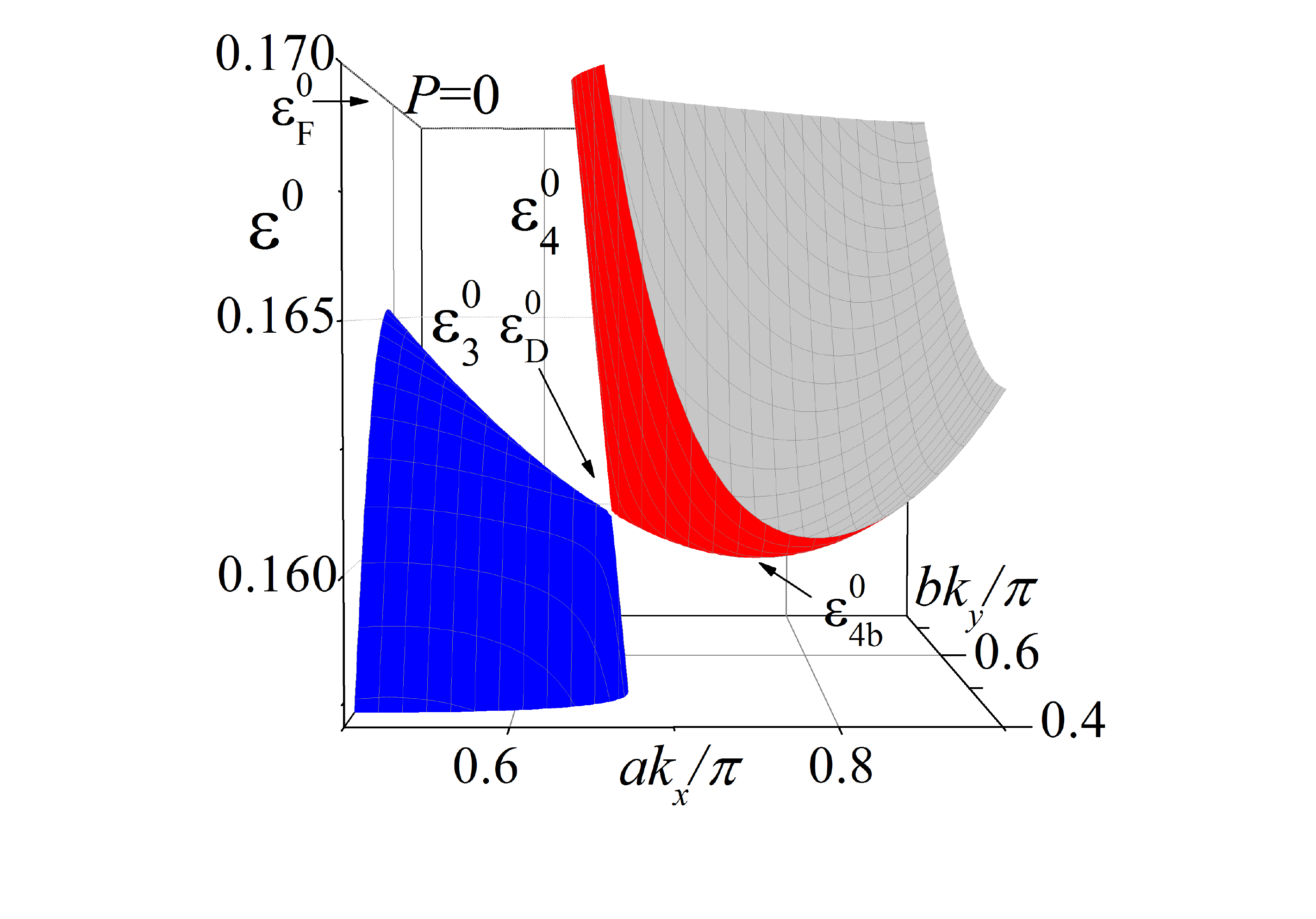}\vspace{-0.0cm}
\end{center}
\caption{
(Color online) 
(a) The third and fourth energy bands ($\varepsilon_{\rm 3}^0$ 
and $\varepsilon_{\rm 4}^0$) at $P=0$, 
where $\varepsilon_{\rm 3t}^0\simeq0.17805$, $\varepsilon_{\rm D}^0\simeq0.16094$ 
and $\varepsilon_{\rm 4b}^0\simeq0.16011$. 
(b) is a figure of (a) from a distant view point along the $k_y$ axis. 
(c) is an enlarged figure of (a) near the Dirac point, 
$\mathbf{k}_{\mathrm{D}}$. 
}
\label{fig2}
\end{figure}
\begin{figure}[bt]
\begin{center}
\begin{flushleft} \hspace{0.5cm}(a) \end{flushleft}\vspace{-0.0cm}
\includegraphics[width=0.46\textwidth]{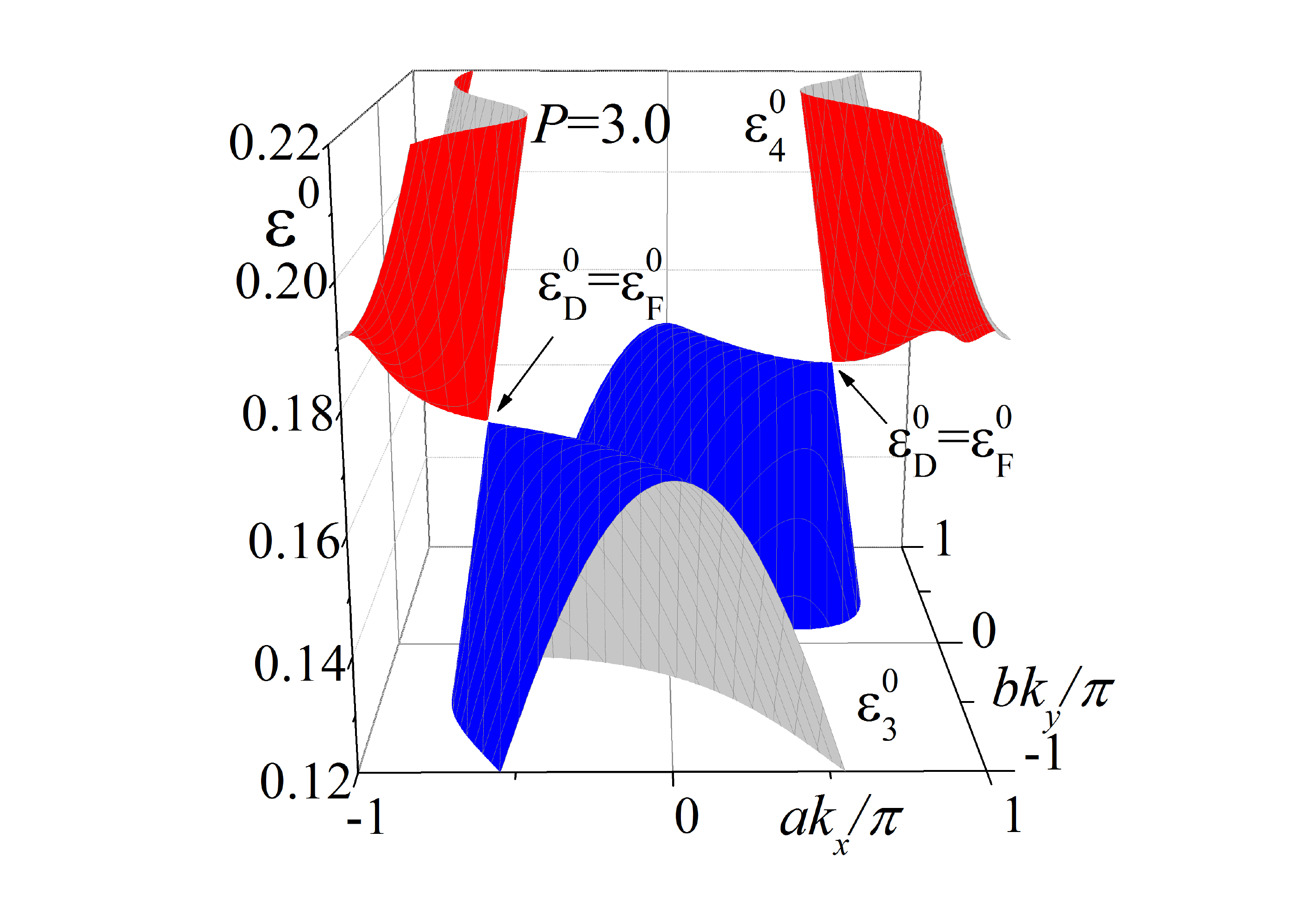}\vspace{-0.0cm}
\begin{flushleft} \hspace{0.5cm}(b) \end{flushleft}\vspace{-0.0cm}
\includegraphics[width=0.46\textwidth]{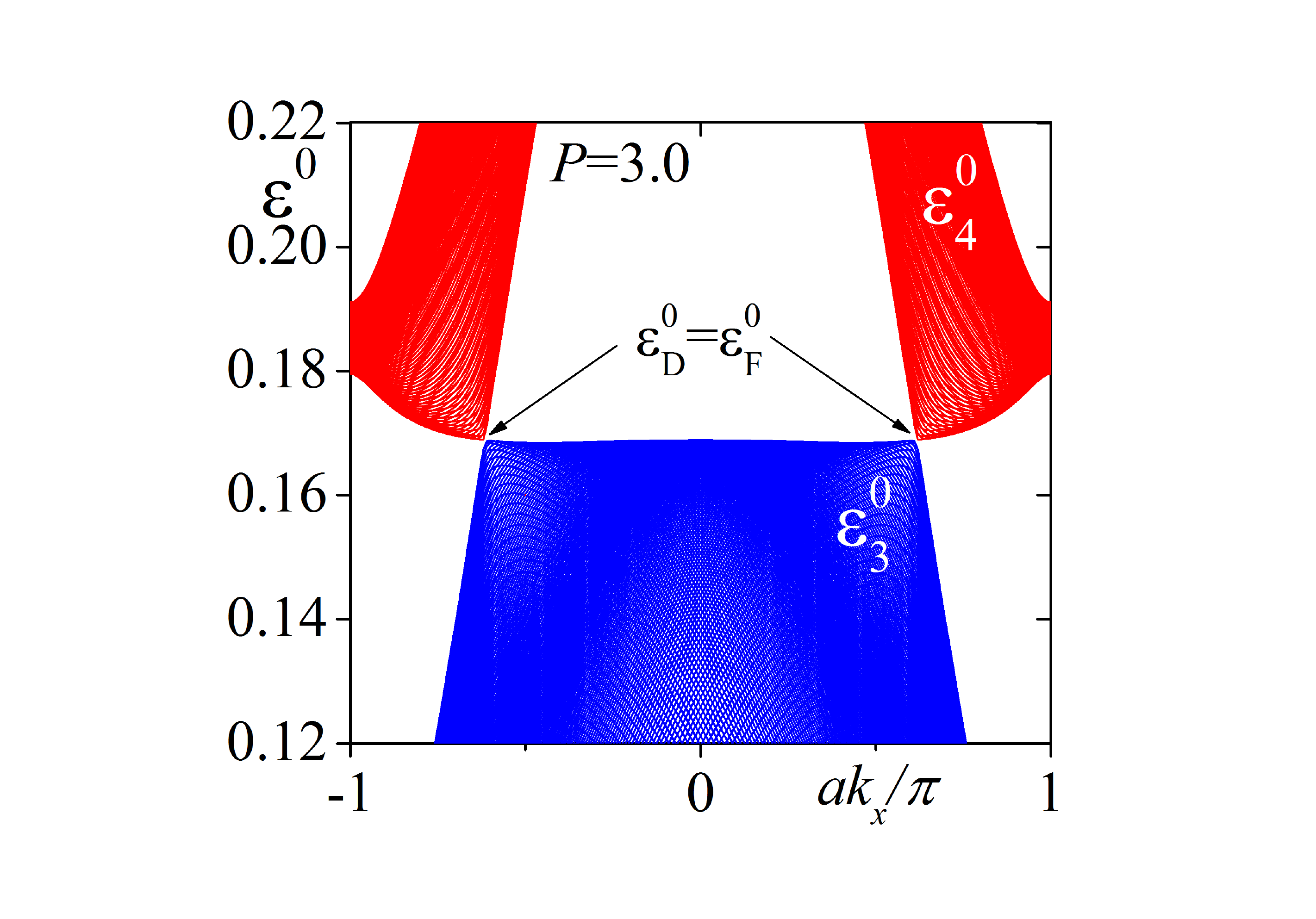}\vspace{-0.0cm}
\begin{flushleft} \hspace{0.5cm}(c) \end{flushleft}\vspace{-0.0cm}
\includegraphics[width=0.46\textwidth]{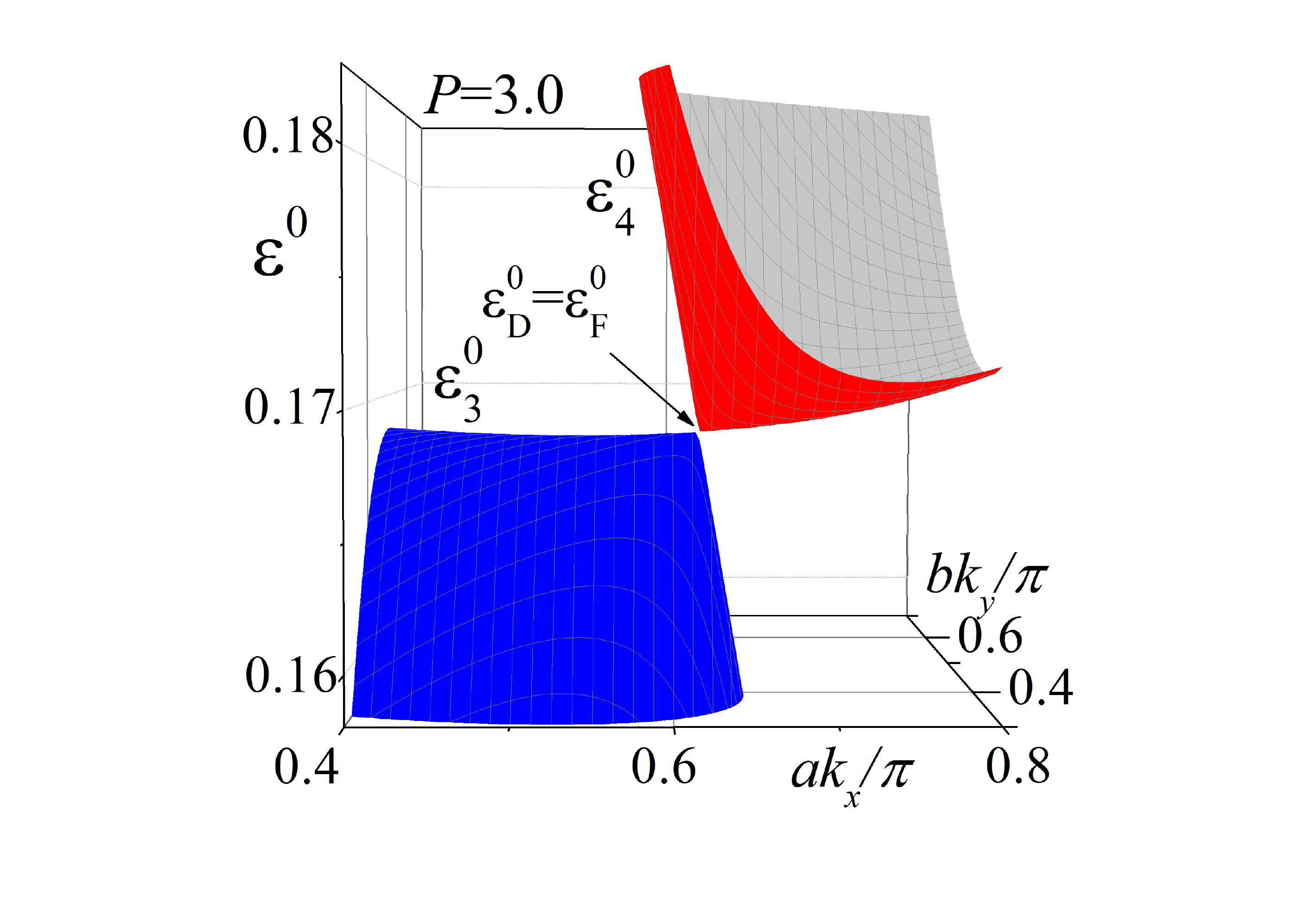}\vspace{-0.0cm}
\end{center}
\caption{
(Color online) 
The same figures of Fig. \ref{fig2} except for $P=3.0$, where 
$\varepsilon_{\rm 3t}^0=\varepsilon_{\rm 4b}^0=\varepsilon_{\rm D}^0=\varepsilon^0_{\rm F}\simeq 0.16887$.
}
\label{fig4}
\end{figure}
\begin{figure}[bt]
\begin{center}
\begin{flushleft} \hspace{0.5cm}(a) \end{flushleft}\vspace{-0.0cm}
\includegraphics[width=0.46\textwidth]{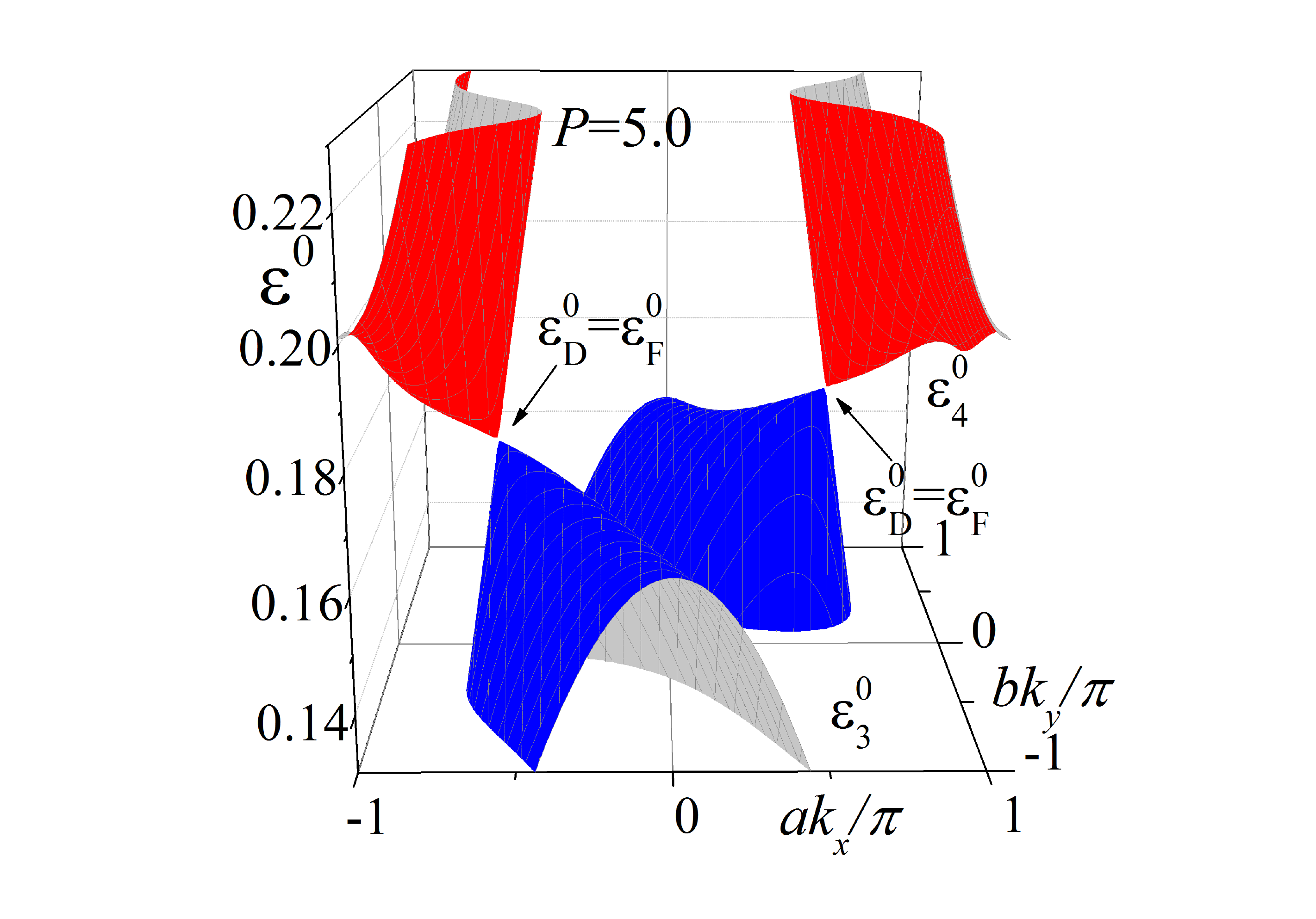}\vspace{-0.0cm}
\begin{flushleft} \hspace{0.5cm}(b) \end{flushleft}\vspace{-0.0cm}
\includegraphics[width=0.46\textwidth]{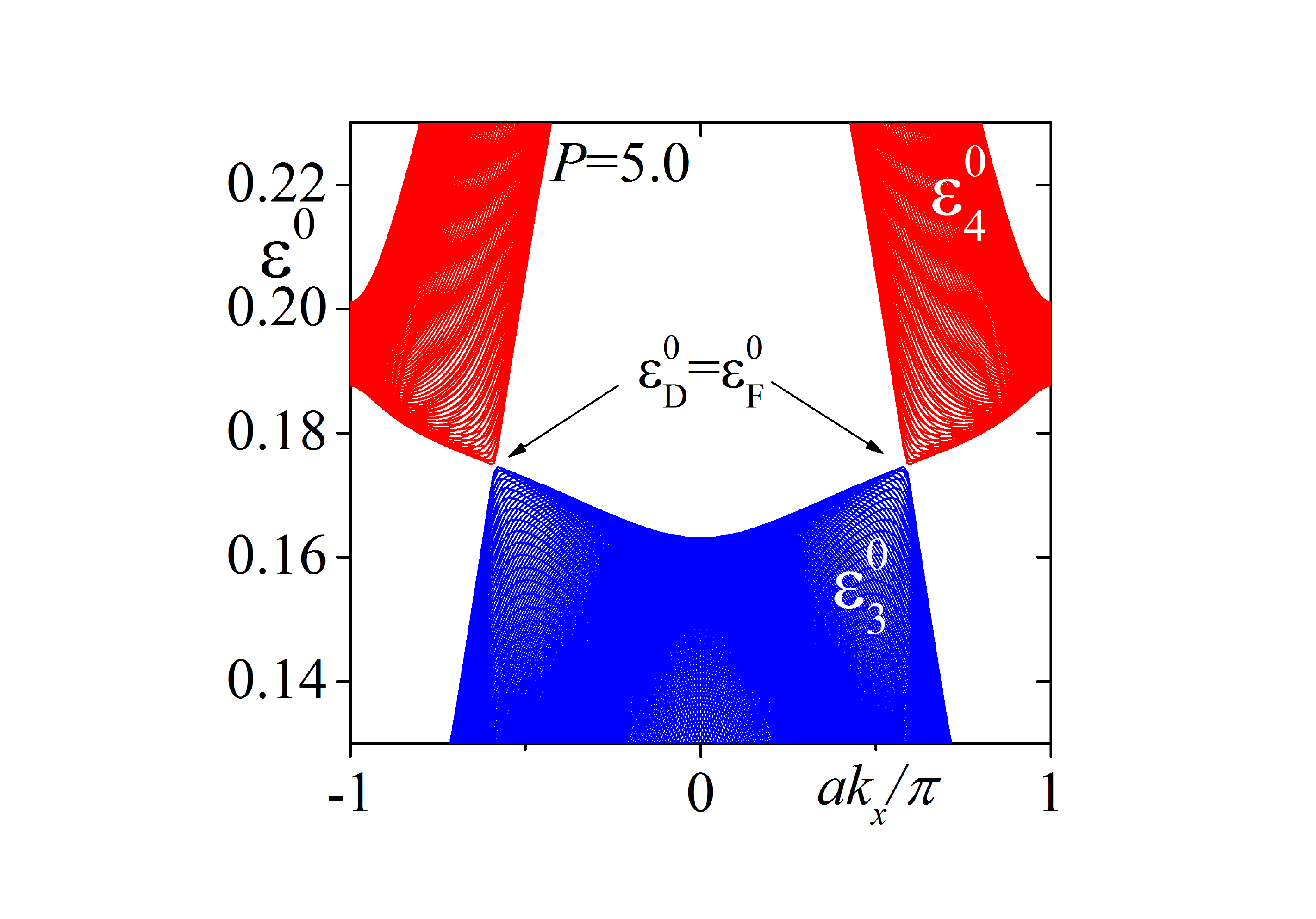}\vspace{-0.0cm}
\begin{flushleft} \hspace{0.5cm}(c) \end{flushleft}\vspace{-0.0cm}
\includegraphics[width=0.46\textwidth]{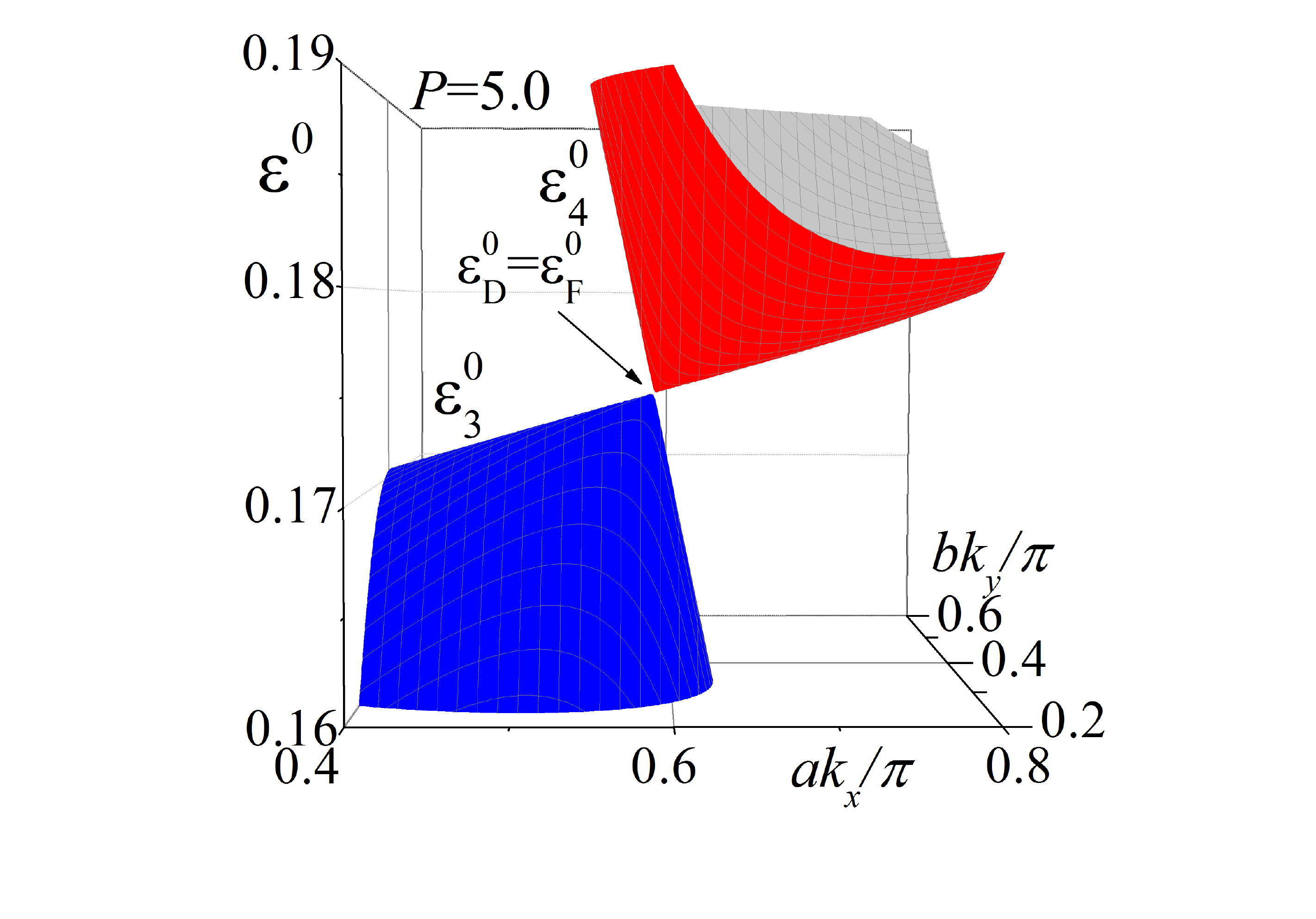}\vspace{-0.0cm}
\end{center}
\caption{
(Color online) 
The same figures of Fig. \ref{fig2} except for $P=5.0$, where 
$\varepsilon_{\rm 3t}^0=\varepsilon_{\rm 4b}^0=\varepsilon_{\rm D}^0=\varepsilon^0_{\rm F}\simeq 0.17479$.
}
\label{fig6}
\end{figure}

\begin{figure}[bt]
\begin{center}
\includegraphics[width=0.46\textwidth]{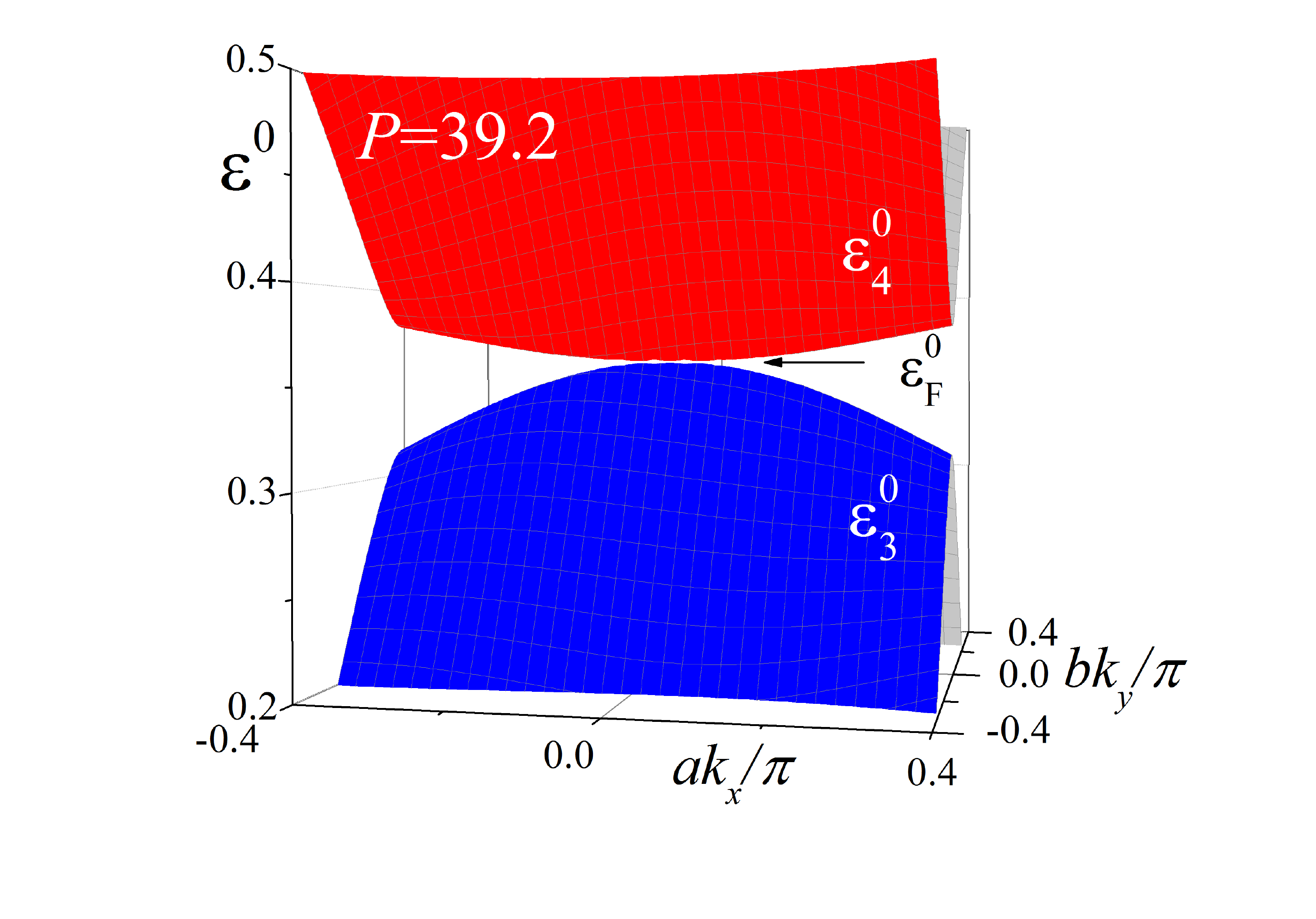}\vspace{-0.5cm}
\end{center}
\caption{
(Color online) 
The third and fourth energy bands at $P=39.2$. 
Two Dirac points merge at $\Gamma$ point.}
\label{fig10}
\end{figure}
\begin{figure}[bt]
\begin{center}
\begin{flushleft} \hspace{0.5cm}
 \end{flushleft}\vspace{-0.0cm}
\includegraphics[width=0.46\textwidth]{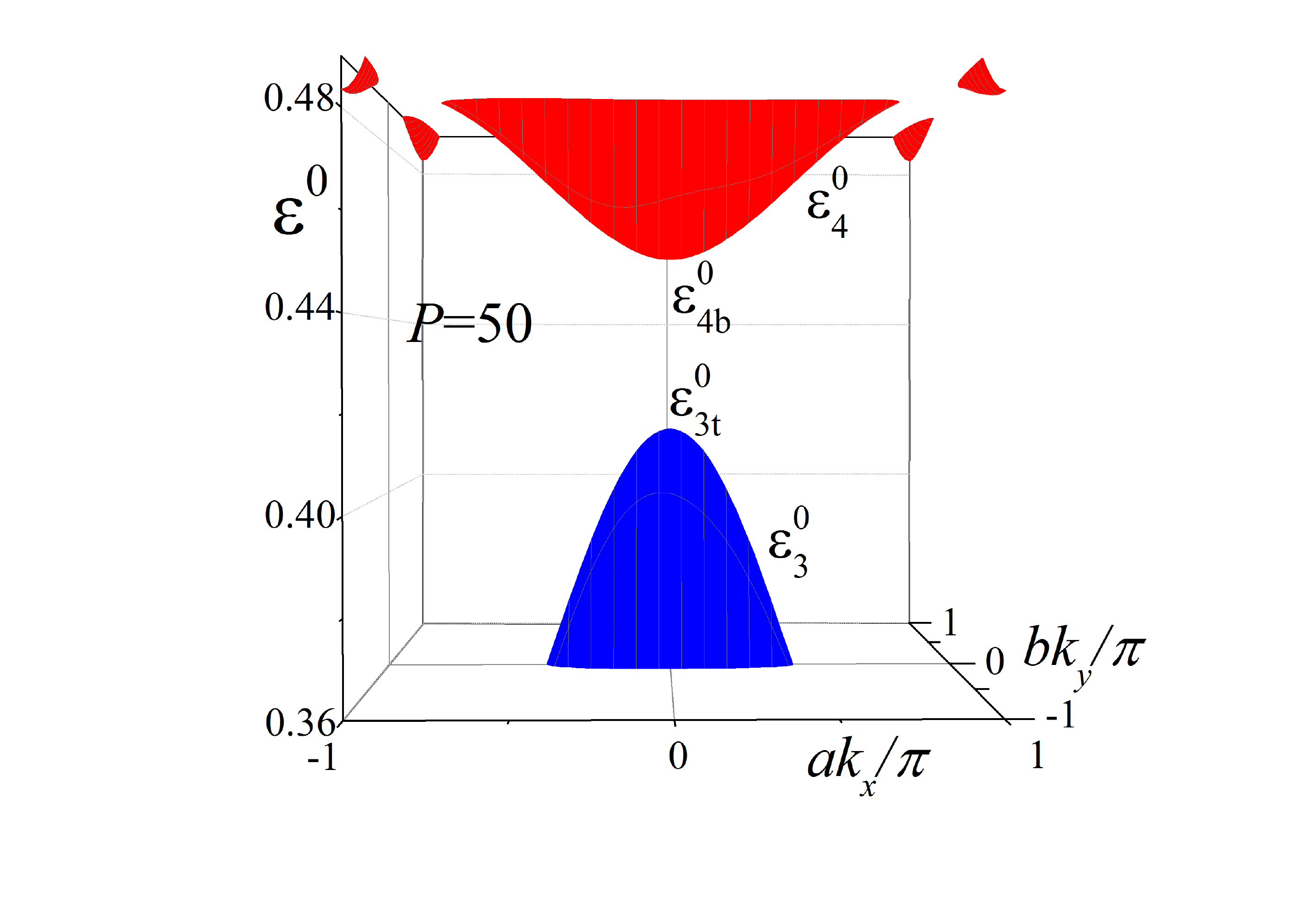}\vspace{-0.5cm}
\end{center}
\caption{
(Color online) 
The third and fourth energy bands at $P=50$. 
The top of the third band and the bottom of the fourth band are obtained as
$\varepsilon_{\rm 3t}^0\simeq 0.41471$ and $\varepsilon_{\rm 4b}^0\simeq 0.45378$, respectively. 
}
\label{fig11}
\end{figure}

\section{Energy band and uniaxial pressure effect}
\label{h=0}

The energies of $\alpha$-(BEDT-TTF)$_2$I$_3$ are described by the two-dimensional 
tight-binding model. The transfer integrals are taken between neighboring sites as shown in 
Fig.~\ref{fig1} and they are given as functions of pressure as
the interpolation formulas
\cite{Mori1984,Katayama2006,Mori2010,Suzumura2013,Kondo2009,Kondo2005}. 
In this study, we use the following interpolation formula\cite{Mori1984,Mori2010,Suzumura2013},
(hereafter, we employ eV and kbar as the units of transfer integrals and the pressure, respectively.)
\begin{equation}
\begin{split}
t_{a1} &= -0.028  (1.0 + 0.089P), \\
t_{a2} &= -0.048  (1.0 + 0.167P), \\
t_{a3} &=   0.020  (1.0-0.025P), \\ 
t_{b1} &=   0.123,                    \\
t_{b2} &=   0.140  (1.0 + 0.011P),  \\
t_{b3} &=   0.062  (1.0 + 0.032P),  \\
t_{b4} &=   0.025,
\end{split} 
\label{eqhoppings}
\end{equation}
where $P$ is the uniaxial strain along the $y$ axis. 
The Hamiltonian in this tight-binding model is explained in Appendix \ref{appendixA} for $h=0$ and in Appendix \ref{AppendixB} for $h\neq 0$.

\begin{figure}[bt]
\begin{center}
\begin{flushleft} \hspace{0.5cm}(a) \end{flushleft}\vspace{-0.5cm}
\includegraphics[width=0.55\textwidth]{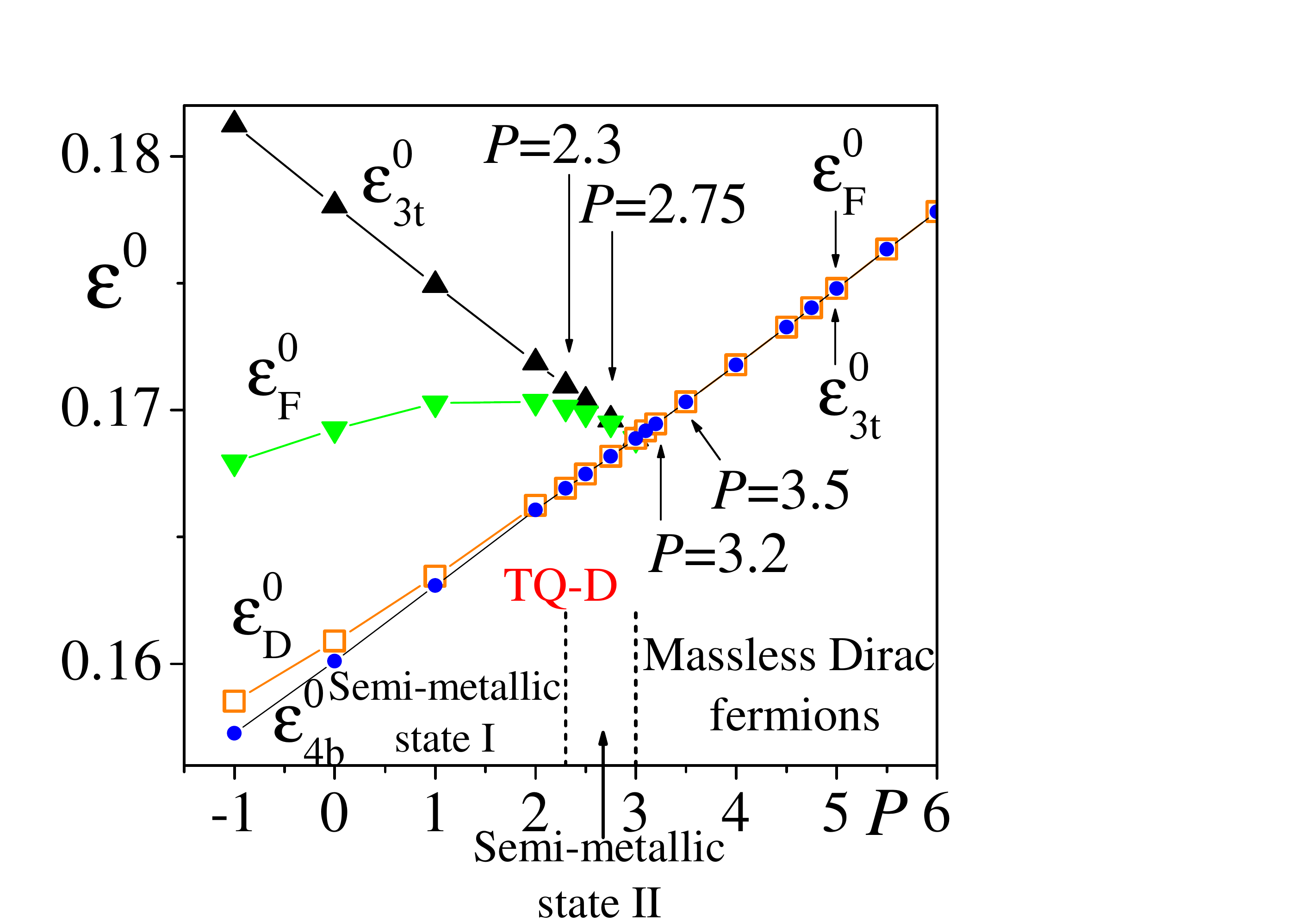}\vspace{-0.5cm}
\begin{flushleft} \hspace{0.5cm}(b) \end{flushleft}\vspace{-0.5cm}
\includegraphics[width=0.55\textwidth]{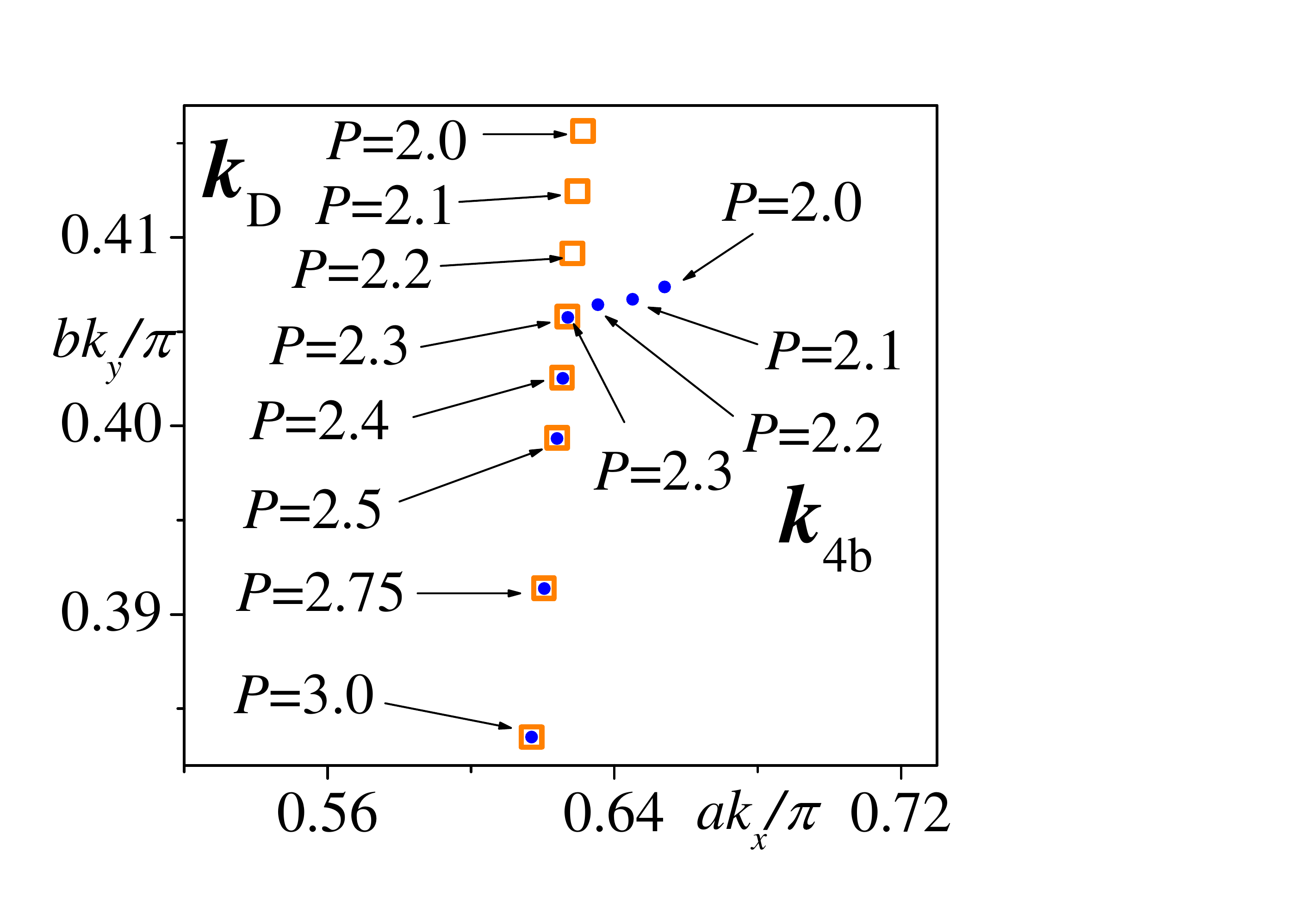}\vspace{-0.5cm}
\end{center}
\caption{
(Color online) 
(a) Pressure dependences of the energies at the Dirac point ($\varepsilon_{\rm D}^0$, orange open squares), 
the top of the third band ($\varepsilon_{\rm 3t}^0$, black filled triangels), the bottom of the fourth band ($\varepsilon_{\rm 4b}^0$, blue filled circles) and the Fermi energy for the 3/4-filled ($\varepsilon_{\rm F}^0$, green filled inverse triangles). 
All of the energies $\varepsilon_{\rm 3t}^0$, $\varepsilon_{\rm 4b}^0$, $\varepsilon_{\rm D}^0$ and 
$\varepsilon_{\rm F}^0$ become the same values at $P\gtrsim 3.0$. At $P=2.3$ 
the Dirac points become {\it ``three-quarter''-Dirac points} (TQ-D). 
(b) Pressure dependences of the wave numbers of the
Dirac point (${\bf k}_{\rm D}$, orange open squares) and the bottom of the fourth band (${\bf k}_{\rm 4b}$, 
blue filled circles). 
}
\label{fig9}
\end{figure}

By using the pressure-dependent hoppings (Eq.~(\ref{eqhoppings})) we show the third band and the fourth band at $P=0$, 
$3.0$, $5.0$, 39.2 and $50$ in 
Figs.~\ref{fig2}, \ref{fig4}, \ref{fig6}, \ref{fig10} 
and \ref{fig11}. These contour plots except for the case of $P=50$ are shown in Figs.~\ref{fig3}, \ref{fig5}, \ref{fig7} and \ref{fig10bc}. 
Katayama, Kobayashi and Suzumura\cite{Katayama2006} have shown that 
at $P \geq 3.0$ the third band and the fourth band touch each other 
at two Dirac points ($\pm \mathbf{k}_{\textrm{D}}$) with the energy ($\varepsilon_{\textrm{D}}^0$) which are the same as the tops of the third band ($\varepsilon_{\textrm{3t}}^0$) 
at $\mathbf{k}=\pm\mathbf{k}_{\textrm{3t}}$ 
and the bottoms of the fourth band ($\varepsilon_{\textrm{4b}}^0$) at $\mathbf{k}=\pm\mathbf{k}_{\textrm{4b}}$. The 
Fermi energy for the 3/4-filled ($\varepsilon^0_{\mathrm{F}}$) is 
equal to $\varepsilon^0_{\mathrm{D}}$, as shown in Fig. \ref{fig9} (a). 
This is supported from the first-principle band calculations by Kino and Miyazaki\cite{Kino} and Alemany, Pouget and Canadell\cite{Alemany2012}. 
It has been also known that the system is semi-metallic at $P<3.0$, where the 
Fermi surfaces are shown in Fig. \ref{fig8}. 
There are a hole pocket centered at $\mathbf{k}_{\textrm{3t}}=(0, \pi/b)$ and 
an electron pocket enclosing two Dirac points and $\mathbf{k} = (\pi/a,0)$. An electron pocket separates into 
two small electron pockets with the same area at $0.2 \lesssim P <3.0$, as shown in Fig. \ref{fig8}.

%
\begin{figure}[bt]
\begin{center}
\begin{flushleft} \hspace{0.5cm}(a) \end{flushleft}\vspace{-0.0cm}
\includegraphics[width=0.46\textwidth]{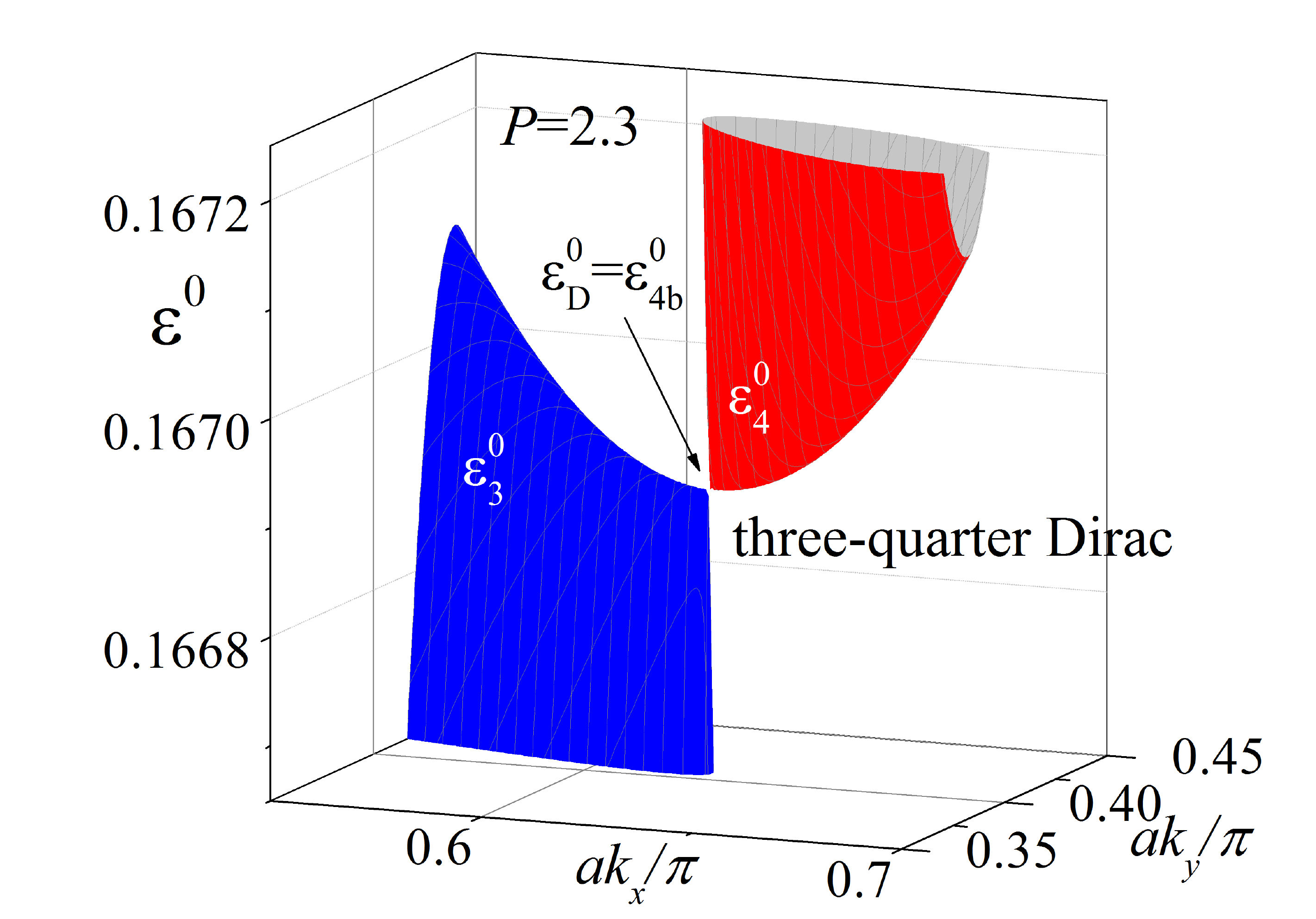}\vspace{-0.0cm} 
\begin{flushleft} \hspace{0.5cm}(b) \end{flushleft}\vspace{-0.0cm}
\includegraphics[width=0.46\textwidth]{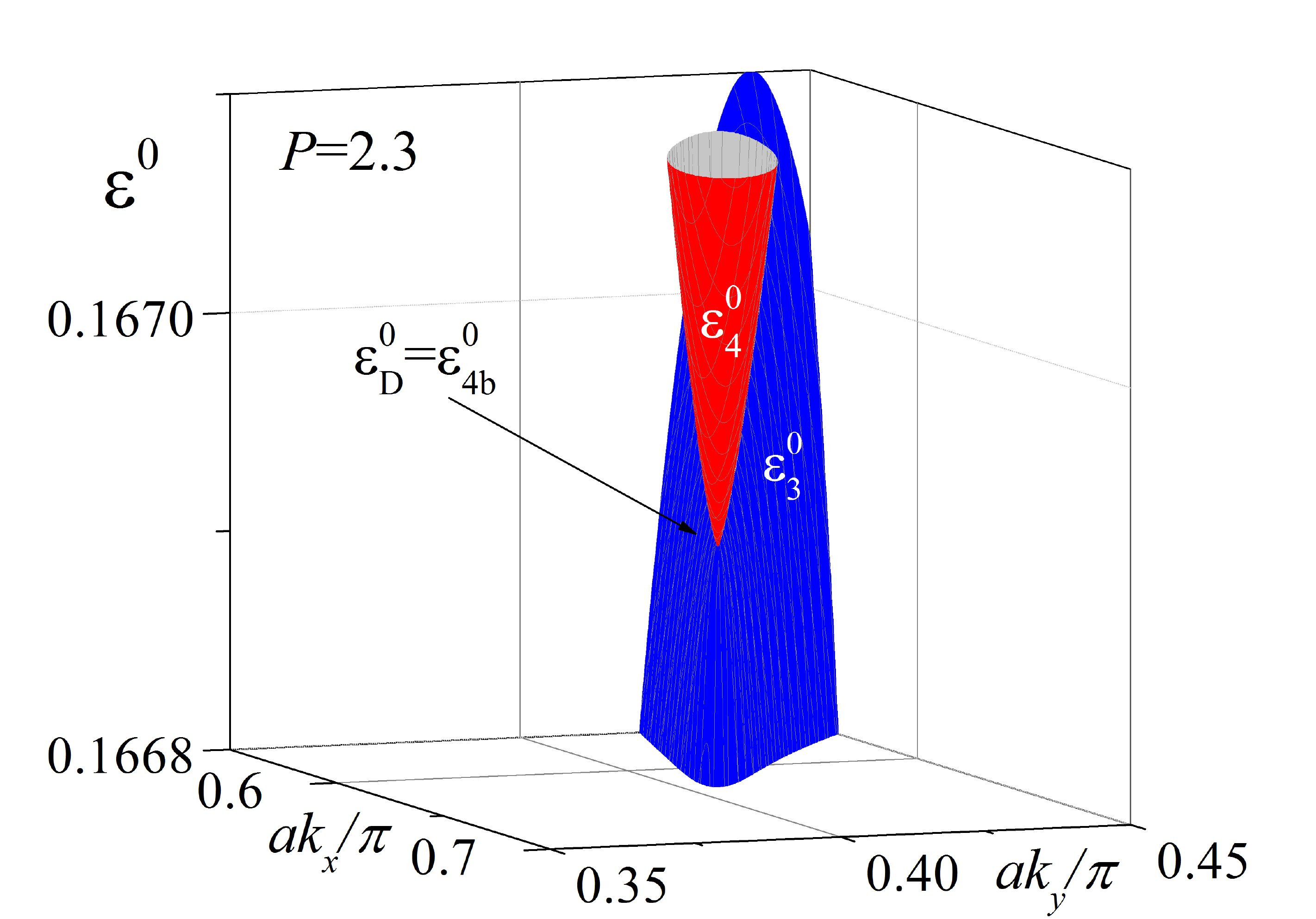}\vspace{-0.0cm}
\begin{flushleft} \hspace{0.5cm}(c) \end{flushleft}\vspace{-0.0cm}
\includegraphics[width=0.46\textwidth]{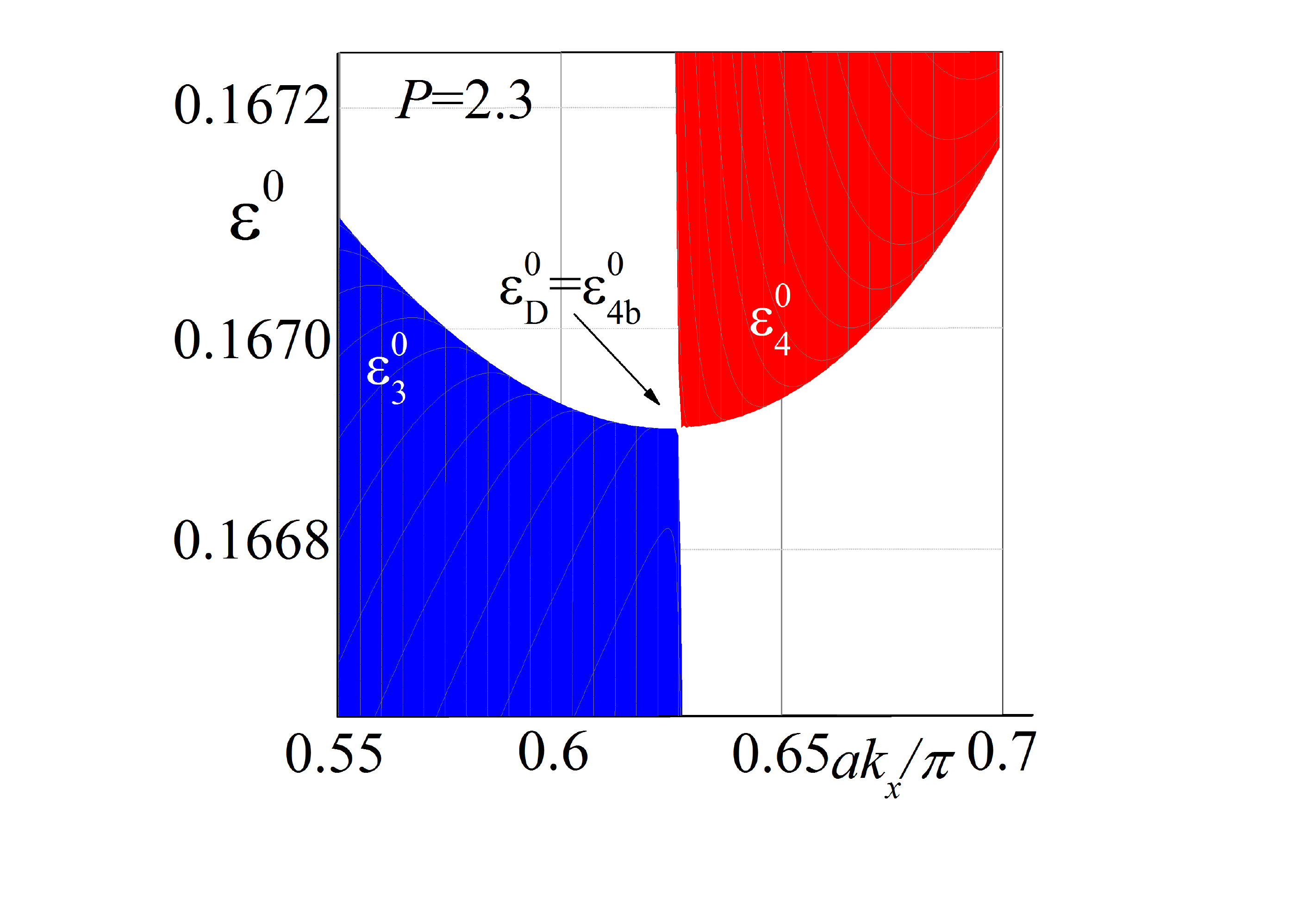}\vspace{-0.0cm}
\end{center}
\caption{
 (Color online)  Energy dispersion near the Dirac point 
({\it ``three-quarter''-Dirac point}) at $P=2.3$ from different three view points ((a), (b) and (c)). 
In (c), $0.35\pi/b\leq k_y<0.45\pi/b$. 
}
\label{fig9i}
\end{figure}

\begin{figure}[bt]
\begin{center}
\begin{flushleft} \hspace{0.5cm} (a) \end{flushleft}\vspace{-0.0cm}
\includegraphics[width=0.46\textwidth]{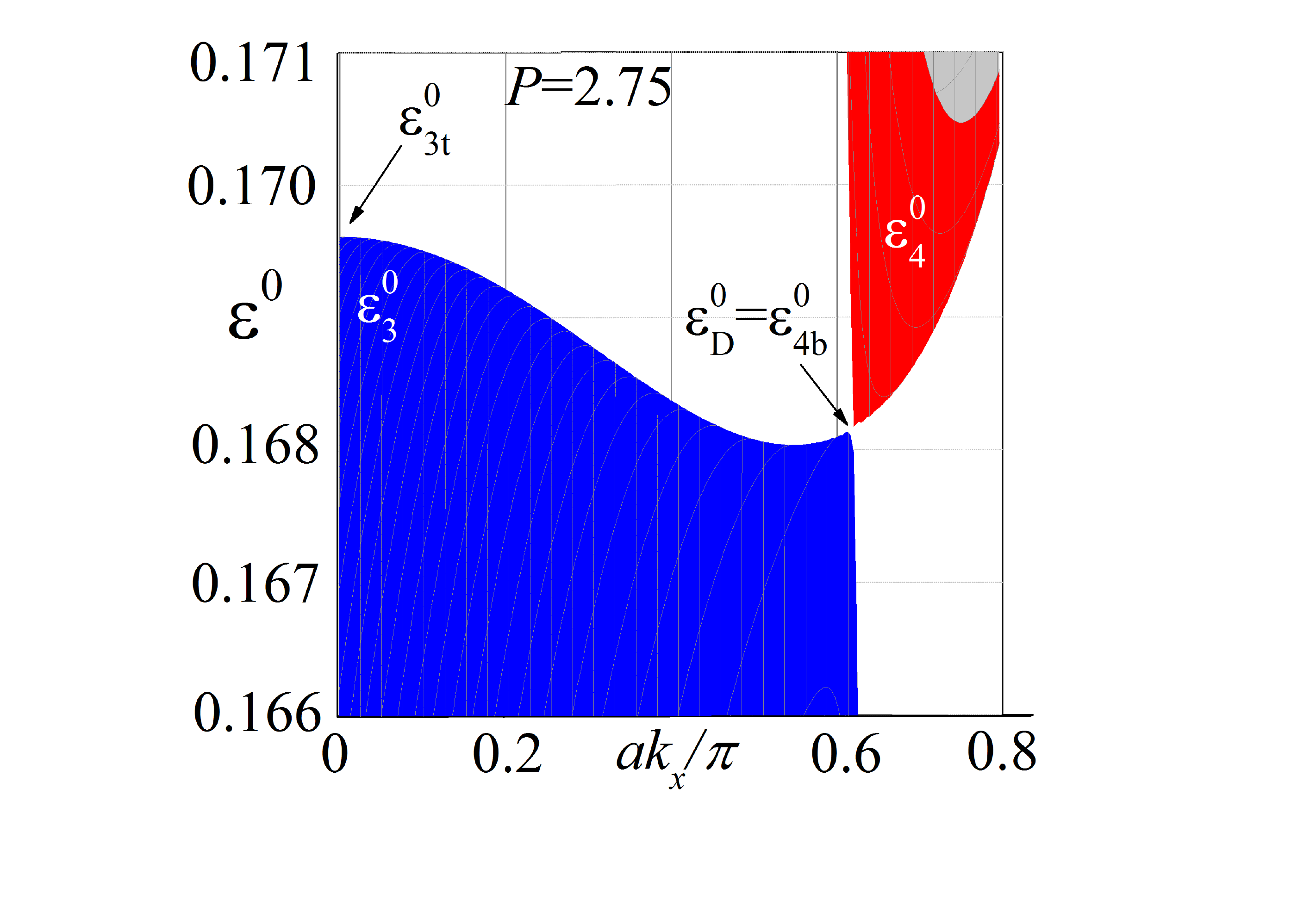}\vspace{-0.0cm}
\begin{flushleft} \hspace{0.5cm} (b) \end{flushleft}\vspace{-0.0cm}
\includegraphics[width=0.46\textwidth]{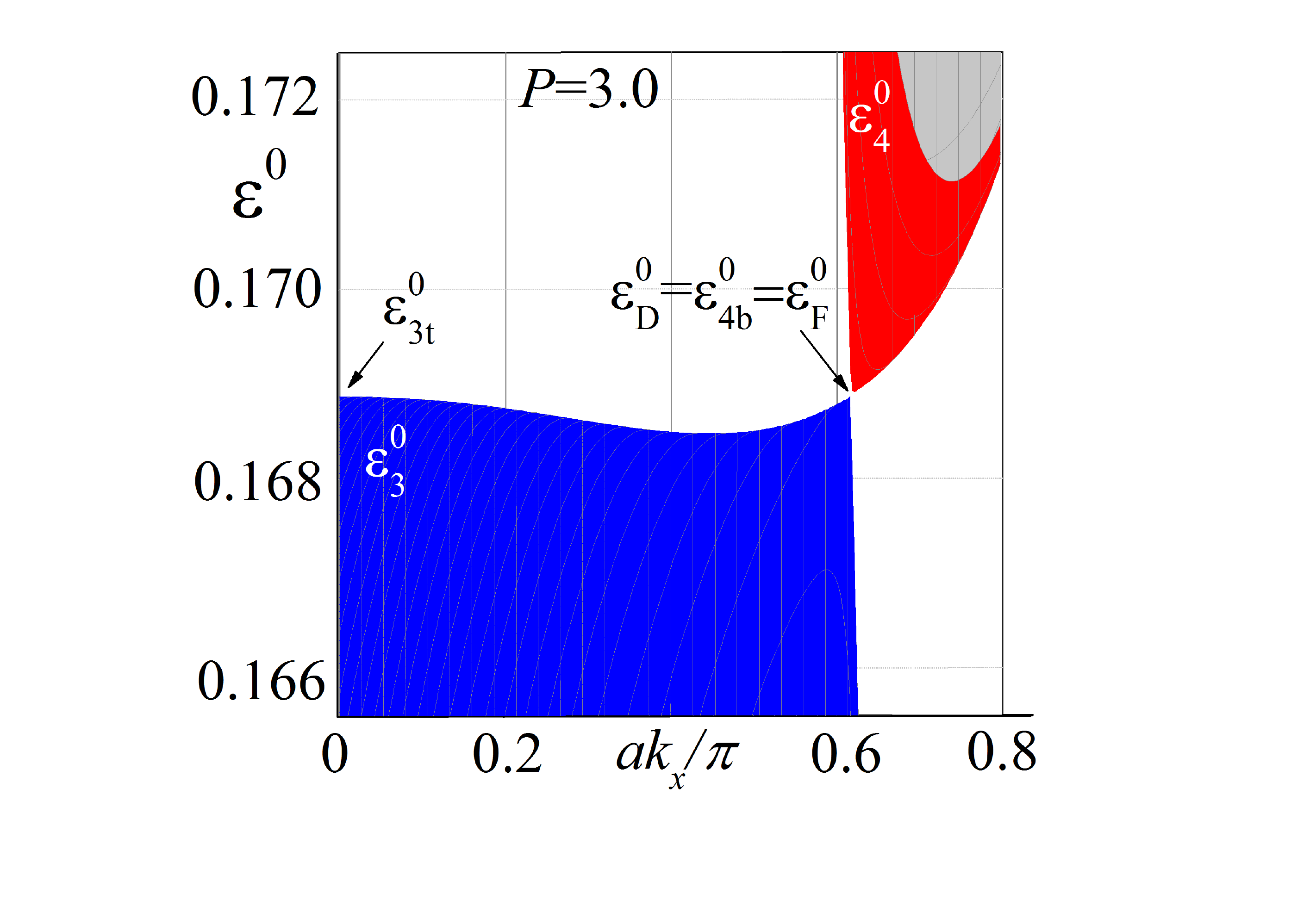}\vspace{-0.0cm}
\begin{flushleft} \hspace{0.5cm} (c) \end{flushleft}\vspace{-0.0cm}
\includegraphics[width=0.46\textwidth]{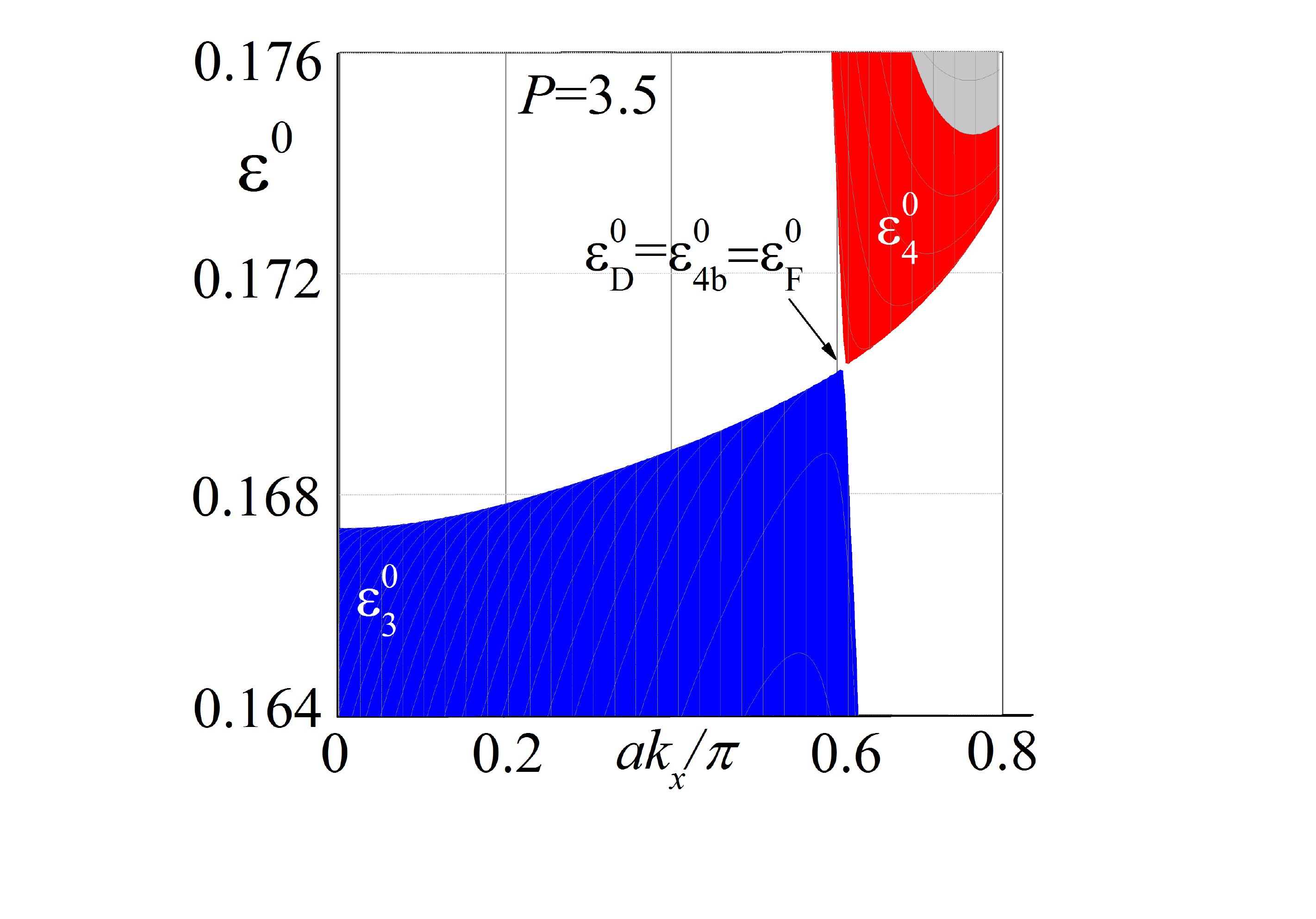}\vspace{-0.5cm}
\end{center}
\caption{
(Color online) 
Energy dispersion near the Dirac point from a view point
along the $k_y$ axis at $P=2.75$ (a) and $P=3.0$ (b), where $0.3\pi/b\leq k_y<\pi/b$, and at $P=3.5$ (c), where $0.25\pi/b\leq k_y<\pi/b$. 
}
\label{fig9f}
\end{figure}

\begin{figure}[bt]
\begin{flushleft} \hspace{0.0cm}
 \end{flushleft}\vspace{0.0cm}
\hspace{-0.0cm}\includegraphics[width=0.45\textwidth]{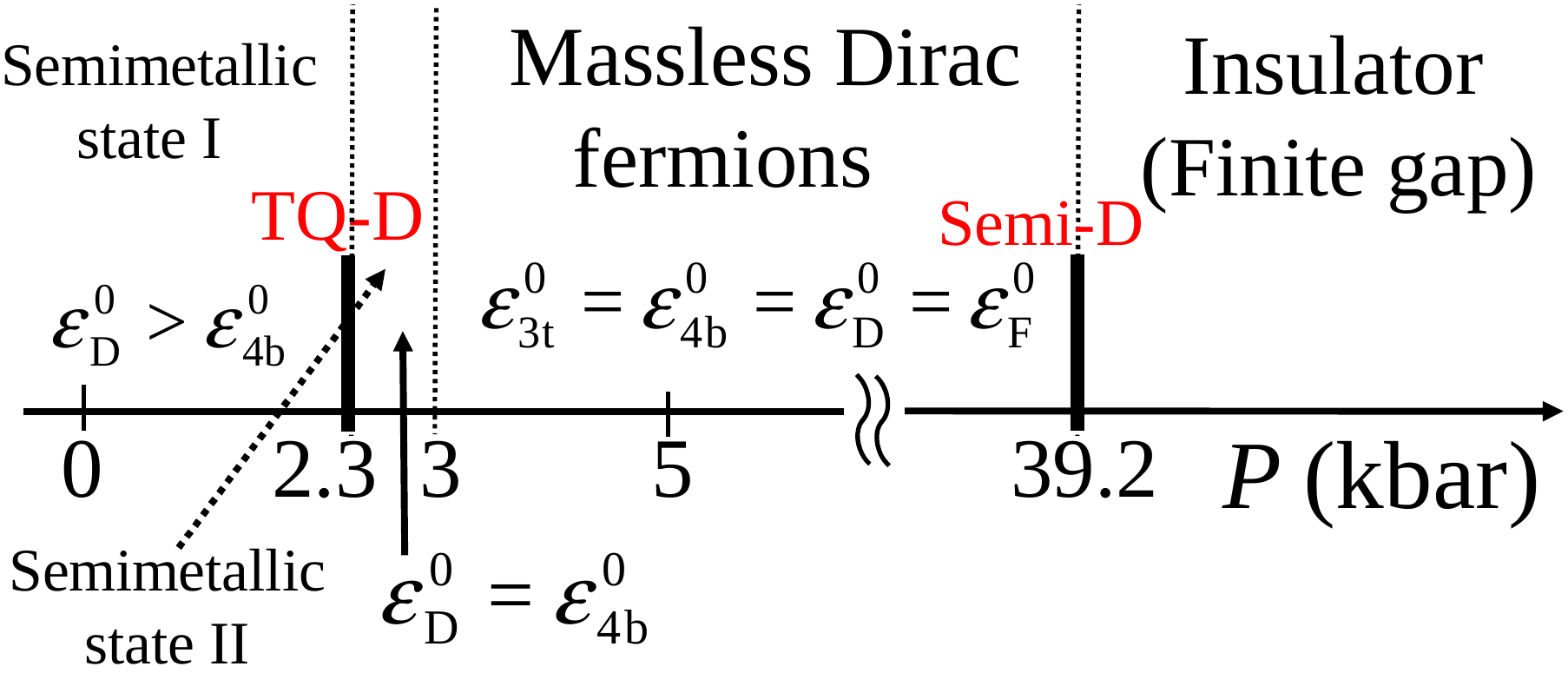}\vspace{0.0cm}
\caption{
(Color online) 
A schematic phase diagram as a function of $P$. 
In a semi-metallic phase I $(P<2.3)$, $\varepsilon_{\textrm{D}}^0>\varepsilon_{\textrm{4b}}^0$. 
In another semi-metallic phase II $(2.3< P<3.0)$, $\varepsilon_{\textrm{D}}^0=\varepsilon_{\textrm{4b}}^0$. 
}
\label{fig11b}
\end{figure}

We find interesting features of the third and fourth bands near the Fermi energy at $P\lesssim 3.0$. When $P<2.3$, the Dirac cones 
are overtilted (for example, see Fig. \ref{fig2} at $P=0$), where $\varepsilon_{\textrm{D}}^0$ at $\mathbf{k}_{\textrm{D}}$ is 
larger than $\varepsilon_{\textrm{4b}}^0$ at $\mathbf{k}_{\textrm{4b}}$, 
which can be also seen in Fig. \ref{fig9} (a). 
As $p$ increases, $\mathbf{k}_{\textrm{4b}}$ and $\mathbf{k}_{\textrm{D}}$ move on $k_x$-$k_y$ plane and 
these wave numbers coincide at $P=2.3$, as shown in Fig. \ref{fig9} (b). 
In this case we have to take into account of higher order terms in energy dispersion at Dirac points, and 
the quadratic term in one direction makes $\varepsilon_{\textrm{D}}^0$ at the Dirac points 
to be the global minima of the  
fourth band (i.e., $\varepsilon_{\textrm{D}}^0=\varepsilon_{\textrm{4b}}^0$, see Figs.~\ref{fig9} (a) and ~\ref{fig9i}). 
On the other hand, $\varepsilon_{\textrm{D}}^0$ is not the local maximum of the third band, as shown in Fig.~\ref{fig9i}. 
At $P=2.3$ the Dirac cones are critically tilted, which 
have a quadratic dispersion in one direction and linear dispersions in the other three directions. In this sense, we name the Dirac cones at $P=2.3$ \textit{``three-quarter''-Dirac cones} and 
these touching points 
\textit{``three-quarter''-Dirac points} 
[$\pm\mathbf{k}_{\textrm{tq}}=\pm\mathbf{k}_{\textrm{D}}\simeq\pm(0.6270\pi/a, 0.4058\pi/b)$]. At $2.3<P<3.0$, $\varepsilon_{\textrm{D}}^0$ is the global minimum of the fourth band and the local maximum of the third band, as shown in Fig. \ref{fig9f} (a) at $P=2.75$. 
At $P=3.0$ the Dirac cone of the third band is almost laid, as shown in Figs. \ref{fig4} and \ref{fig9f} (b). 
Since the density of states near the Dirac points are proportional to 
$|\varepsilon^0 -\varepsilon_{\textrm{D}}^0|$ 
and the density of states near the global maximum of the third band are constant, we obtain at $2.3 < P < P_c=3.0$ (see Appendix \ref{AppendixB2}) 
\begin{equation}
\varepsilon_{\textrm{3t}}^0-\varepsilon_{\textrm{F}}^0 \propto (P_c-P)^2, \label{P_c}
\end{equation}
which can be seen in Fig.~\ref{fig9} (a).

At $3.0 < P <39.2$, $\varepsilon^{0}_{\textrm{D}}$ is the global minimum of the fourth band and the global maximum of the third band, i.e., massless Dirac fermions are realized\cite{Katayama2006}, as shown in Fig.~\ref{fig9f} (c) at $P = 3.5$ and Fig.~\ref{fig6} at $P = 5.0$. 
Three bands from the bottom are fully occupied and the fourth band is completely empty at $T=0$.

Two Dirac points move and merge\cite{Suzumura2013} at a semi-Dirac point ($\Gamma$ point) at $P=39.2$, as shown in Fig. \ref{fig10}. 
At $P>39.2$, 
the energy gap becomes finite\cite{Suzumura2013}. The top of the third band and the bottom of the fourth band are approximately given by the anisotropic parabolic bands\cite{Montambaux2009_prb,Montambaux2009}, where massive Dirac fermions are realized, as shown in Fig. \ref{fig11} at $P=50$.

Based on these results, we give a schematic phase diagram as a function of $P$ in Fig. \ref{fig11b}. The semi-metallic state is divided to two phases (I and II) at $P<2.3$ 
and at $2.3<P<3.0$. 

\section{Energy in magnetic field}

We obtain the energy in the magnetic field as eigenvalues of a $4q\times 4q$ matrix, when the magnetic flux in the unit cell ($\Phi$) 
is a rational number in the unit of the flux quantum ($\phi_0=2\pi\hbar c/e\simeq 4.14\times 10^{-15}$Tm$^2$), i.e., 
\begin{equation}
h=\frac{\Phi}{\phi_0}=\frac{p}{q},
\label{eq_pq}
\end{equation}
where $p$ and $q$ are integers. This is explained in Appendix \ref{AppendixB}. 
Hereafter, we represent the magnetic field by $h$. 
Since $a\simeq 9.211$~\AA \ and 
$b\simeq 10.85$\AA \ in $\alpha$-(BEDT-TTF)$_2$I$_3$\cite{review}, 
$h=1$ corresponds to $H\simeq 4.14\times 10^{3}$~T. 
The lowest magnetic field studied in this paper
is $h=2/1901$ i.e., $H\simeq 4.36$~T.

\begin{figure}[bt]
\begin{flushleft} \hspace{0.0cm}(a) \end{flushleft}\vspace{-0.8cm}
\includegraphics[width=0.42\textwidth]{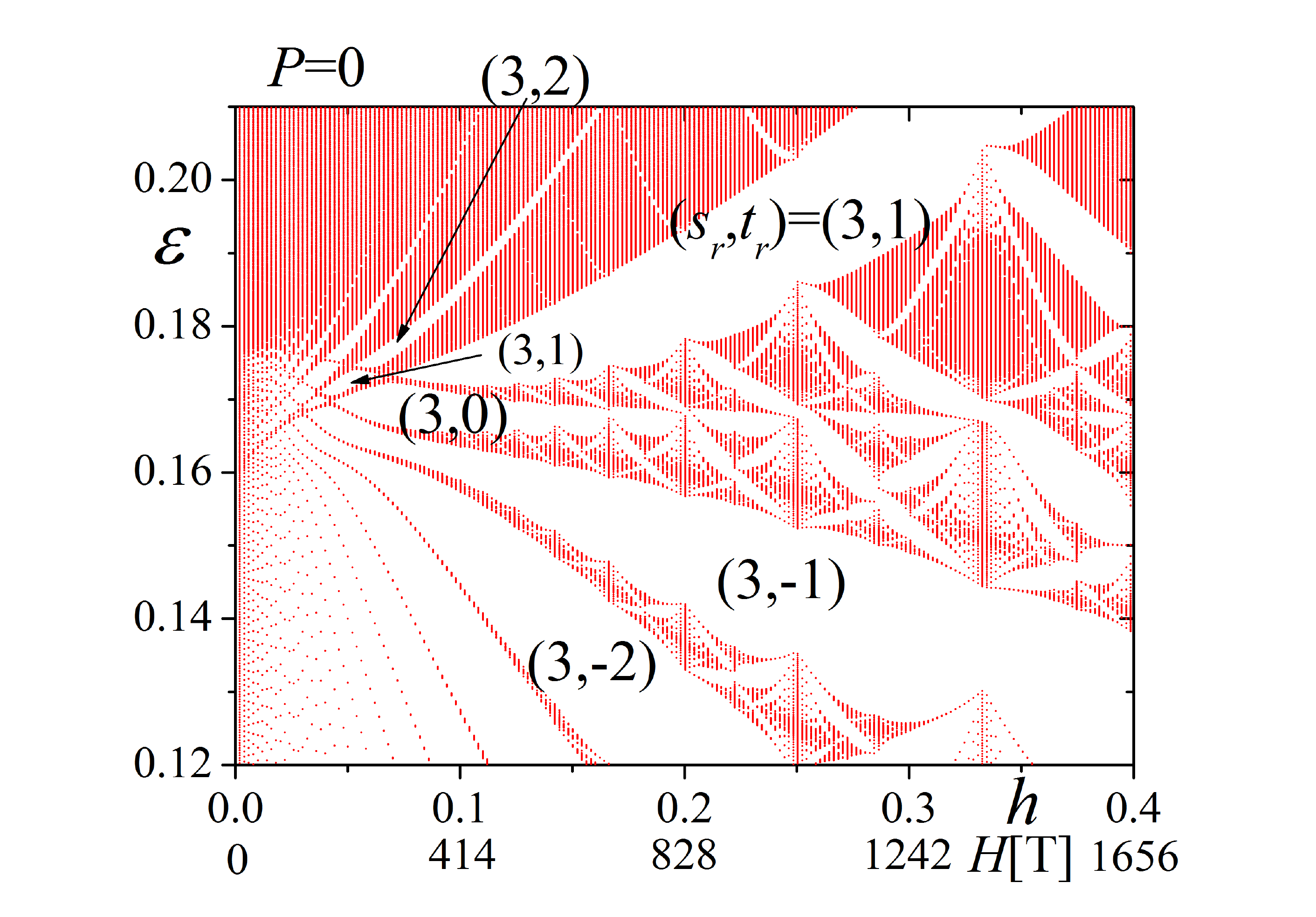}\vspace{-0.5cm}
\begin{flushleft} \hspace{0.0cm}(b) \end{flushleft}\vspace{-0.8cm}
\includegraphics[width=0.42\textwidth]{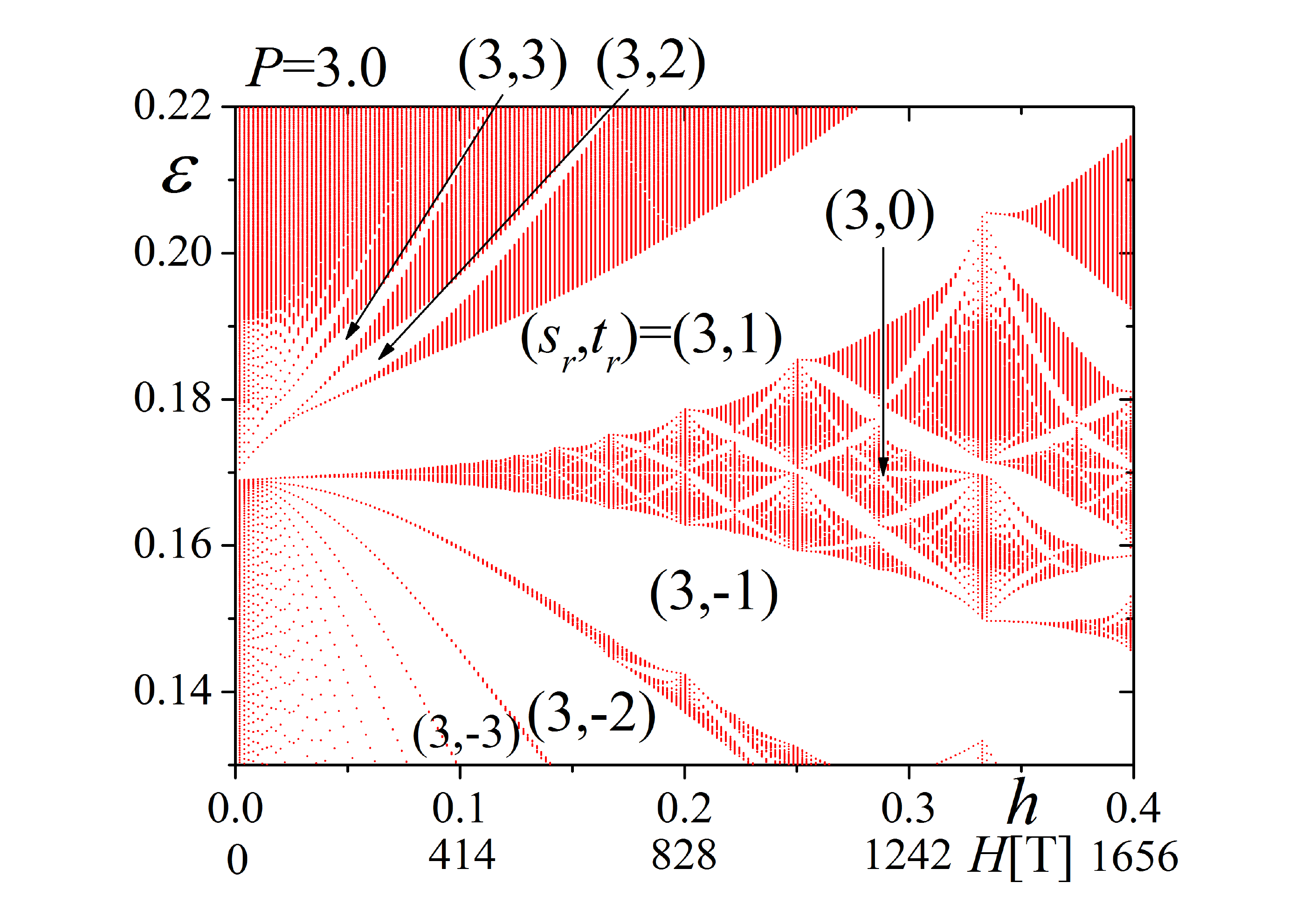}\vspace{-0.5cm}
\begin{flushleft} \hspace{0.0cm}(c) \end{flushleft}\vspace{-0.8cm}
\includegraphics[width=0.42\textwidth]{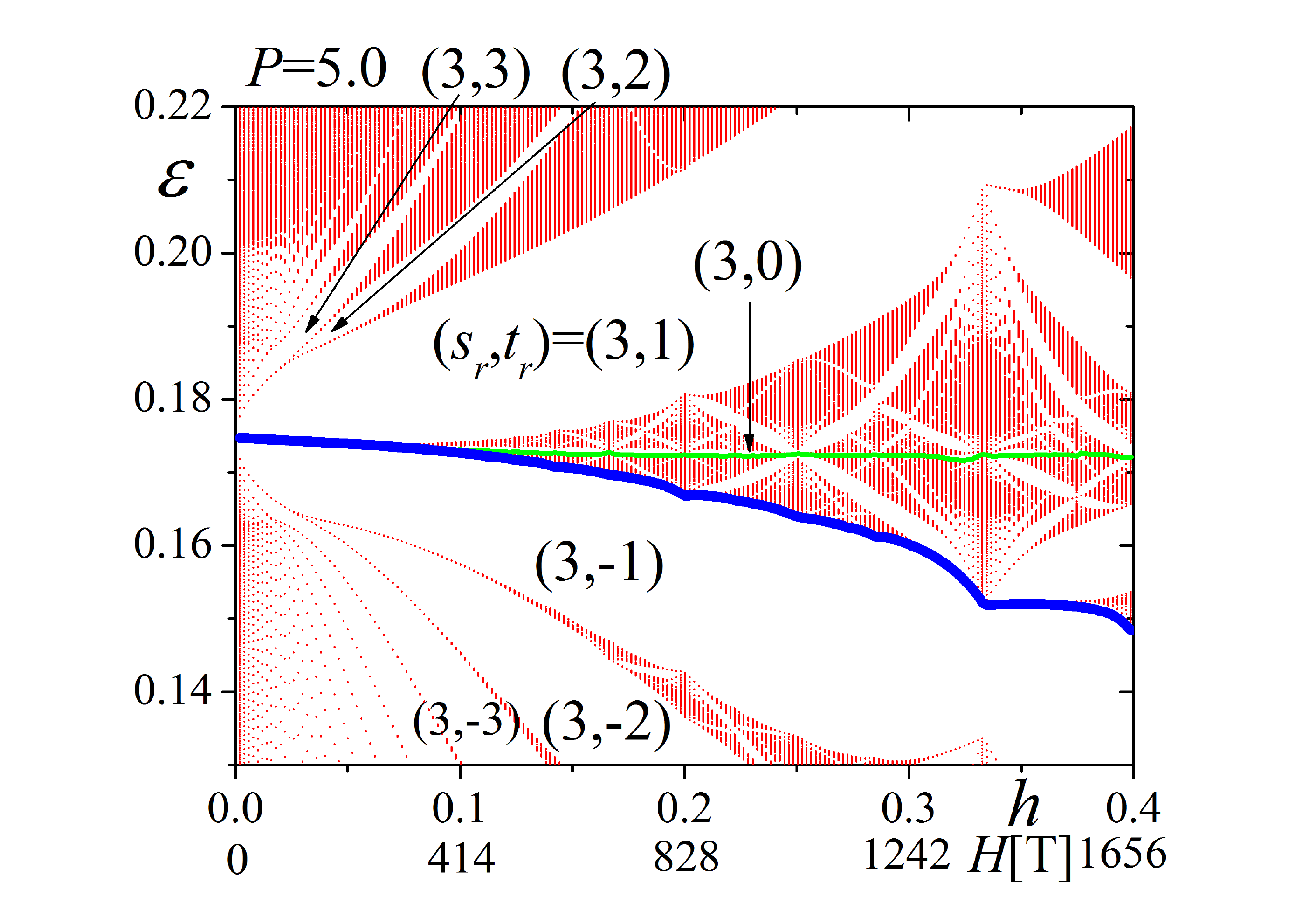}\vspace{-0.5cm}
\begin{flushleft} \hspace{0.0cm}(d) \end{flushleft}\vspace{-0.8cm}
\includegraphics[width=0.42\textwidth]{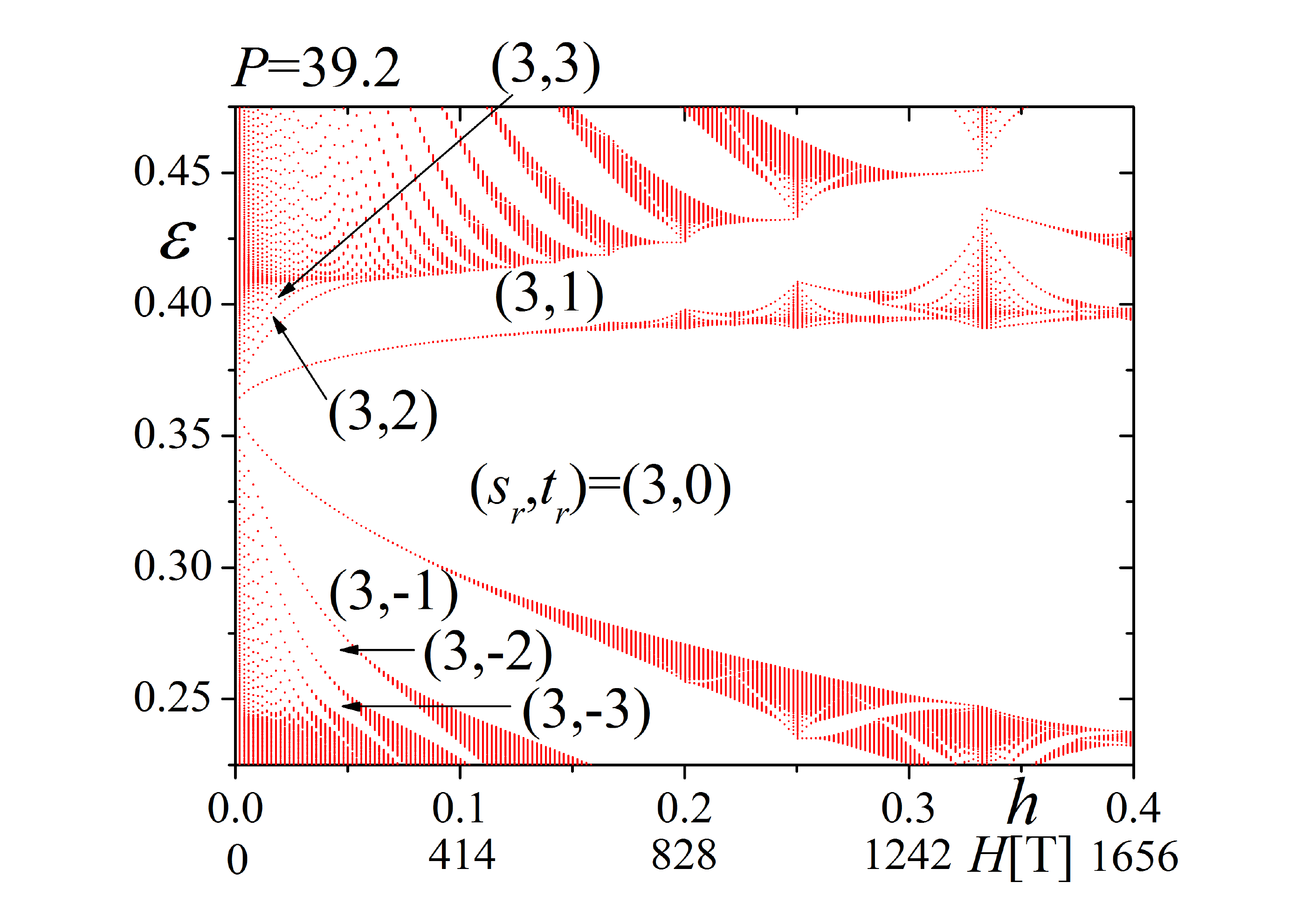}\vspace{-0.5cm}
\caption{
Energies as a function of $h$ for $P=0$ (a), $P=3.0$ (b), $P=5.0$ (c) and $P=39.2$ (d). 
We take $h=\frac{2m}{499}$ ($m=1, 2, 3, \cdots, 99$) and $h=\frac{2(2m-1)}{998}$ 
 ($m=1, 2, 3, \cdots, 100$), namely, where $p/q=2/998, 2/499, 6/998, 4/499, 
\cdots, 198/499, 398/998$. 
}
\label{fig13}
\end{figure}

We show the energies as a function of $h$ (the Hofstadter butterfly diagrams) at $P=0$, $5.0$ and $39.2$ in Fig. \ref{fig12} in Appendix \ref{AppendixB}. 
The energies near the Fermi energy at $P=0$, $3.0$, $5.0$ and $39.2$ are shown in Fig. \ref{fig13}. If $q$ is small, each band may be broadened, and we have to 
consider the $\mathbf{k}$-dependence of the energy. 
If $q$ is large, the widths of $4q$ bands become narrow, and the
$\mathbf{k}$-dependences of each band can be neglected, as long as the contour line of the energy 
in the wave-number space is closed at $h=0$. When the contour line of the energy 
in the wave-number space is open, which is the case for $\varepsilon^0 \simeq 0.175$ at $P=0$ 
(Fig.~\ref{fig3} (a)), we have to consider the $\mathbf{k}$-dependences in each band. In fact, the energies are broadening above $\varepsilon\gtrsim 0.175$, as shown in Figs.~\ref{fig13} (a) and \ref{fig15}. There are $4q$ bands, some of which may overlap each other.

When the chemical potential 
is in the energy gap in the magnetic field, Hall conductance is quantized. The quantized value is obtained as a first Chern number\cite{TKNN,Kohmoto_1985,Kohmoto_1989}. 
It is also given as a solution of the Diophantine equation\cite{Kohmoto_1985,Kohmoto_1989},
\begin{eqnarray}
 r = q s_r + p t_r, 
\label{eqDiophantine}
\end{eqnarray}
where $p$ and $q$ are given in Eq.~(\ref{eq_pq}), 
$r$ is the number of energy bands below the chemical potential, and $s_r$ and $t_r$ are integers obtained in this Diophantine equation. Although $s_r$ and $t_r$ are not given uniquely from Eq.~(\ref{eqDiophantine}), we can uniquely assign integers ($s_r$ and $t_r$) in the energy gaps in the Hofstadter butterfly diagrams. 
In this system, $s_r$ and $t_r$ are shown in Fig.~\ref{fig13}.

%

\subsection{Semi-metallic state I at $P< 2.3$}

\begin{figure}[bt]
\includegraphics[width=0.53\textwidth]{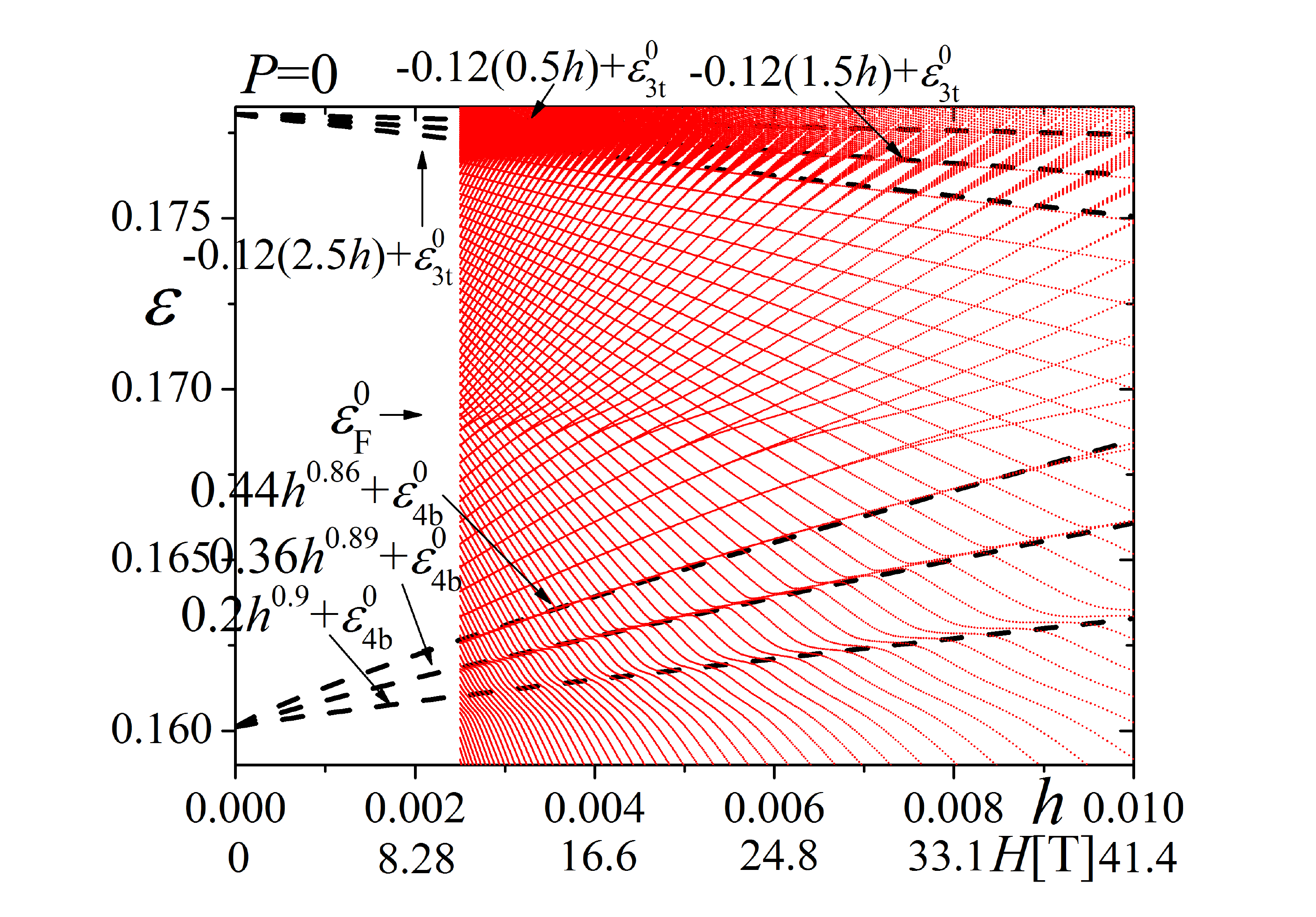}\vspace{-0.0cm}
\caption{
Energies near the Fermi energy as a function of $h$ at $P=0$, 
where $\varepsilon_{\rm F}^0\simeq  0.16925$. 
We take $h=2/q$ ($q=200, 201, \cdots$, $799, 800$). 
}
\label{fig15}
\end{figure}

The energy near the Fermi energy at $P=0$ at the relatively low magnetic field is shown in Fig. \ref{fig15}. We fit the energy levels for the fourth band starting from $h=0$ and $\varepsilon=\varepsilon^0_{\textrm{4b}}$ as
\begin{equation}
  \varepsilon_n-\varepsilon^0_{\textrm{4b}} \propto h^{\delta_n}, \label{delta_n}
\end{equation}
where $\delta_n=0.9$, $0.89$ and $0.86$, as shown in Fig. \ref{fig15}. 
Those Landau levels are not linear in $h$. If a fitting could be performed 
at very low magnetic fields, $\delta_n=1$ would be obtained 
due to the parabolic dispersion of the fourth band around $\varepsilon^0_{\textrm{4b}}$ (see Fig. \ref{fig2} (c)). 
However, $h$ is not sufficiently low in Fig. \ref{fig15}. Therefore, the deviation from the parabolic dispersion around $\varepsilon^0_{\textrm{4b}}$ 
makes the fitting parameter $\delta_n$ to be smaller than 1.

Two upward-sloping Landau levels starting from $\varepsilon^0_{\textrm{4b}}$ in Fig. \ref{fig15} are almost degenerate at low $h$ and below $\varepsilon^0_{\textrm{F}}$. They are smoothly separated near $\varepsilon^0_{\textrm{F}}$. The lift of the degeneracy of the Landau levels around $\varepsilon^0_{\textrm{F}}$ (Fig. 12) is understood 
semiclassically as follows. The fourth band has minima $\varepsilon^0_{\textrm{4b}}$ at 
$\pm \mathbf{k}_{4\textrm{b}}$ (see Fig. \ref{fig2} and Fig. \ref{fig3} (a)). 
When the energy is located between $\varepsilon^0_{\textrm{4b}}$ and the energy at the saddle point ($\mathbf{k}=(\pi/a,0)$) of the fourth band, as seen in Fig. \ref{fig3} (a), the contour line of energy in the fourth band consists of two closed regions (two electron pockets). Two minima are considered to be independent, resulting in the degenerated Landau levels. When the energy is larger than that at the saddle point, the contour line of energy in the fourth band is one closed loop, making no degeneracy of Landau levels. The energy at the saddle point is close to $\varepsilon^0_{\textrm{F}}$. 
The similar situation has been studied by Montambaux, Piechon, Fuchs and Goerbig\cite{Montambaux2009_prb,Montambaux2009}.

The Landau levels for the third band are fitted by
\begin{equation}
  \varepsilon_n = -0.12 \left( n+ \frac{1}{2} \right) h + \varepsilon^0_{\textrm{3t}}, \ n=0, 1, 2, \cdots 
\end{equation}
which are depicted by black broken lines in Fig.~\ref{fig15}. 
These Landau levels are understood as the Landau quantization for a free hole pocket centered at 
$\mathbf{k}=\mathbf{k}_{\textrm{3t}}$.

\begin{figure}[bt]
\begin{flushleft} \hspace{0.0cm}(a) \end{flushleft}\vspace{-0.0cm}
\includegraphics[width=0.49\textwidth]{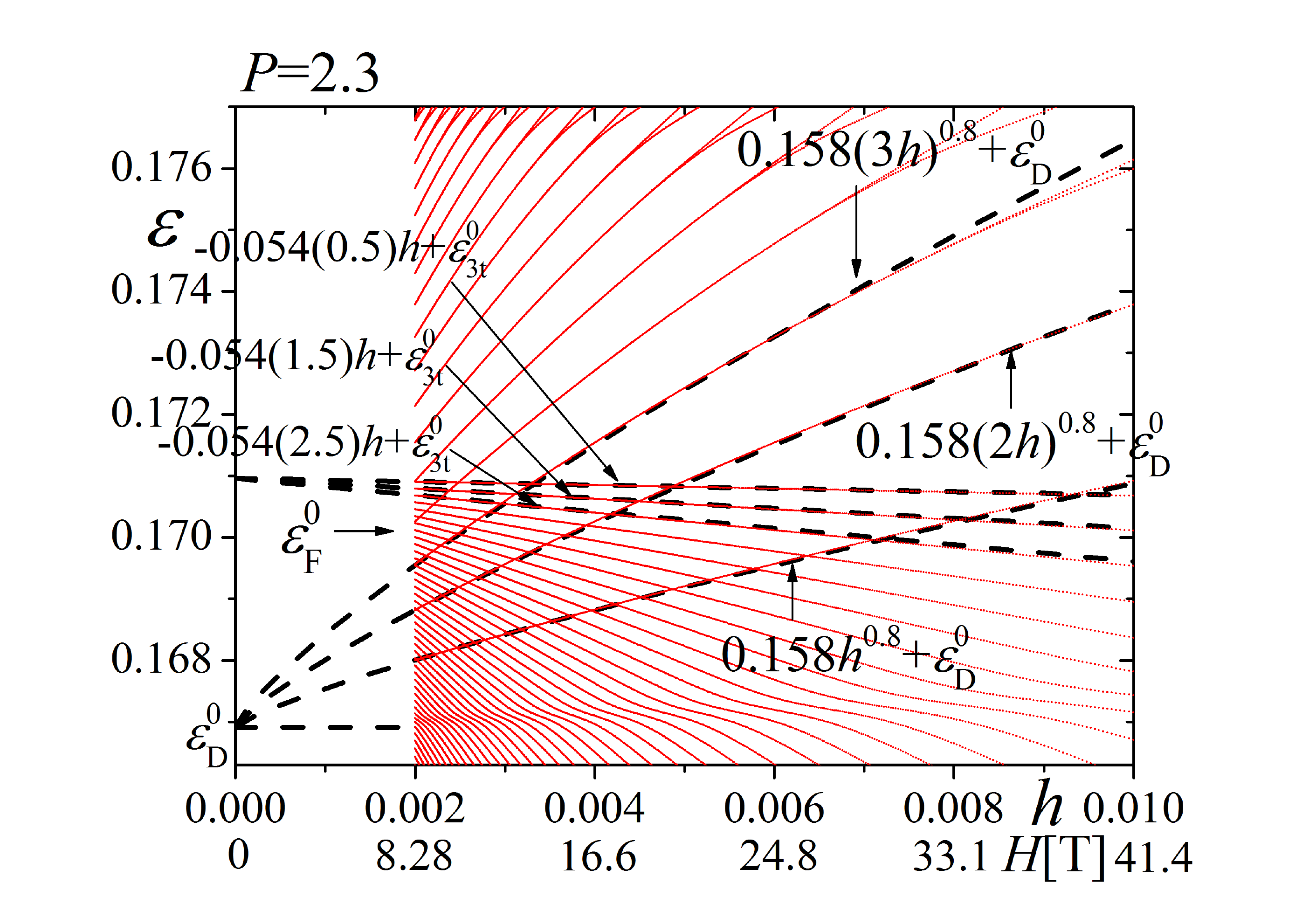}\vspace{-0.2cm}
\begin{flushleft} \hspace{0.0cm}(b) \end{flushleft}\vspace{-0.0cm}
\includegraphics[width=0.49\textwidth]{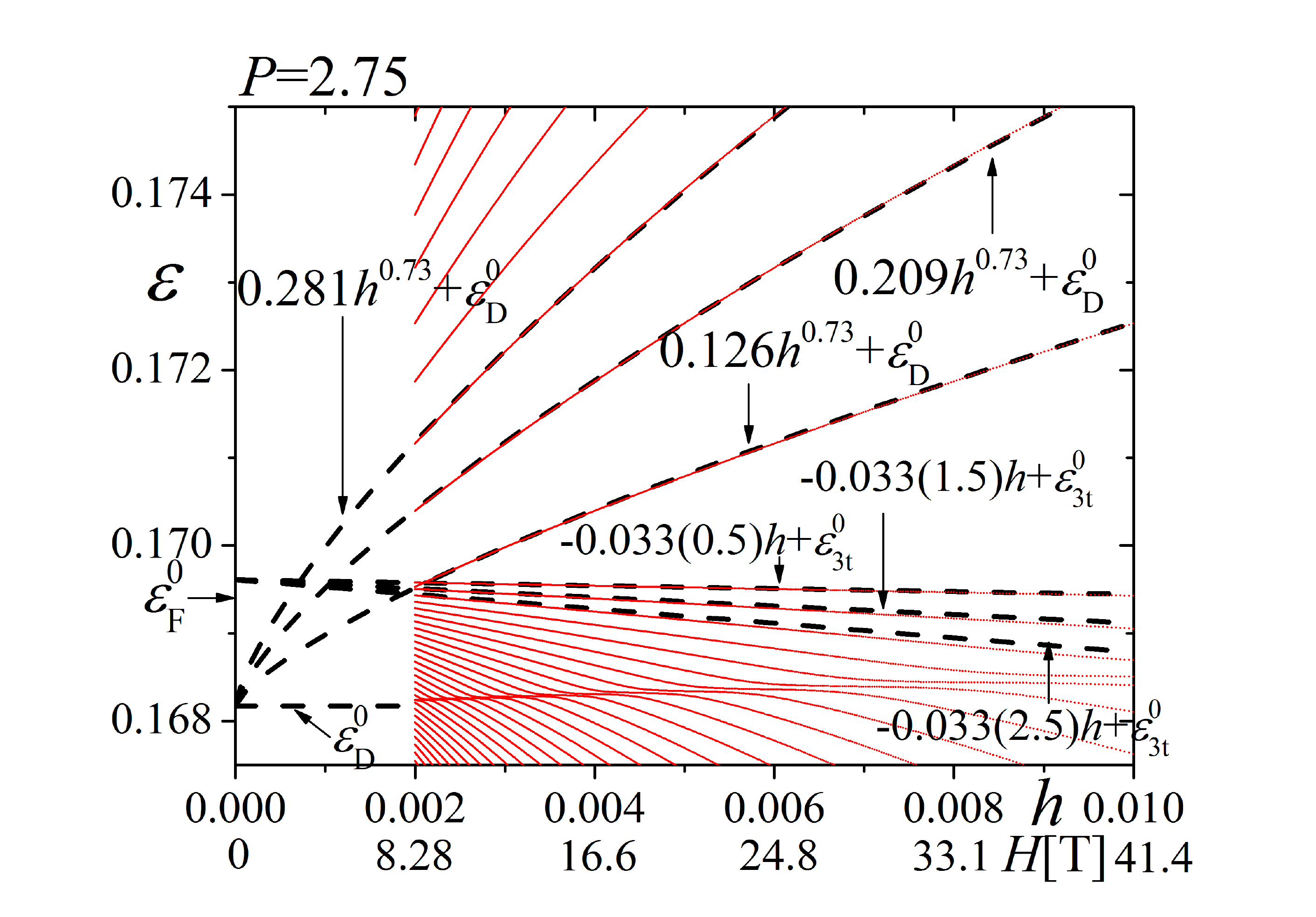}\vspace{-0.2cm}
\begin{flushleft} \hspace{0.0cm}(c) \end{flushleft}\vspace{-0.0cm}
\includegraphics[width=0.49\textwidth]{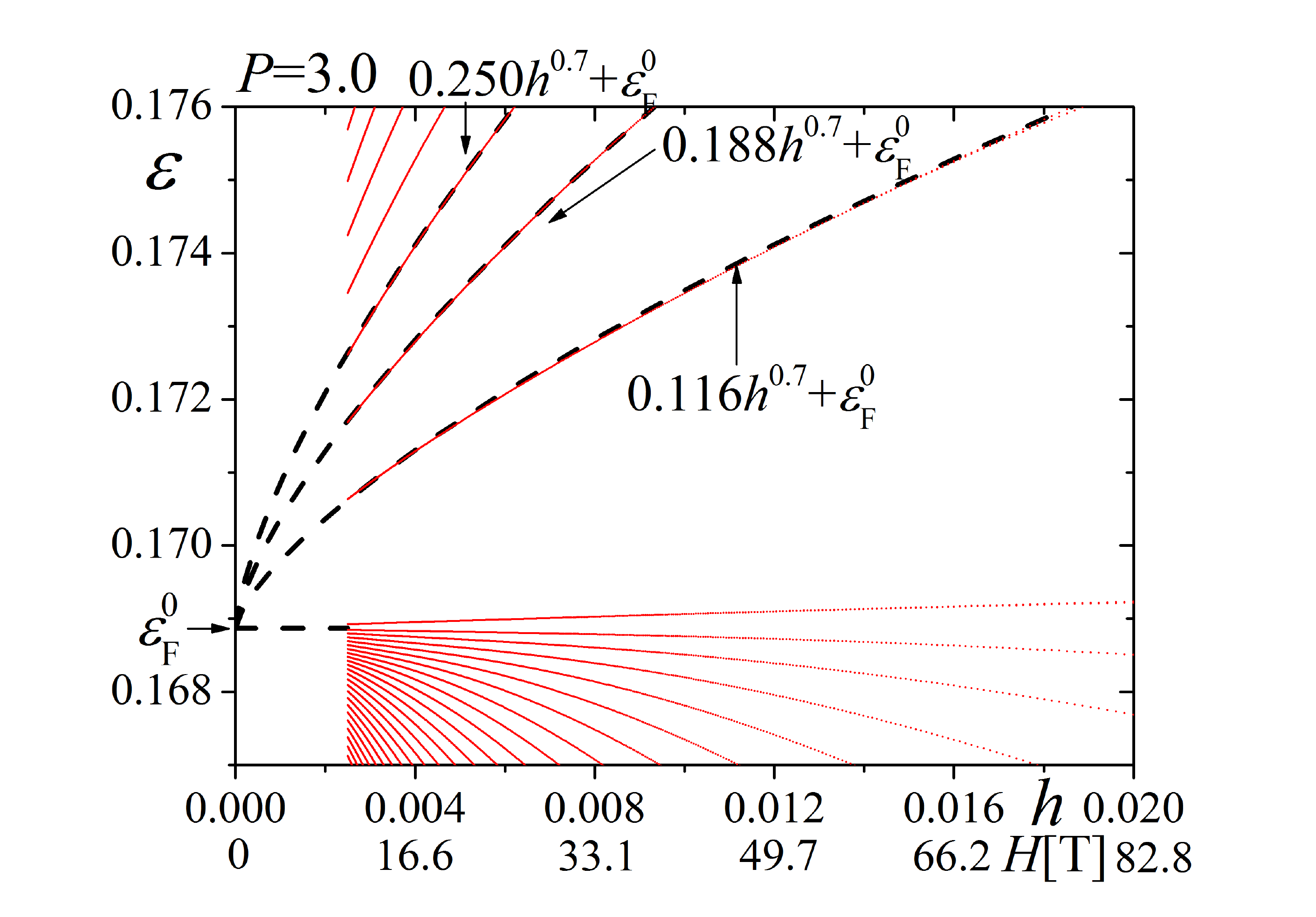}\vspace{-0.2cm}
\caption{
Energies near the Fermi energy as a function of $h$ for $P=2.3$ (a), $P=2.75$ (b) and $P=3$ (c). We choose $p=2$ and $200\leq q \leq 1000$ ($q=200, 201, \cdots, 999, 1000$) for (a) and (b) and $p=2$ and $100\leq q \leq 800$ ($q=100, 101, \cdots, 799, 800$) for (c), respectively. 
}
\label{fig16}
\end{figure}

\subsection{``three-quarter''-Dirac points at $P=2.3$}

In order to write the energy near {\it ``three-quarter''-Dirac points} at $P=2.3$  
we take a model (see Appendix \ref{Appendix_area}) as
\begin{equation}
{\cal H}^0_{\rm tqD}= \left( \begin{array}{cc}
-w_xq_x+\alpha^{\prime}_2q_x^2& w_xq_x+\alpha^{\prime\prime}_2q_x^2-iw_yq_y \\
w_xq_x+\alpha^{\prime\prime}_2q_x^2+iw_yq_y& \ -w_xq_x+\alpha^{\prime}_2q_x^2 \\
 \end{array} \right), \label{HM_0a}
\end{equation}
where $\mathbf{q}=0$ is corresponding to $\mathbf{k}_{\rm tq}$. 
The eigenvalues are obtained as 
\begin{align}
\varepsilon^0_{\rm tqD\pm} (\mathbf{q})
&= -w_xq_x+ \alpha^{\prime}_2 q_x^2 \nonumber \\
&\pm \sqrt{(w_xq_x+\alpha^{\prime\prime}_2q_x^2)^2+(w_yq_y)^2}.\label{tq-D}
\end{align}
The fourth band and the third band correspond to 
$\varepsilon^0_{\rm tqD+}(\mathbf{q})$ and $\varepsilon^0_{\rm tqD-}(\mathbf{q})$, respectively. 
The energy $\varepsilon^0_{\rm tqD+}(\mathbf{q})$ around $\mathbf{q}=(0,0)$ is linear in three directions 
$\mathbf{q}=\pm(0,|q_y|)$ and $\mathbf{q}=(-|q_x|,0)$ but quadratic in one direction $\mathbf{q}=(|q_x|,0)$, when $w_x>0$. Therefore this model represents the dispersion near a {\it ``three-quarter''-Dirac point}, as shown in Fig. \ref{fig9i}. 
We obtain the area enclosed by the constant energy line at 
$\varepsilon^0_{\rm tqD+} (\mathbf{q}) = \varepsilon$ to be
\begin{equation}
  A(\varepsilon) \simeq \frac{\sqrt{2 w_x}\pi}{4 w_y} \alpha_2^{-\frac{3}{4}} \varepsilon^{\frac{5}{4}}, \label{area}
\end{equation}
where $\alpha_2=\alpha^{\prime}_2+\alpha^{\prime\prime}_2$, in the limit of $\varepsilon \to +0$. Eq. (\ref{area}) is derived in Appendix \ref{Appendix_area}. By using Eq. (\ref{area}) and 
the semiclassical quantization rule of Eq. (\ref{eqquantization0}) with $\gamma=0$, 
we obtain semiclassically the Landau levels for {\it ``three-quarter''-Dirac cones} in the fourth band as 
\begin{equation}
 \varepsilon_n^{\textrm{(three-quarter-Dirac)}}-\varepsilon_{\textrm{D}}^0 \propto \left( n h \right)^{\frac{4}{5}}.\label{4/5}
\end{equation}

The Landau levels starting from $\varepsilon=\varepsilon^0_{\textrm{D}}$ at $h=0$ 
are fitted as 
\begin{align}
\varepsilon_0 &
=\varepsilon^0_{\textrm{D}}, \label{epsilon_0}\\
 \varepsilon_1 &= 0.158 h^{0.8}+\varepsilon^0_{\textrm{D}}, \label{epsilon_1}\\
 \varepsilon_2 &= 0.158 (2 h)^{0.8}+\varepsilon^0_{\textrm{D}},  \label{epsilon_2}\\
 \varepsilon_3 &= 0.158 (3 h)^{0.8}+\varepsilon^0_{\textrm{D}}, \label{epsilon_3}
\end{align}
as shown in Fig.~\ref{fig16} (a), which are consistent with the semiclassical quantization of 
the energy (Eq. (\ref{4/5})). 
The level, $\varepsilon_0$, is not as clearly seen as $\varepsilon_1$, $\varepsilon_2$, and $\varepsilon_3$. 
The reason for the ambiguous energy levels of $n=0$ in Fig.~\ref{fig16} (a) might be the mixing of the $n=0$ Landau level for the fourth band and the Landau levels for the third band with a negligible tunneling barrier at {\it ``three-quarter''-Dirac points}.

When the magnetic field is low, the Landau levels for the third band are approximately written by 
\begin{equation}
  \varepsilon_n = -0.054 \left( n+ \frac{1}{2} \right) h + \varepsilon^0_{\textrm{3t}}, \ n=0, 1, 2
\end{equation}
which comes from a hole pocket centered at $\mathbf{k}=\mathbf{k}_{\textrm{3t}}$.

\subsection{Semi-metallic state II at $2.3 < P< 3.0$}

At $2.3 < P < 3.0$, $\varepsilon^0_{\textrm{D}}$ is the global minimum of the fourth band but only the local maximum of the third band. The global maximum of the third band, $\varepsilon^0_{\textrm{3t}}$, is obtained at $\mathbf{k}_{3\textrm{t}}=(0, \pi/a)$. The Fermi energy, $\varepsilon^0_{\textrm{F}}$, is between $\varepsilon^0_{\textrm{D}}$ 
and $\varepsilon^0_{\textrm{3t}}$. We defined this state as semi-metallic state II (see Fig. \ref{fig11b}).

At $P=2.75$ the Landau levels for the fourth band are fitted by 
\begin{equation}
\varepsilon_n-\varepsilon_{\textrm{D}}^0 \propto h^{0.73}, \label{0.73}
\end{equation}
as shown in Fig.~\ref{fig16} (b). The fitting parameter (the power of $h$) is obtained to be 0.73, which is different from 0.8 expected in the case of the {\it ``three-quarter''-Dirac point} at $P=2.3$. The effect of the finite linear term in one direction, which is zero in the case of the {\it ``three-quarter''-Dirac point}, is not large enough 
to make the fitting parameter to be 0.5 in the region of the magnetic field in Fig.~\ref{fig16} (b). 


The Landau levels for the third band (Fig.~\ref{fig16} (b)) are fitted by 
\begin{equation}
  \varepsilon_n = -0.033 \left( n+ \frac{1}{2} \right) h + \varepsilon^0_{\textrm{3t}}, \ n=0, 1, 2, \cdots
\end{equation}
which is understood as the Landau quantization of a free hole pocket.

The energies as a function of a magnetic field are changed smoothly as a pressure $P$ is changed in the semi-metallic state II (Figs.~\ref{fig16} (a), (b) and (c)). The fitting parameters (the power in $h$) for the quantized energy in the fourth band are changed continuously from $4/5$ ({\it ``three quarter''-Dirac point}) to smaller values, while the quantized energies in the third band are well fitted by the Landau levels for a free hole band, as long as the quantized energy is larger than the energy at the Dirac point. The quantization of the energy of the third band at $P=3.0$ is discussed in the following subsection.

\subsection{At the critical pressure $P_c=3.0$}
\label{P=3.0}

\begin{figure}[bt]
\begin{flushleft} \hspace{0.0cm}(a) \end{flushleft}\vspace{-0.0cm}
\includegraphics[width=0.5\textwidth]{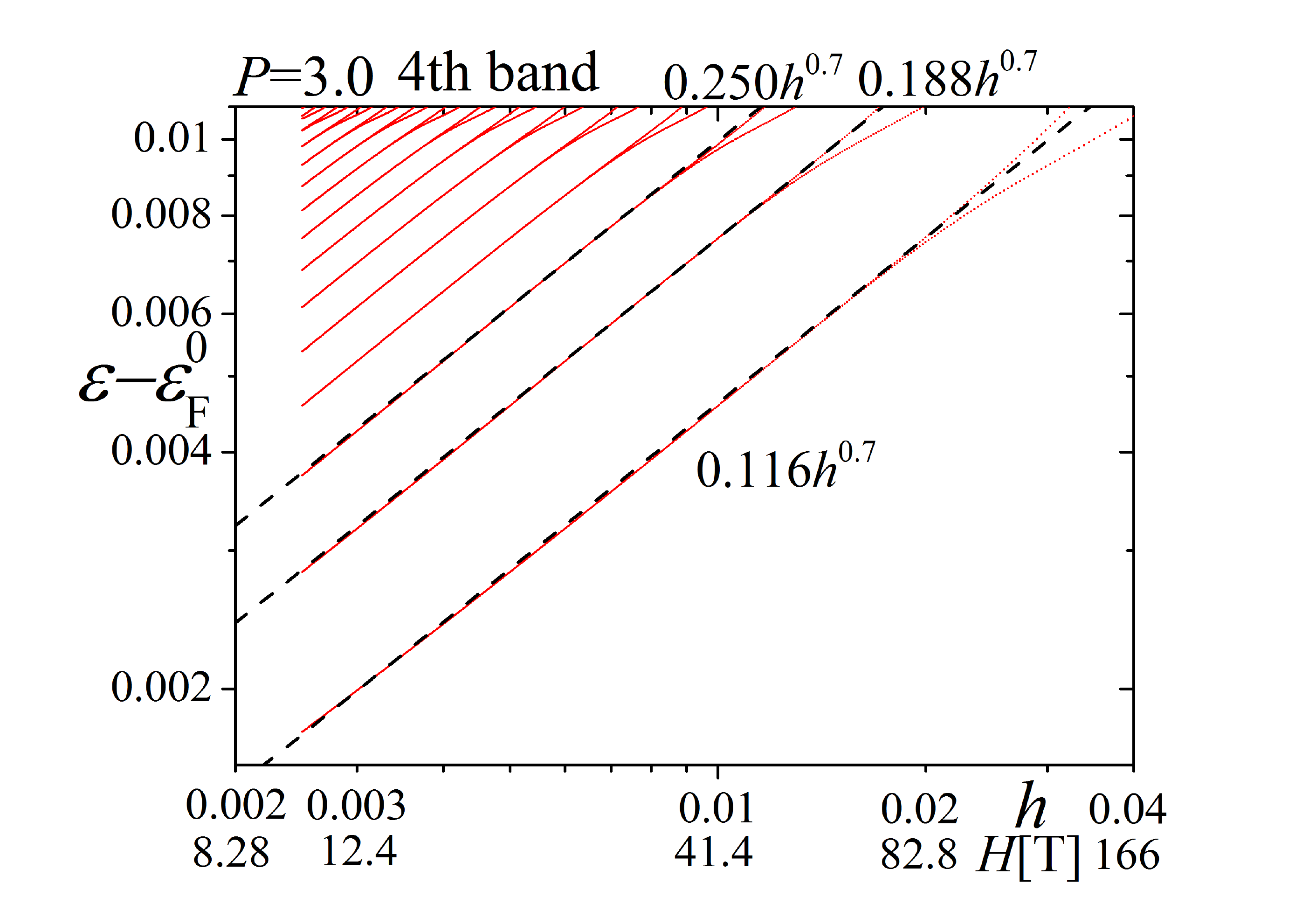}\vspace{-0.0cm}
\begin{flushleft} \hspace{0.0cm}(b) \end{flushleft}\vspace{-0.0cm}
\includegraphics[width=0.5\textwidth]{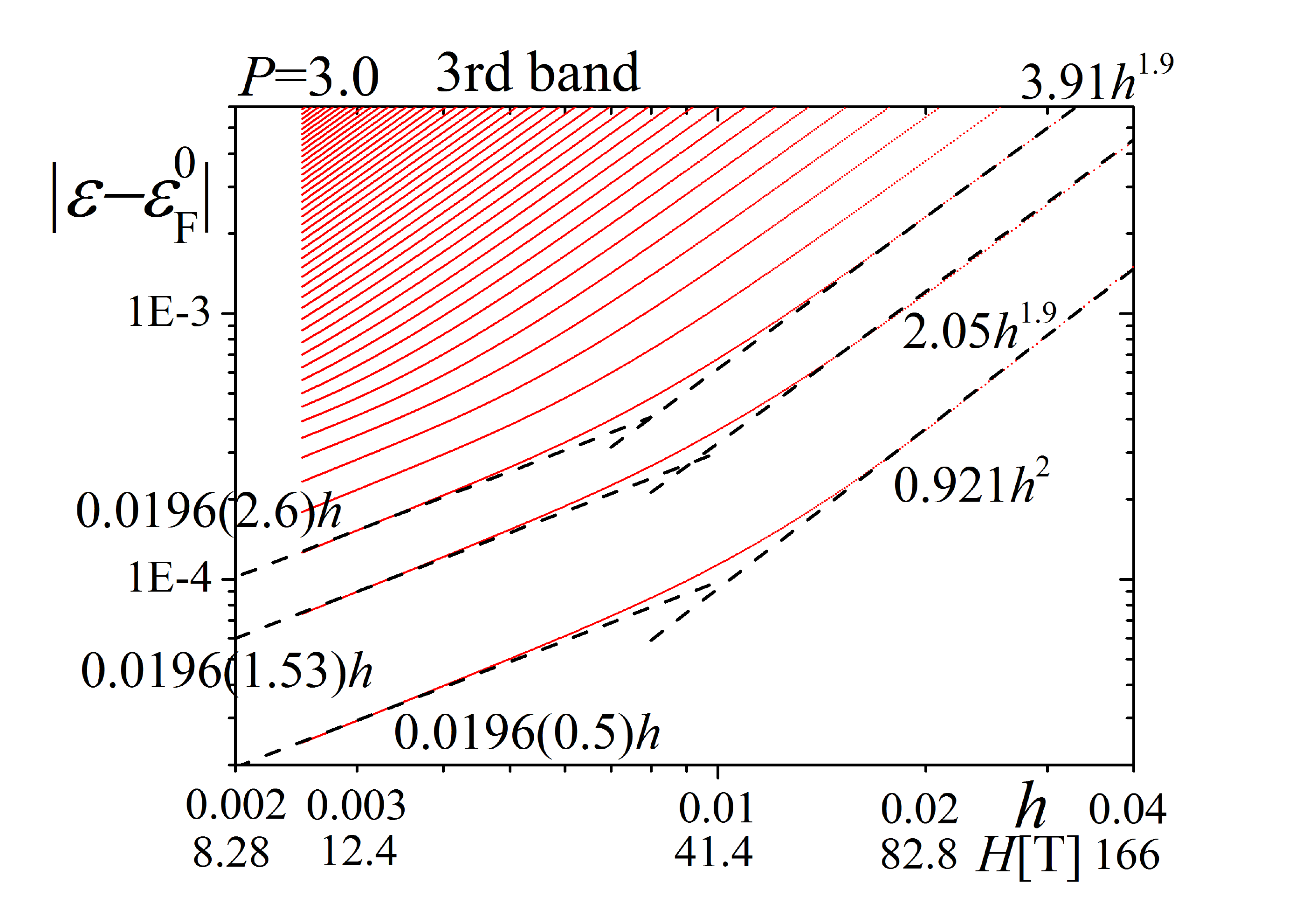}\vspace{-0.0cm}
\caption{
The log-log plot of Fig. \ref{fig16} (c) at $P=3.0$ at 0.025$\leq h\leq0.04$ (10.4 T$\leq H\leq$ 166 T). (a) the energy larger than $\varepsilon_{\textrm{F}}^0$, which corresponds to the fourth band at $h=0$ and (b) the energy smaller than $\varepsilon_{\textrm{F}}^0$, which corresponds to the third band at $h=0$. 
}
\label{fig17}
\end{figure}

The energy, $\varepsilon^0_{\textrm{3t}}$ 
at $\mathbf{k}_{3\textrm{t}}=(0, \pi/a)$ is the same as $\varepsilon^0_{\textrm{D}}$ at $P=3.0$ (see Figs. ~\ref{fig9}(a) and ~\ref{fig9f}(b)). 
Then the third band is almost constant at the line connecting $\varepsilon^0_{\textrm{3t}}$ and $\varepsilon^0_{\textrm{D}}$. 
We calculate the magnetic-field-dependence of the energy (Fig.~\ref{fig16}(c)). The log-log plot near the Fermi energy 
is shown in Fig.~\ref{fig17}. The energies for the fourth band are fitted by 
\begin{equation}
\varepsilon_n-\varepsilon_{\textrm{F}}^0 \propto h^{0.7}. \label{0.7}
\end{equation}
Eq. (\ref{0.7}) is obtained from a fitting at the intermediate magnetic field. 
If we could perform a fitting at the low magnetic field limit, we could obtain $\propto\sqrt{h}$.


For the third band, the quantized energies below $\varepsilon^0_{\textrm{F}}$ are fitted by 
\begin{equation}
\varepsilon_n = -0.0196 \left( n+ \frac{1}{2} \right) h + \varepsilon^0_{\textrm{F}}, \ n=0, 1, 2 \label{3.0}
\end{equation}
for $h \lesssim 0.005$
and
\begin{equation}
  \varepsilon_{\textrm{F}}^0-\varepsilon_n \propto h^2 \label{3.0_2}
\end{equation}
for $h \gtrsim 0.01$ as shown in Fig.~\ref{fig17}(b).

The magnetic-field-dependences of Eqs. (\ref{3.0}) and (\ref{3.0_2}) can be understood as follows.
When the magnetic field is weak ($h \lesssim 0.005$), the energy is quantized as 
the Landau levels for a free hole pocket around $\mathbf{k}_{\textrm{3t}}$. 
On the other hand, when $h \gtrsim 0.01$, we can neglect the small curvature around $\mathbf{k}_{\textrm{3t}}$
and very small regions of local maxima around $\pm \mathbf{k}_{\textrm{D}}$.
Then, an almost flat ridge from $\mathbf{k}_{\textrm{D}}$ to $- \mathbf{k}_{\textrm{D}}$
via $\mathbf{k}_{\textrm{3t}}$ is quantized in the intermediate value of the magnetic field.
We consider a model for this situation as 
\begin{equation}
 \mathcal{H}^{0, \textrm{ridge}} = \frac{1}{2m} p_x^2 + V (p_y), \label{H^0}
\end{equation}
where 
\begin{equation}
V(p_y)=
 \left\{ \begin{array}{ll}
 0  & \mbox{if $|p_y| < p_0$} \\
 \infty & \mbox{otherwise}
 \end{array} \right. \label{vpy}
\end{equation}
, where $p_0$ is the length of the ridge, i.e., 
$p_0 \simeq 2 \hbar | \mathbf{k}_{\rm D}-\mathbf{k}_{\textrm{3t}}|$. 
In the presence of the magnetic field, $\mathbf{p}$ is replaced by
\begin{equation}
 -i \hbar \nabla-\frac{e}{c}\mathbf{A},
\end{equation}
where $\mathbf{A}$ is vector potential, and 
we take  
\begin{equation}
 \mathbf{A} = (0, H x, 0).
\end{equation}
Then the eigenvalue $\varepsilon^{\textrm{ridge}}$ is obtained by the equation,
\begin{equation}
\left\{ \frac{-\hbar^2}{2m} \frac{\partial^2}{\partial x^2} + V (-i \hbar \frac{\partial}{\partial y} -\frac{e}{c}Hx) \right\}
\Psi (x,y) = \varepsilon^{\textrm{ridge}} \Psi(x,y).
\end{equation}
The eigenstates are obtained as
\begin{equation}
 \Psi(x,y) = e^{ik_y y} \psi(x),
\end{equation}
where $\psi(x)$ is a solution of
\begin{equation}
\left\{ \frac{-\hbar^2}{2m} \frac{d^2}{dx^2} + V (\hbar k_y-\frac{e}{c}Hx) \right\} \psi(x) 
= \varepsilon^{\textrm{ridge}} \psi(x).
\label{eqQuantumwell}
\end{equation}
Since Eq.~(\ref{eqQuantumwell}) is the Schr\"odinger's equation for the one-dimentional quantum well
with width $2c/(eHp_0)$, the eigenvalue is quantized as
\begin{equation}
 \varepsilon^{\textrm{ridge}}_n= \frac{\hbar \pi^2 p_0^2 e^2}{8mc^2} (nH)^2, 
 \label{eq_h2}
\end{equation} 
where $n= 1,2,3, \cdots$. 
In spite of the simple approximation (Eqs. (\ref{H^0}) and (\ref{vpy})), we can explain the $h^2$-dependence seen in Fig.~\ref{fig17}(b). 

\subsection{Dirac fermions system at $3.0\leq P<39.2$, semi-Dirac fermions at $P=39.2$ and massive Dirac fermions system at $P>39.2$}

\begin{figure}[bt]
\begin{flushleft} \hspace{0.0cm}(a) \end{flushleft}\vspace{-0.0cm}
\includegraphics[width=0.49\textwidth]{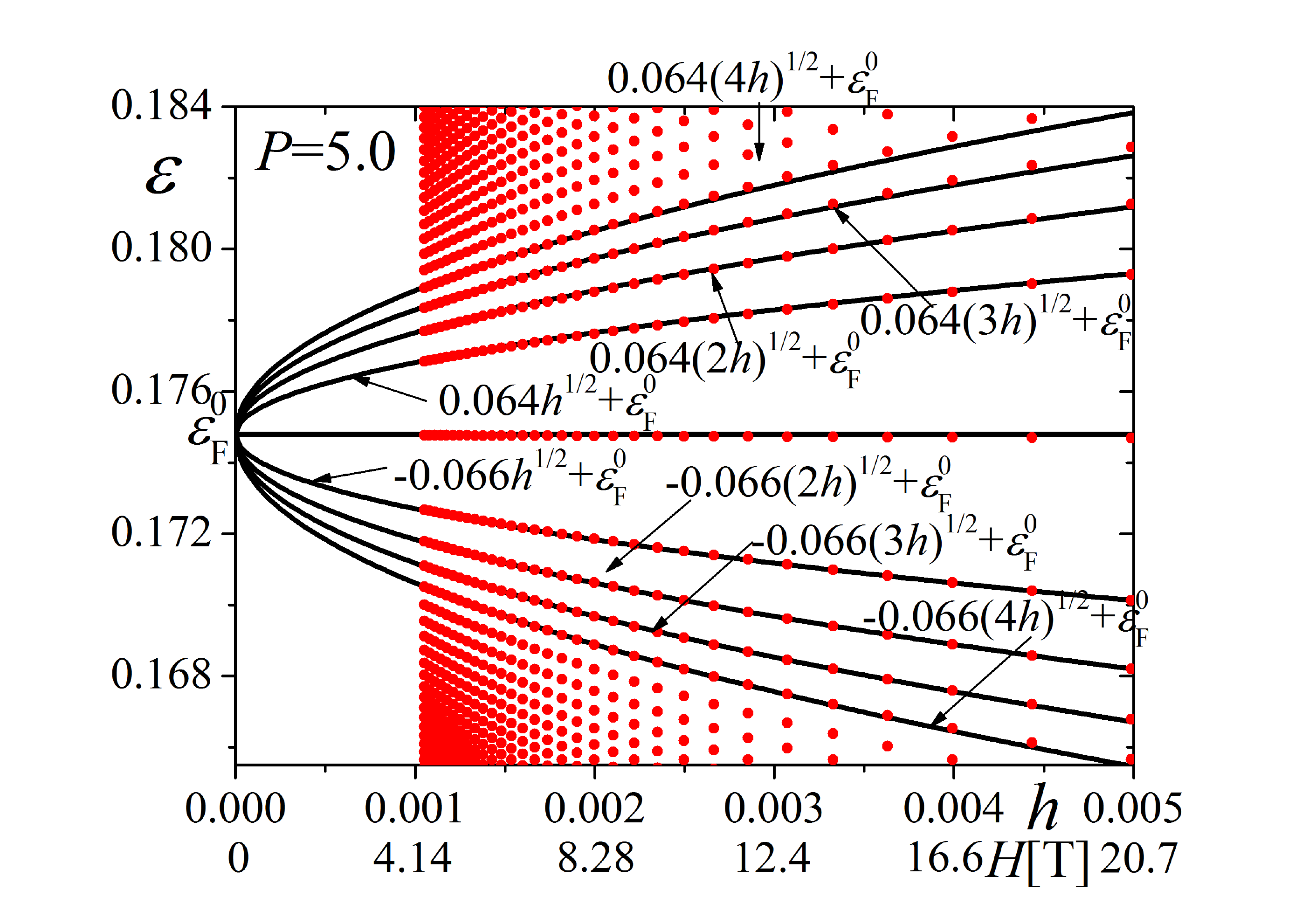}\vspace{-0.2cm}
\begin{flushleft} \hspace{0.0cm}(b) \end{flushleft}\vspace{-0.0cm}
\includegraphics[width=0.49\textwidth]{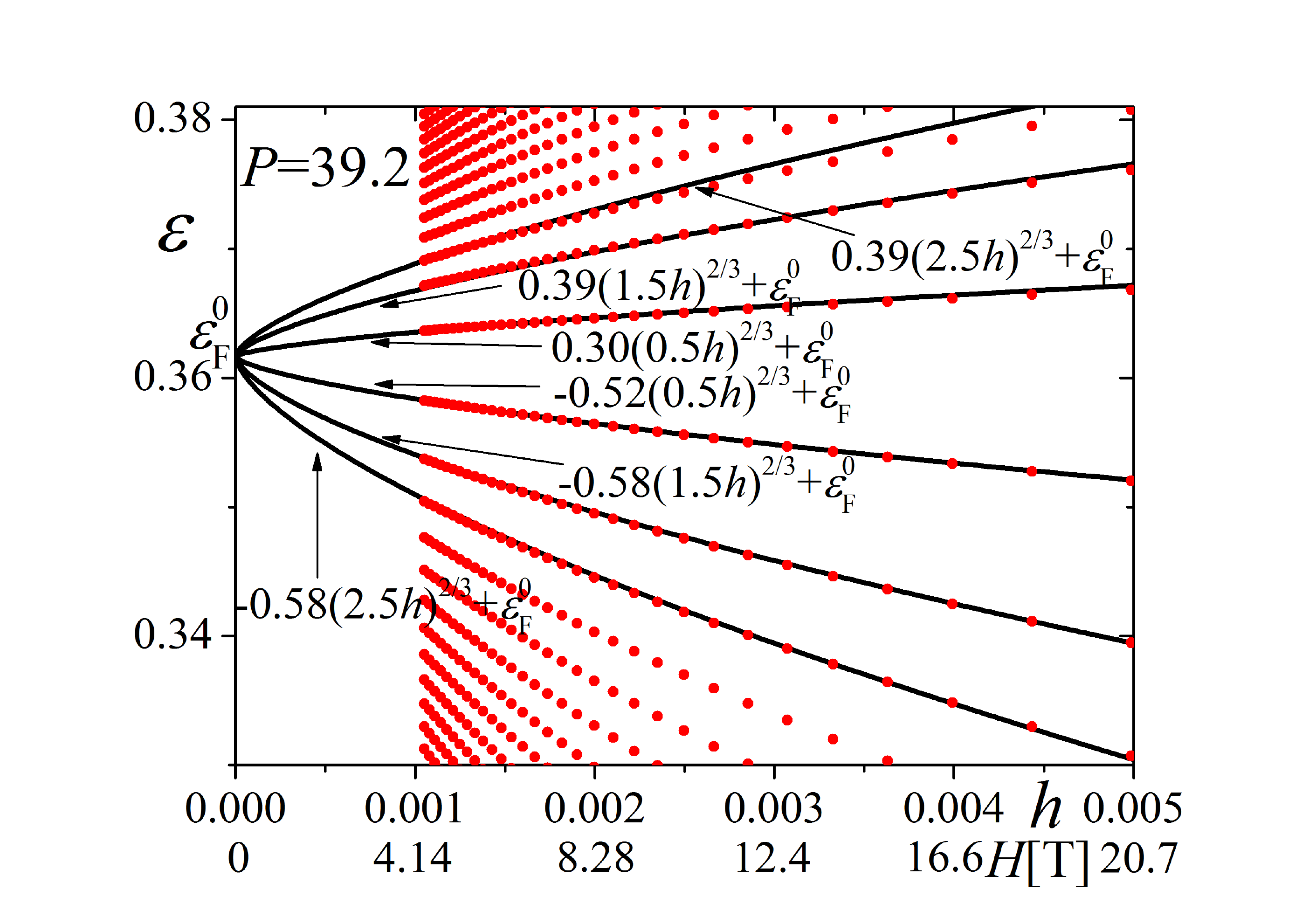}\vspace{-0.2cm}
\begin{flushleft} \hspace{0.0cm}(c) \end{flushleft}\vspace{-0.0cm}
\includegraphics[width=0.49\textwidth]{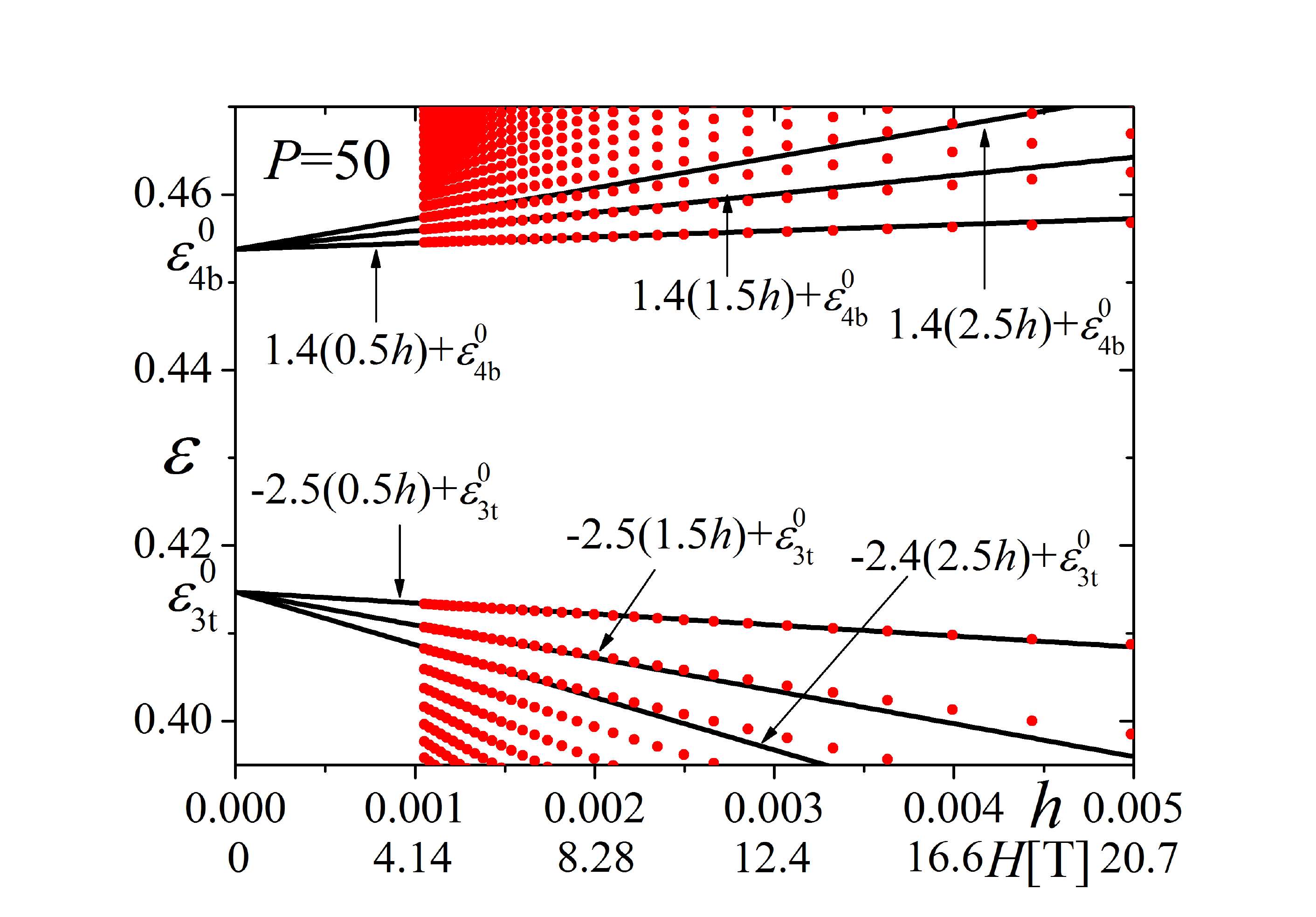}\vspace{-0.2cm}
\caption{
Energies as a function of $h$ for $P=5.0$ (a), $P=39.2$ (b) and $P=50$ (c). 
We choose $p=2$ and $q=1901, 1851, 1801, \cdots, 451, 401$. 
}
\label{fig14}
\end{figure}

We show the energies near the Fermi energy as a function 
of $h$ at $P=5.0$, $39.2$ and $50$ in Fig. \ref{fig14} at the low magnetic field. 
The magnetic-field-dependences of the energies at $P=5.0$ are fitted by 
\begin{eqnarray}
 \varepsilon_n =
\left\{ \begin{array}{ll}
0.064\sqrt{nh}+\varepsilon^0_{\textrm{F}}, \ \ \ \ \  \  n=0, 1, 2, 3, 4\\
-0.066\sqrt{|n|h}+\varepsilon^0_{\textrm{F}}, \ \  n=0, -1, -2, -3, -4
\end{array} \right.
\end{eqnarray}
which is expected in the system with massless Dirac fermions (Eq. (\ref{eq_1x})).

At $P=39.2$ the dispersion is parabolic in two directions and linear in the other two directions at the semi-Dirac point, as shown in Fig.~\ref{fig10}. 
The magnetic-field-dependences of the energies near $\varepsilon_{\textrm{F}}^0$ at the low magnetic field are fitted by 
\begin{align}
\varepsilon_n=
\left\{ \begin{array}{ll}
0.39g_{+}(n)\left[\left(n+\frac{1}{2} \right) h \right]^{\frac{2}{3}}+ \varepsilon^0_{\textrm{F}},  \ \  \ \ n=0, 1, 2\\
-0.58g_{-}(n)\left[\left| n+\frac{1}{2} \right| h \right]^{\frac{2}{3}}+ \varepsilon^0_{\textrm{F}}, \ \   n=0, -1, -2
\end{array} \right. 
\label{e_39.2}
\end{align}
where $g_{+}(0) \simeq 0.769$, $g_{-}(0) \simeq 0.897$ and $g(n)=1$ for $|n| = 1, 2$, as shown in Fig. \ref{fig14} (b). This magnetic-field dependence is expected in the system with the semi-Dirac point (Eq. (\ref{eqmergedDirac})).

At $P=50$, where massive Dirac fermions are realized, the Landau levels are fitted by 
\begin{align}
\varepsilon_n&=-2.5\left( n+ \frac{1}{2} \right) h + \varepsilon^0_{\textrm{3t}}, \label{50_1}\\
\varepsilon_n&=1.4 \left( n+ \frac{1}{2} \right) h + \varepsilon^0_{\textrm{4b}},\label{50_2}
\end{align}
where $n=0, 1, 2$. 
Eqs. (\ref{50_1}) and (\ref{50_2}) are due to the anisotropic parabolic bands.

%
%
%

\begin{figure}[bt]
\begin{flushleft} \hspace{0.5cm}(a) \end{flushleft}\vspace{-0.5cm}
\includegraphics[width=0.42\textwidth]{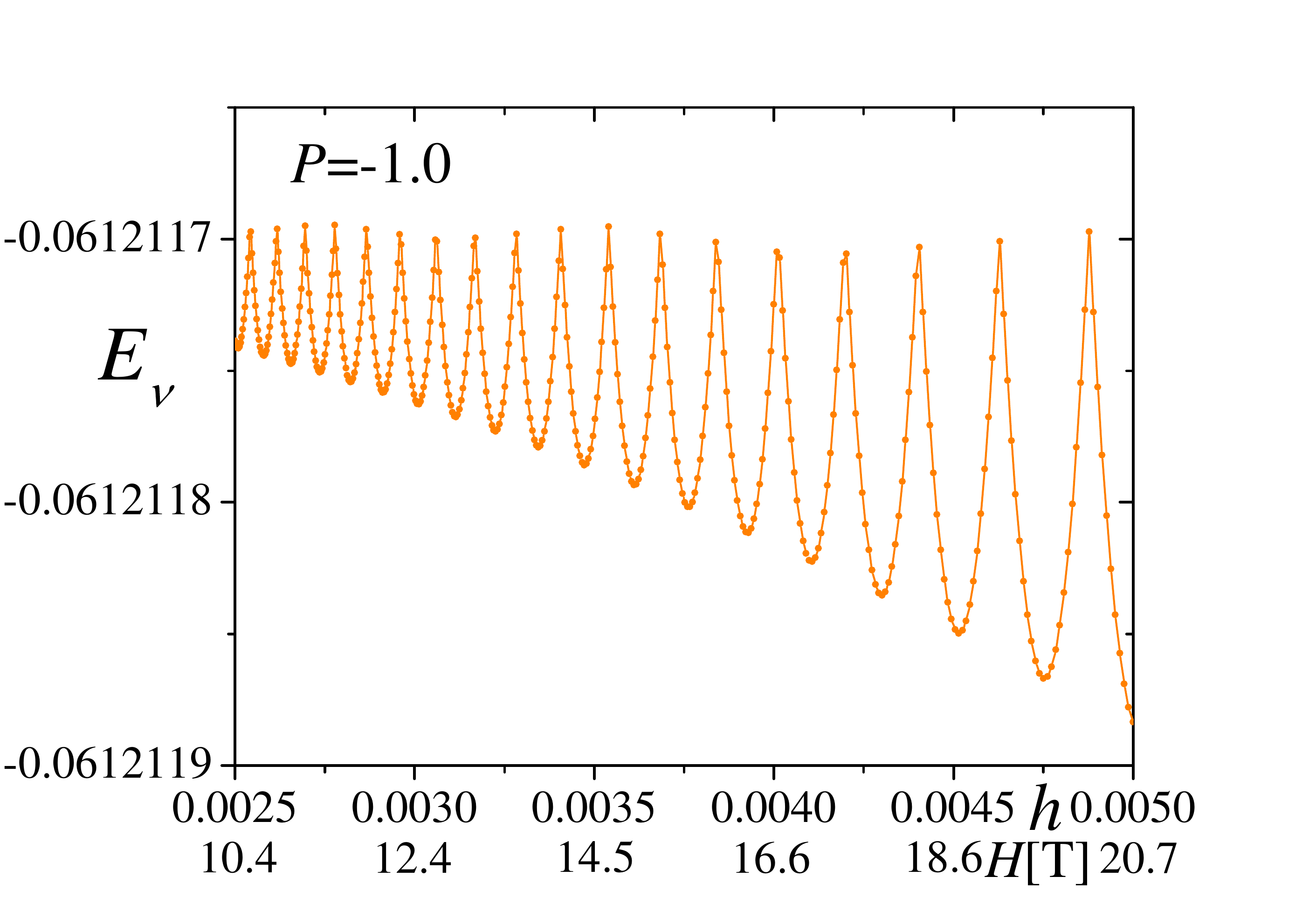}\vspace{-0.5cm}
\begin{flushleft} \hspace{0.5cm}(b) \end{flushleft}\vspace{-0.5cm}
\includegraphics[width=0.42\textwidth]{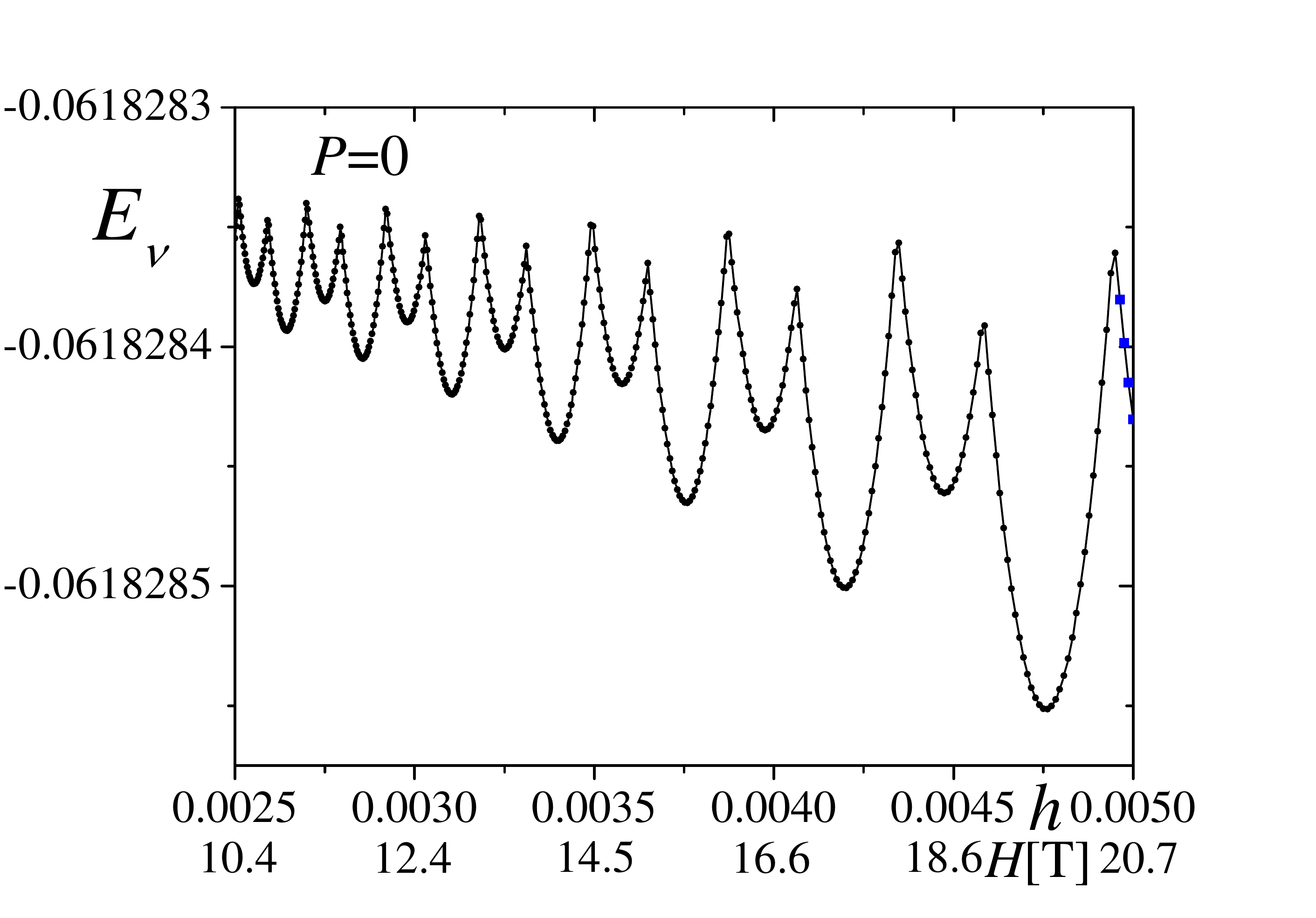}\vspace{-0.5cm}
\begin{flushleft} \hspace{0.5cm}(c) \end{flushleft}\vspace{-0.5cm}
\includegraphics[width=0.42\textwidth]{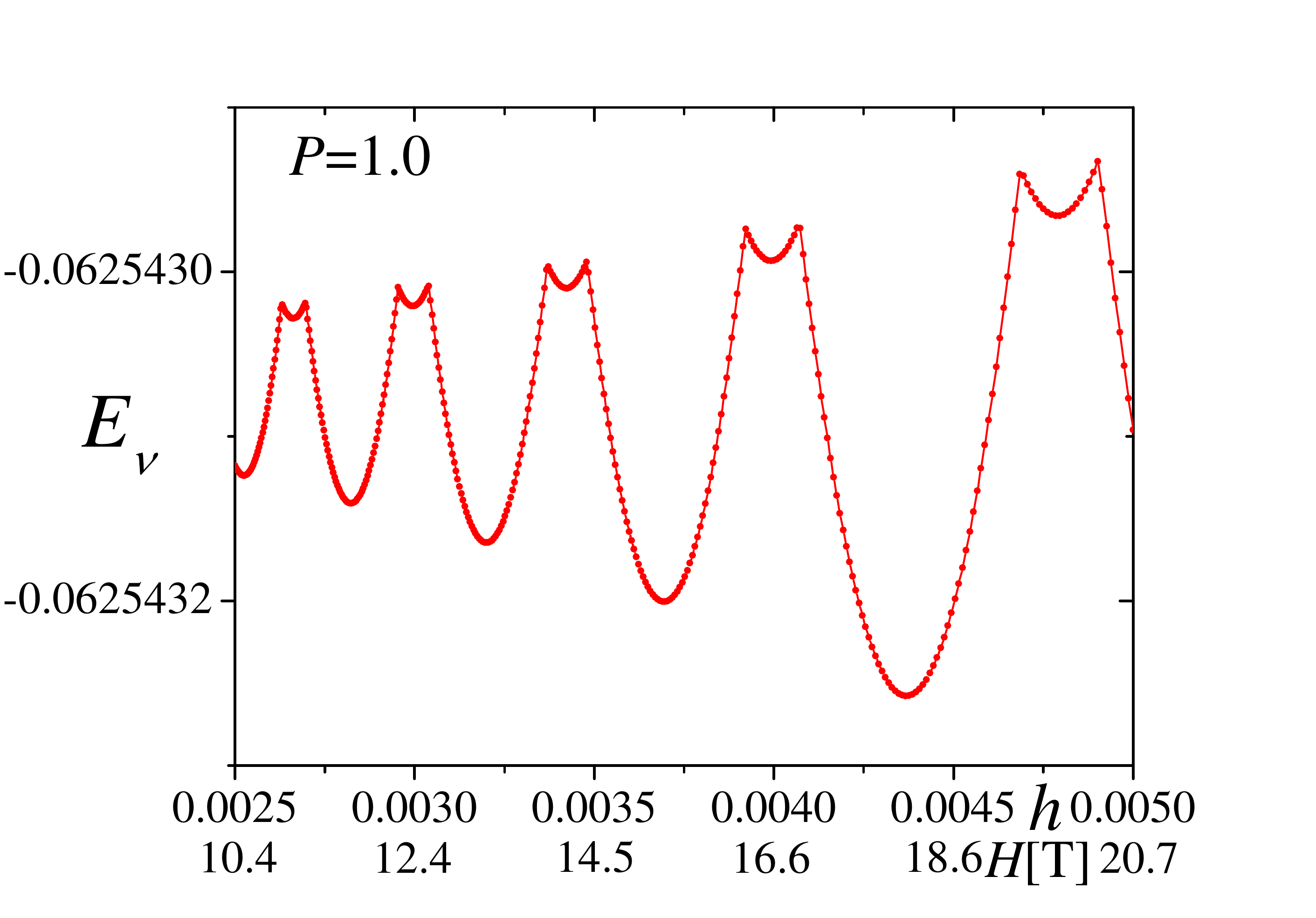}\vspace{-0.5cm}
\begin{flushleft} \hspace{0.5cm}(d) \end{flushleft}\vspace{-0.5cm}
\includegraphics[width=0.42\textwidth]{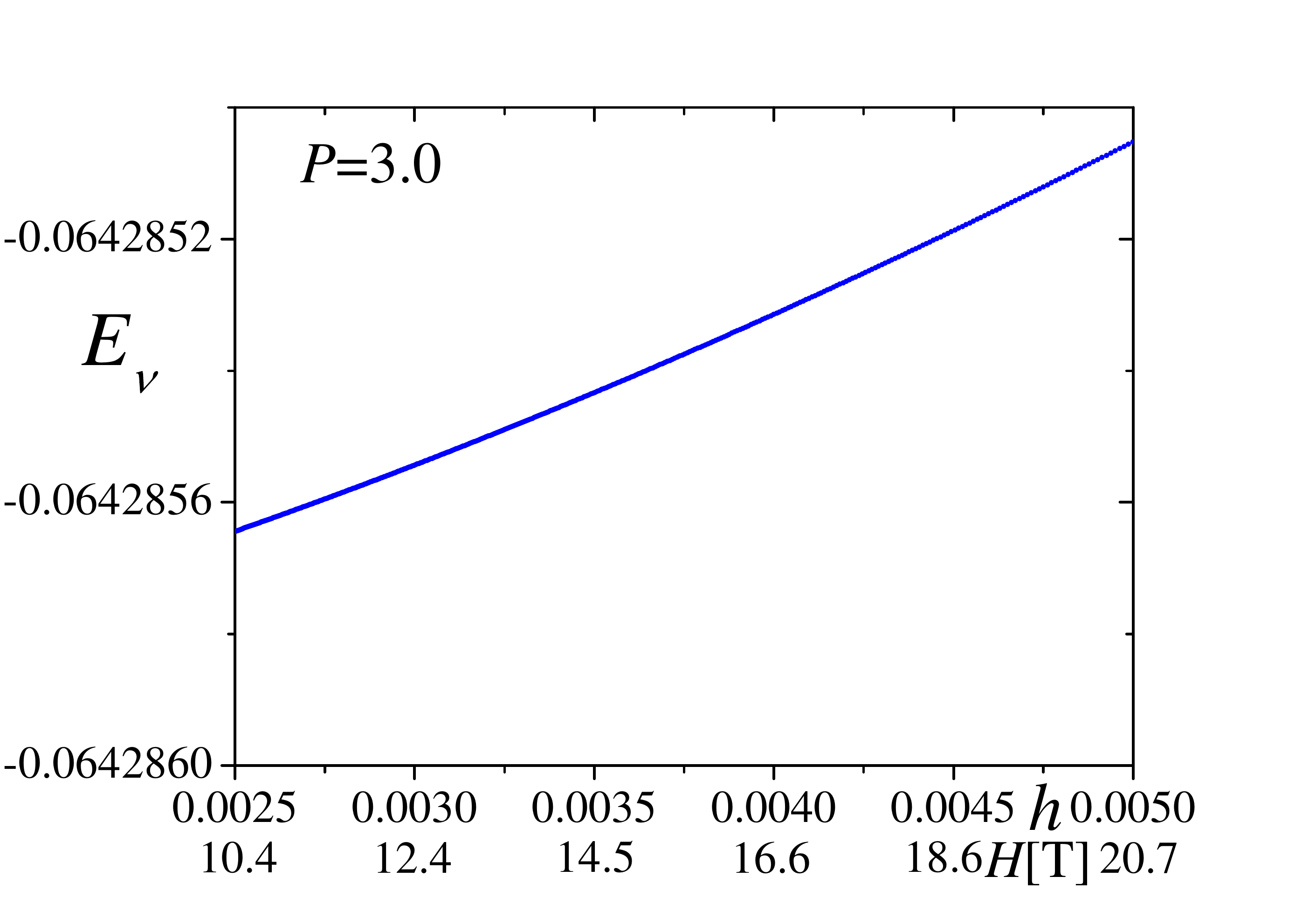}\vspace{-0.1cm}
\caption{
Total energies as a function of $h$ at $P=-1.0$ (a), $P=0$ (b), $P=1.0$ (c) 
and $P=3$ (d) with the fixed $\nu=3/4$. 
The same values of $h=p/q$ are used as those in Fig. \ref{fig18}.
}
\label{fig25}
\end{figure}
\begin{figure}[bt]
\begin{flushleft} \hspace{0.5cm}(a) \end{flushleft}\vspace{-0.5cm}
\includegraphics[width=0.42\textwidth]{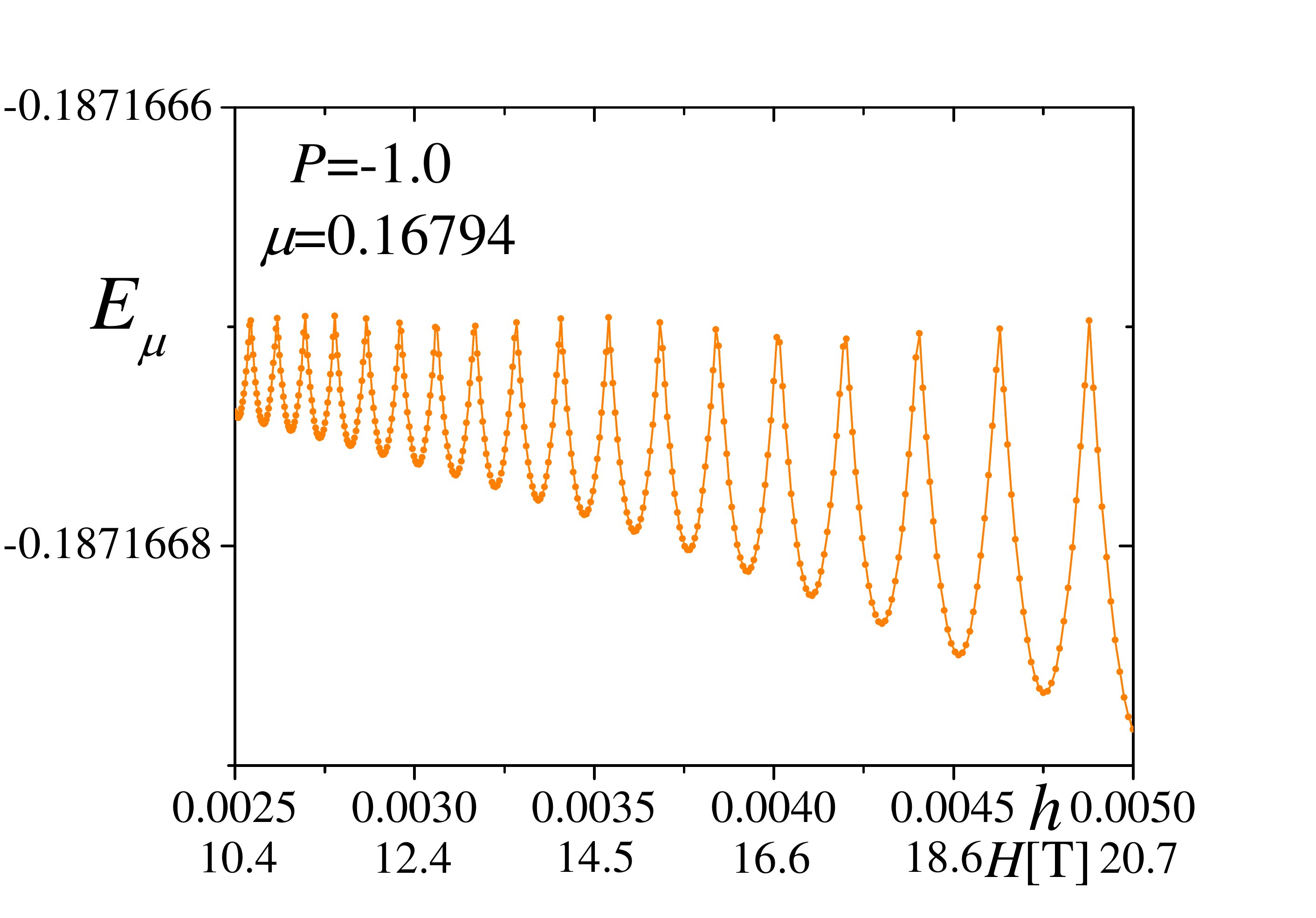}\vspace{-0.5cm}
\begin{flushleft} \hspace{0.5cm}(b) \end{flushleft}\vspace{-0.5cm}
\includegraphics[width=0.42\textwidth]{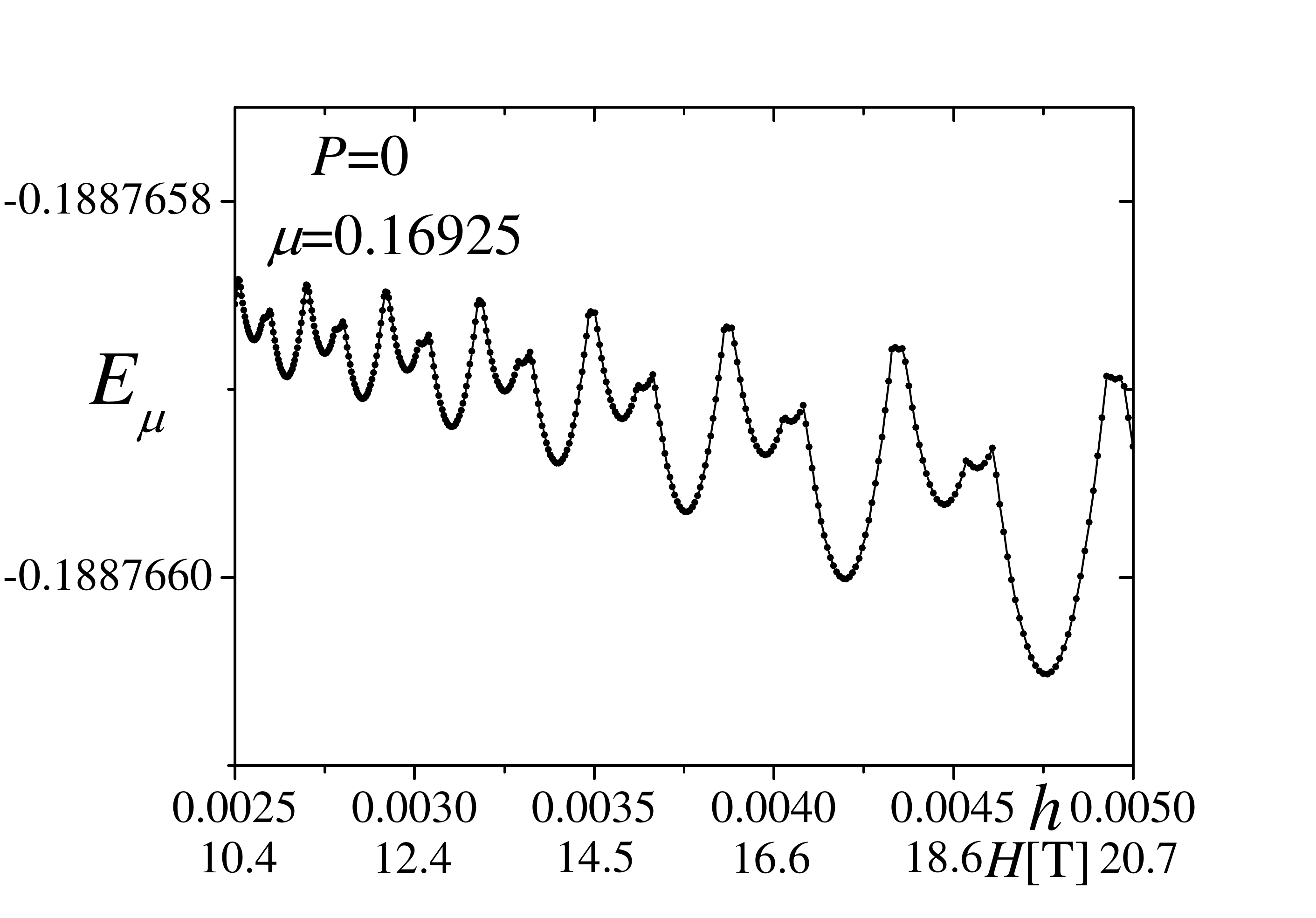}\vspace{-0.5cm}
\begin{flushleft} \hspace{0.5cm}(c) \end{flushleft}\vspace{-0.5cm}
\includegraphics[width=0.42\textwidth]{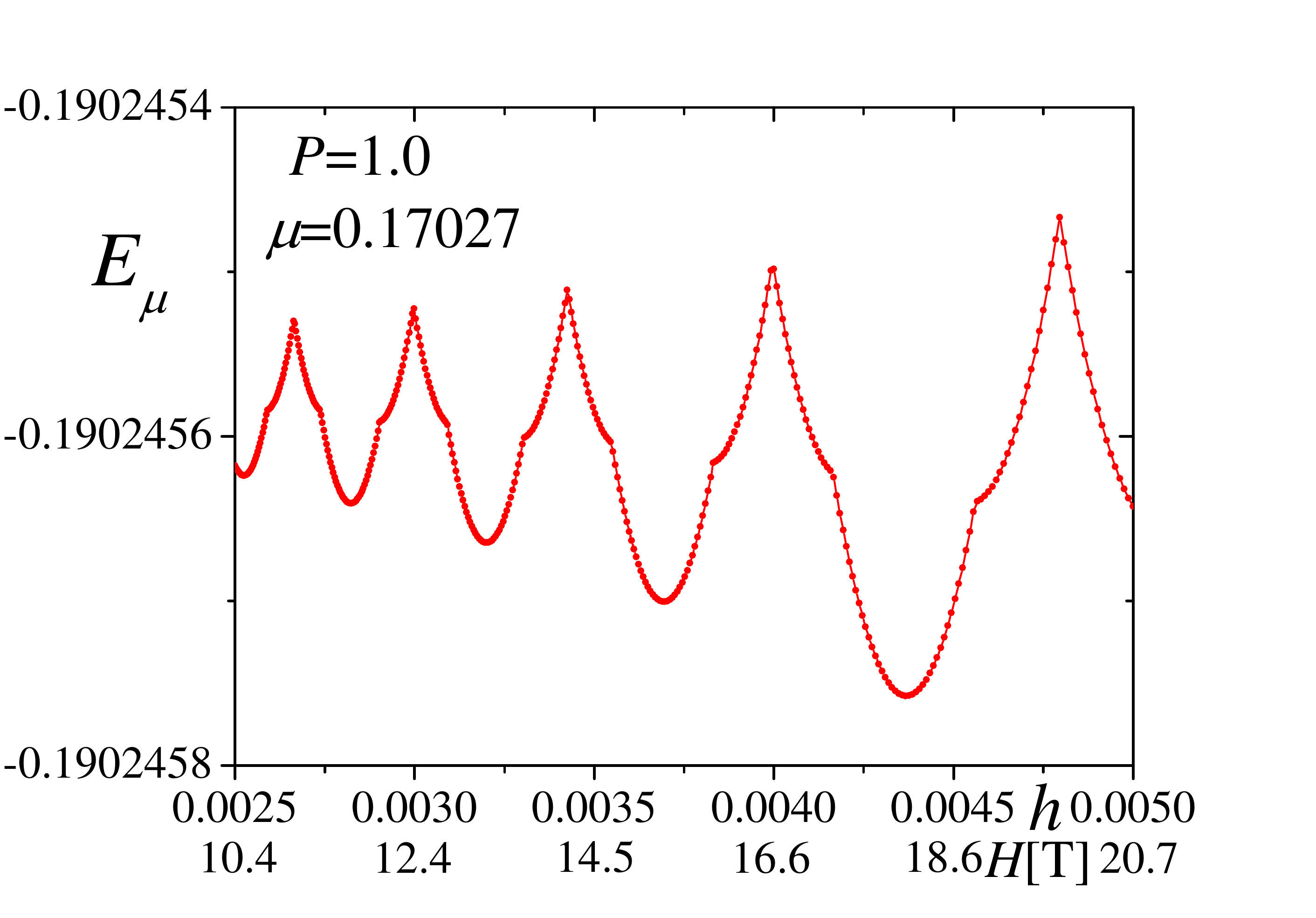}\vspace{-0.5cm}
\begin{flushleft} \hspace{0.5cm}(d) \end{flushleft}\vspace{-0.5cm}
\includegraphics[width=0.42\textwidth]{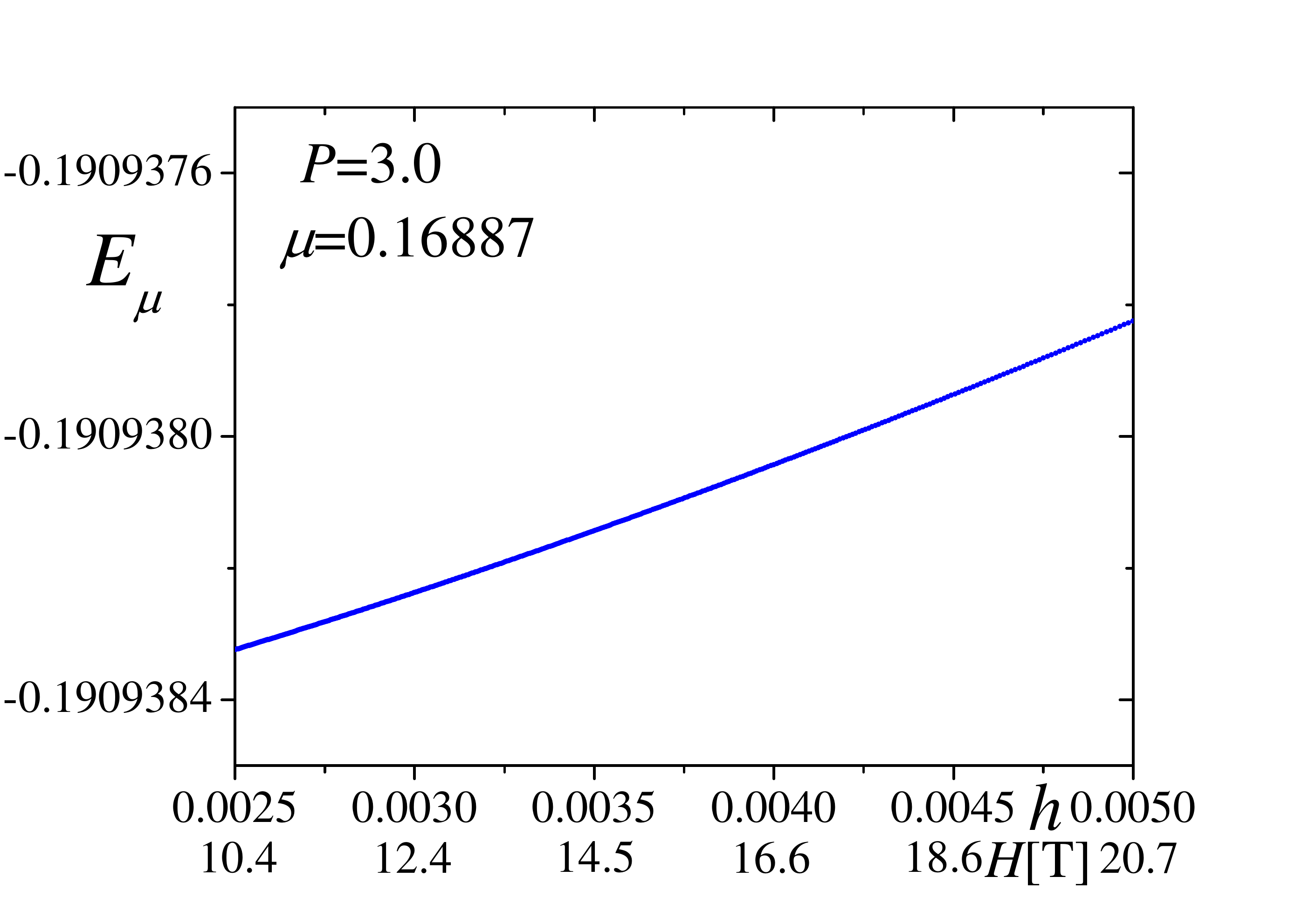}\vspace{-0.1cm}
\caption{
Total energies as a function of $h$ at $P=-1.0$ (a), 
$P=0$ (b), $P=1.0$ (c) and $P=3.0$ (d) with the fixed $\mu = \varepsilon^0_{\textrm{F}}$. 
The same values of $h=p/q$ are used as those in Fig. \ref{fig18}.
}
\label{fig26}
\end{figure}
\begin{figure}[bt]
\begin{flushleft} \hspace{0.5cm}(a) \end{flushleft}\vspace{-0.5cm}
\includegraphics[width=0.42\textwidth]{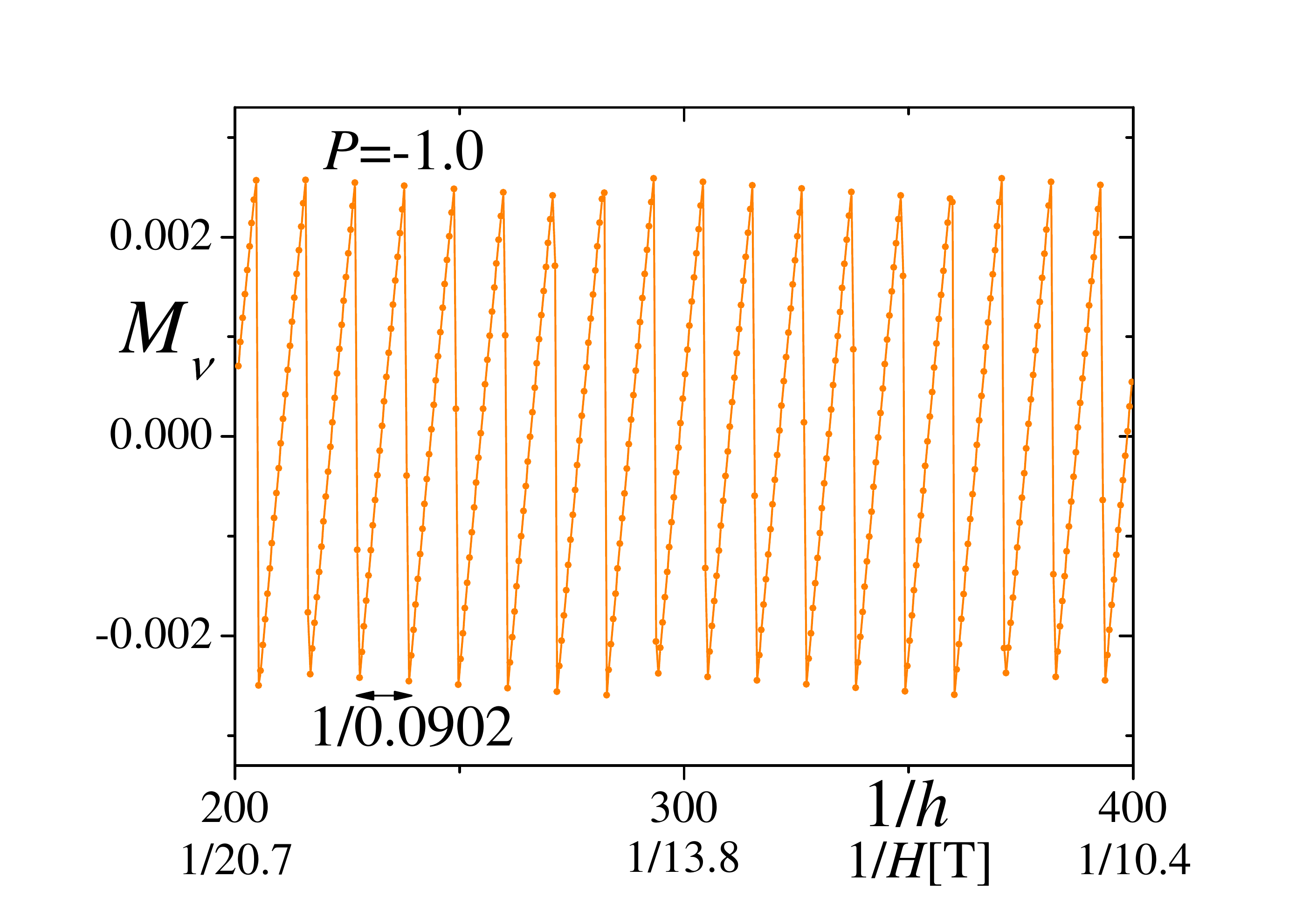}\vspace{-0.5cm}
\begin{flushleft} \hspace{0.5cm}(b) \end{flushleft}\vspace{-0.5cm}
\includegraphics[width=0.42\textwidth]{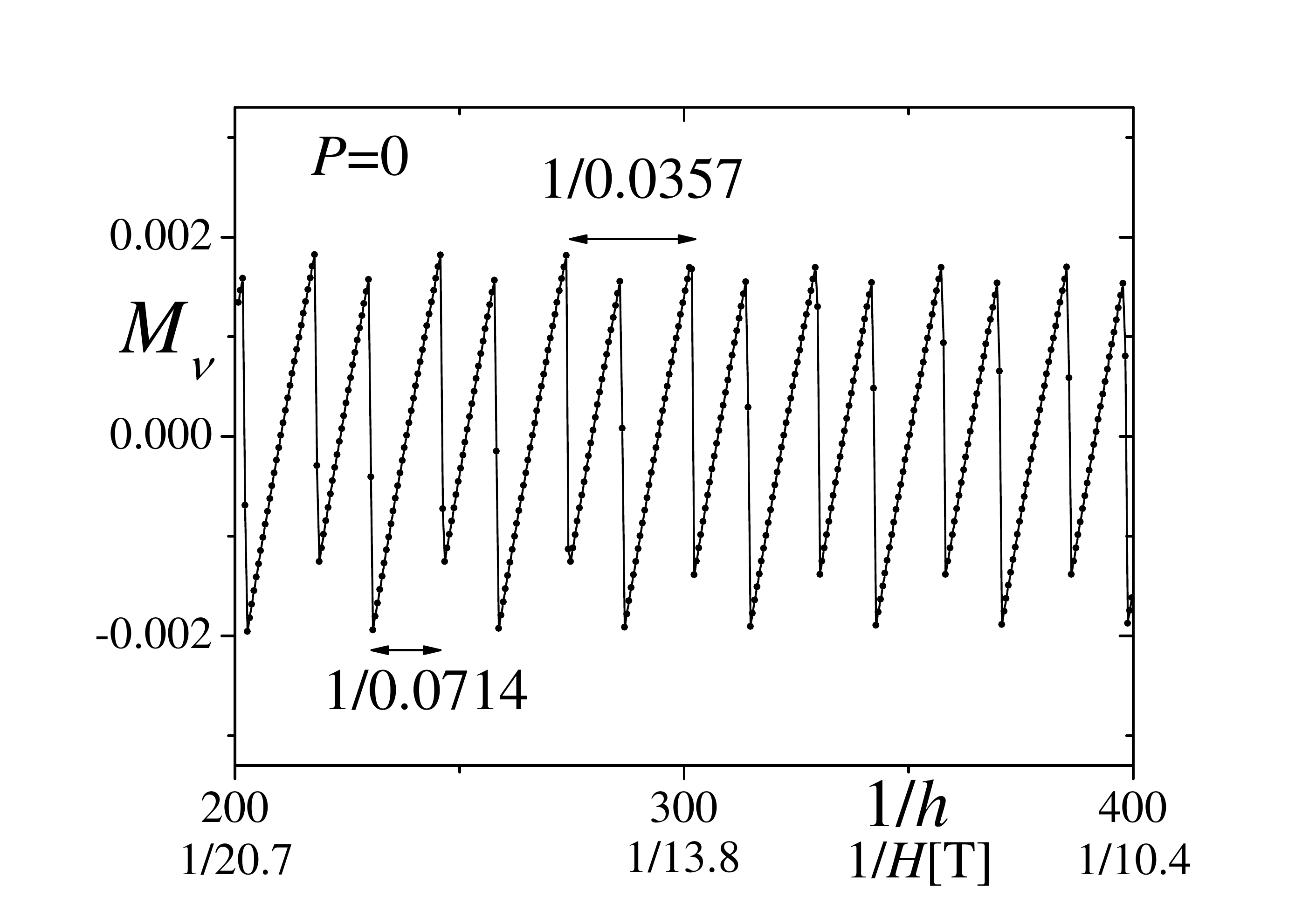}\vspace{-0.5cm}
\begin{flushleft} \hspace{0.5cm}(c) \end{flushleft}\vspace{-0.5cm}
\includegraphics[width=0.42\textwidth]{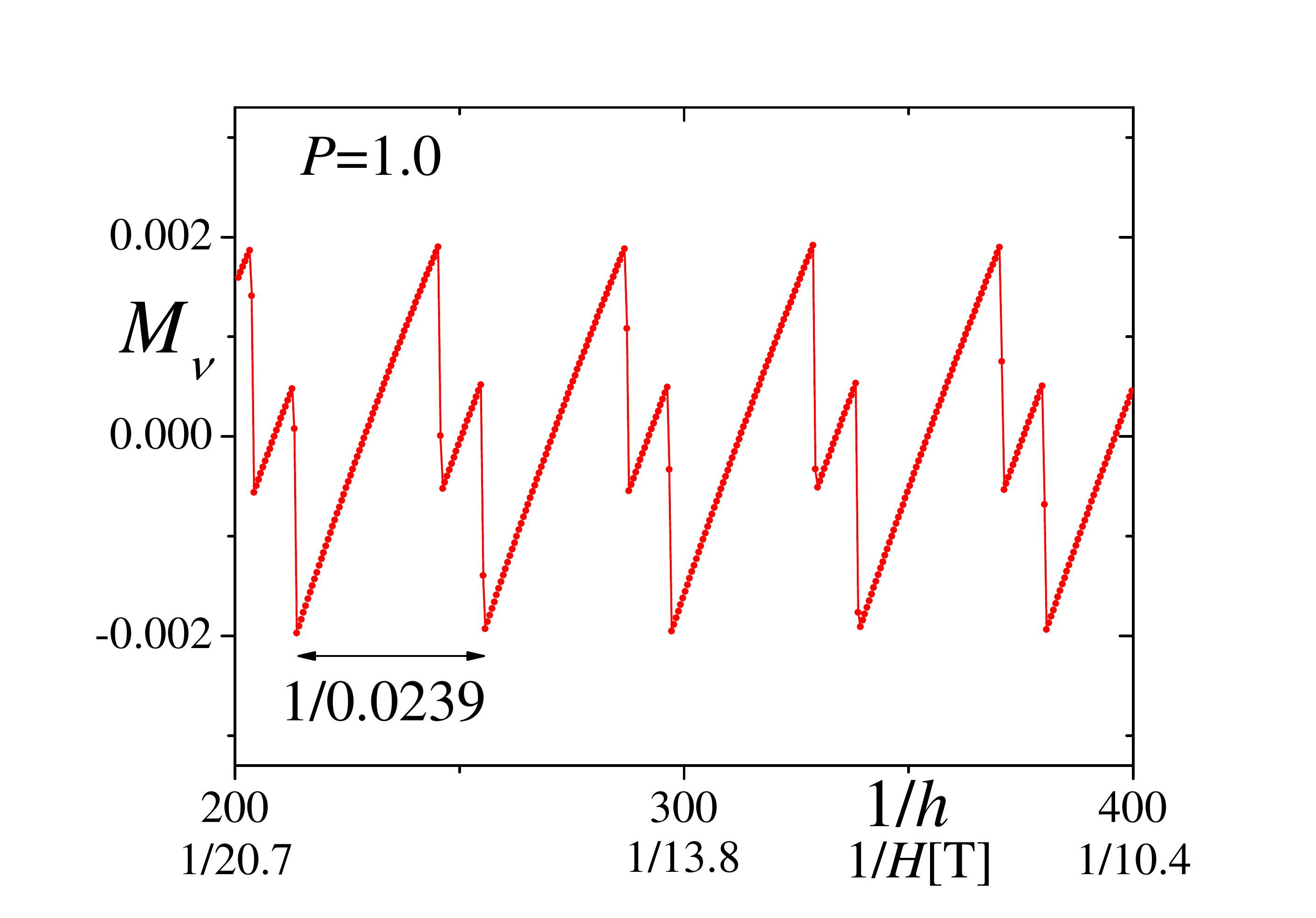}\vspace{-0.5cm}
\begin{flushleft} \hspace{0.5cm}(d) \end{flushleft}\vspace{-0.5cm}
\includegraphics[width=0.42\textwidth]{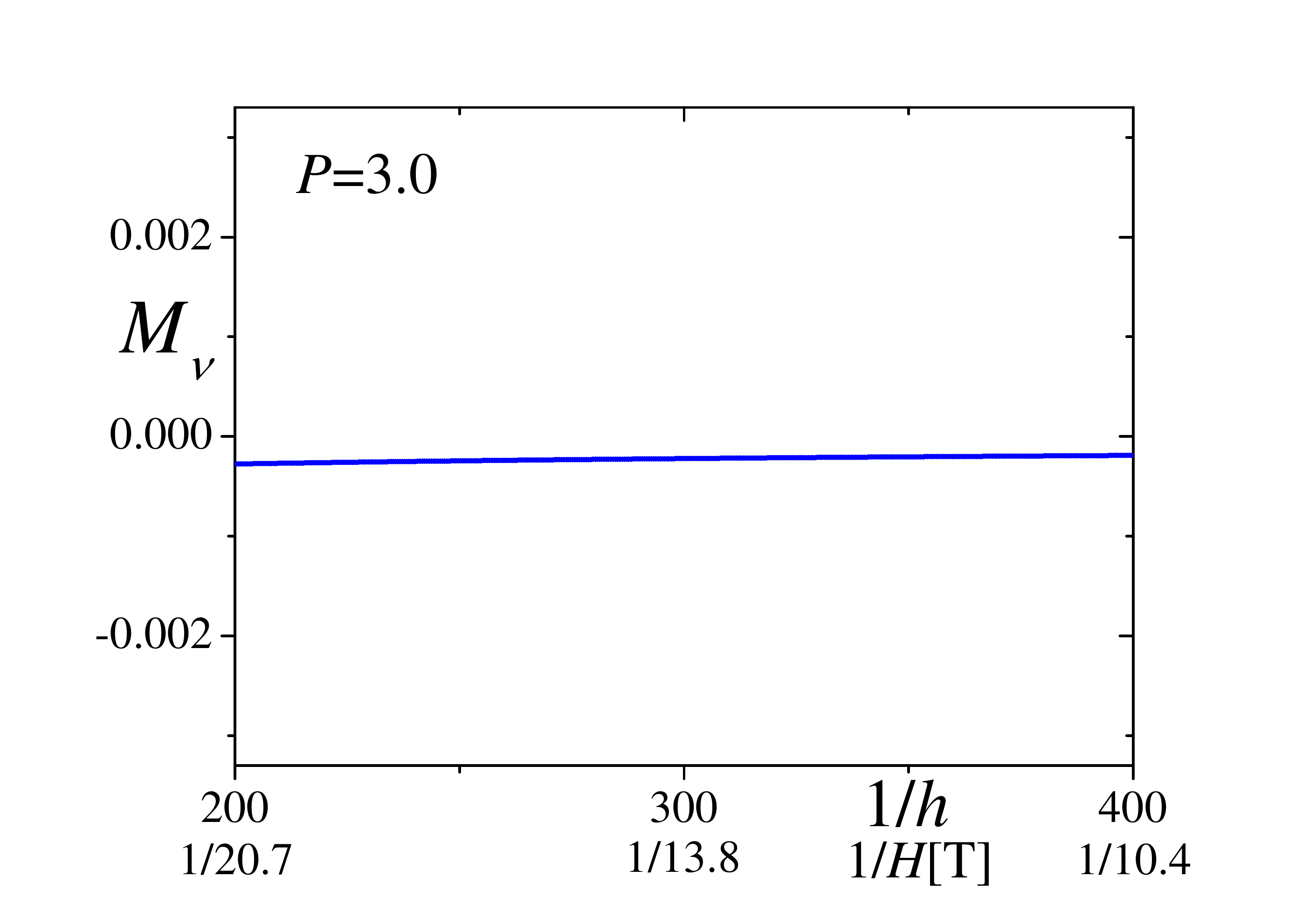}\vspace{-0.1cm}
\caption{
Magnetizations  as a function of $1/h$ at $P=-1.0$ (a), $P=0$ (b), $P=1.0$ (c), and $P=3.0$ (d),
calculated by numerical differentiation of total energies in Fig. \ref{fig25}.
}
\label{fig27}
\end{figure}
\begin{figure}[bt]
\begin{flushleft} \hspace{0.5cm}(a) \end{flushleft}\vspace{-0.5cm}
\includegraphics[width=0.42\textwidth]{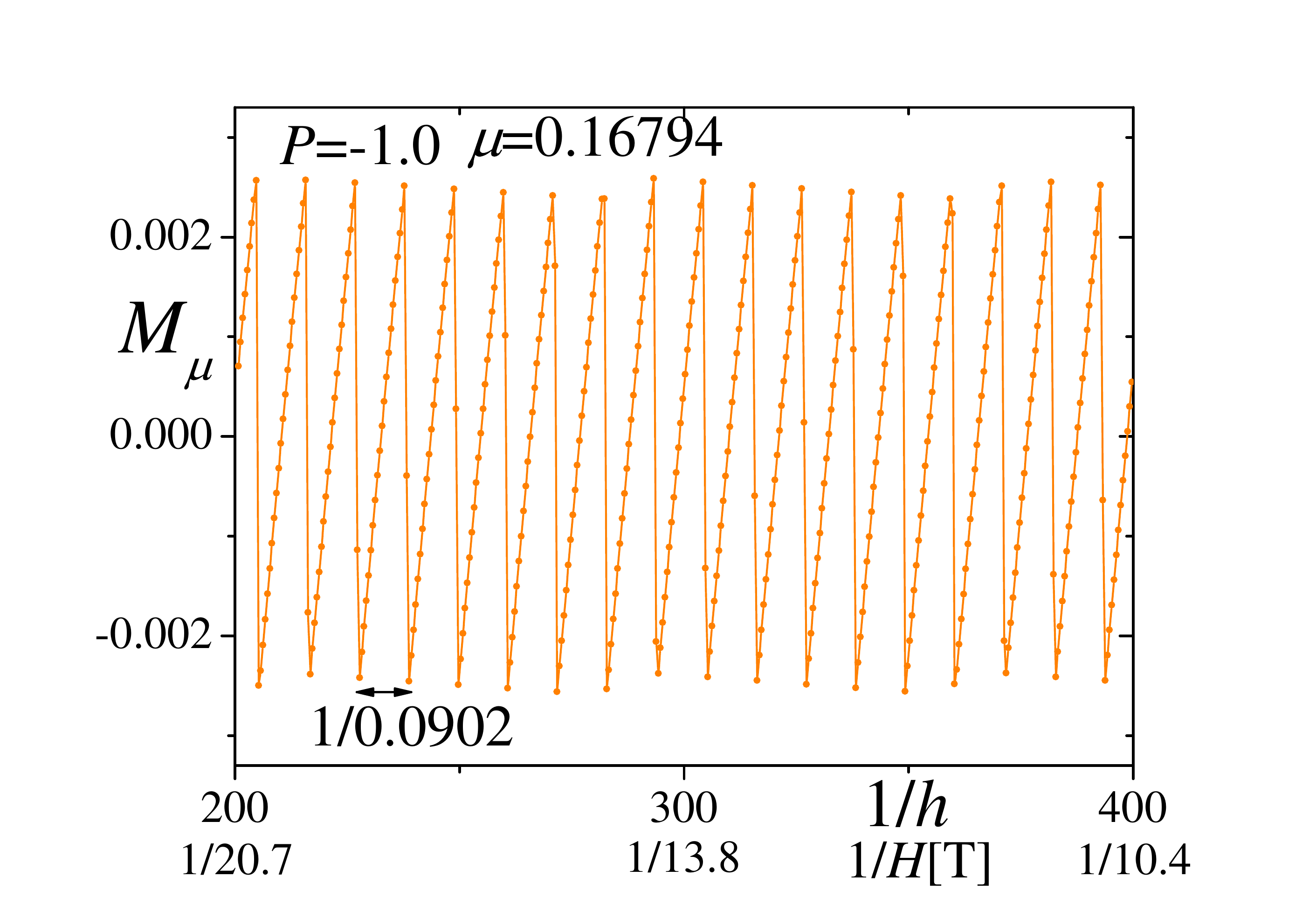}\vspace{-0.5cm}
\begin{flushleft} \hspace{0.5cm}(b) \end{flushleft}\vspace{-0.5cm}
\includegraphics[width=0.42\textwidth]{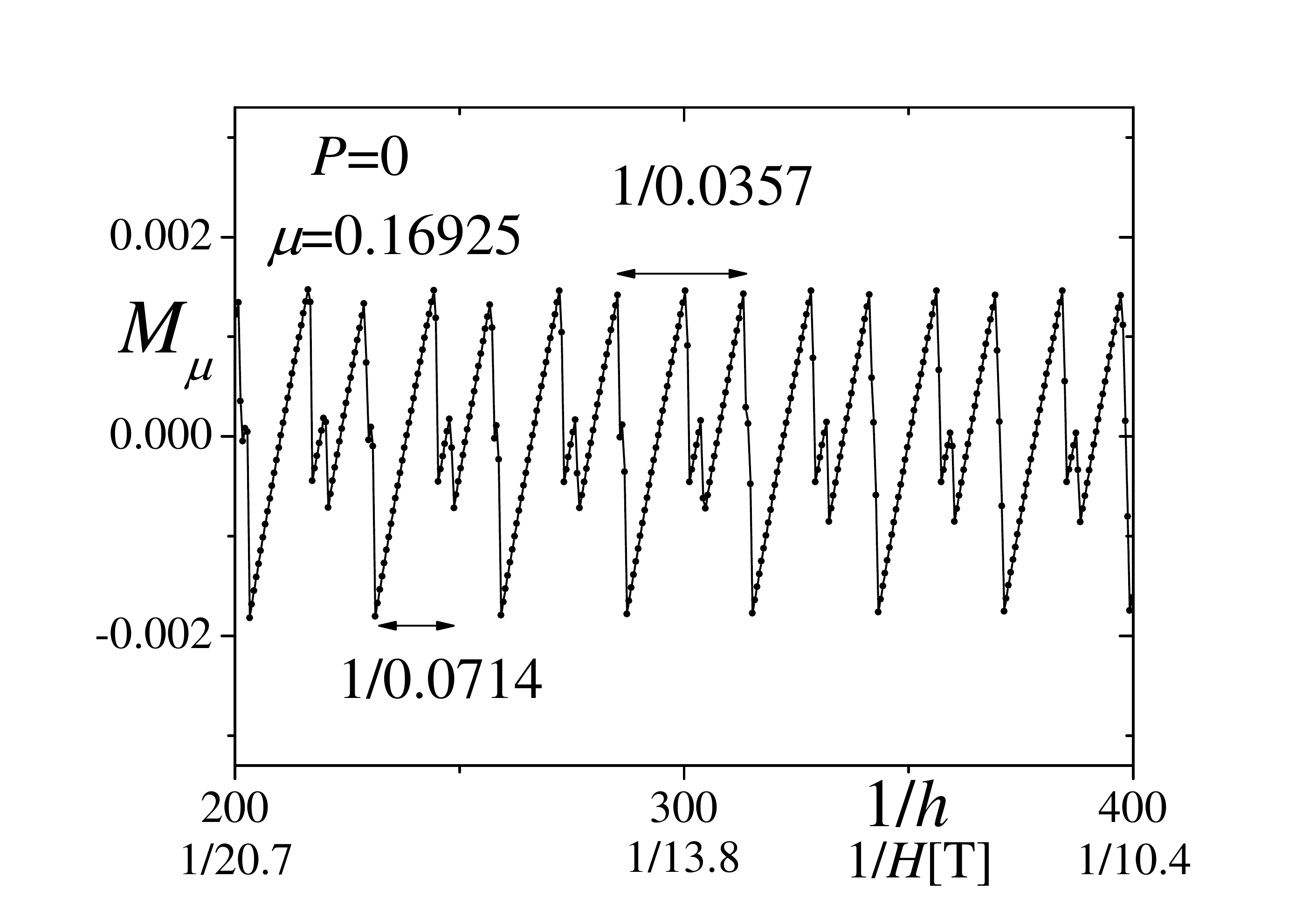}\vspace{-0.5cm}
\begin{flushleft} \hspace{0.5cm}(c) \end{flushleft}\vspace{-0.5cm}
\includegraphics[width=0.42\textwidth]{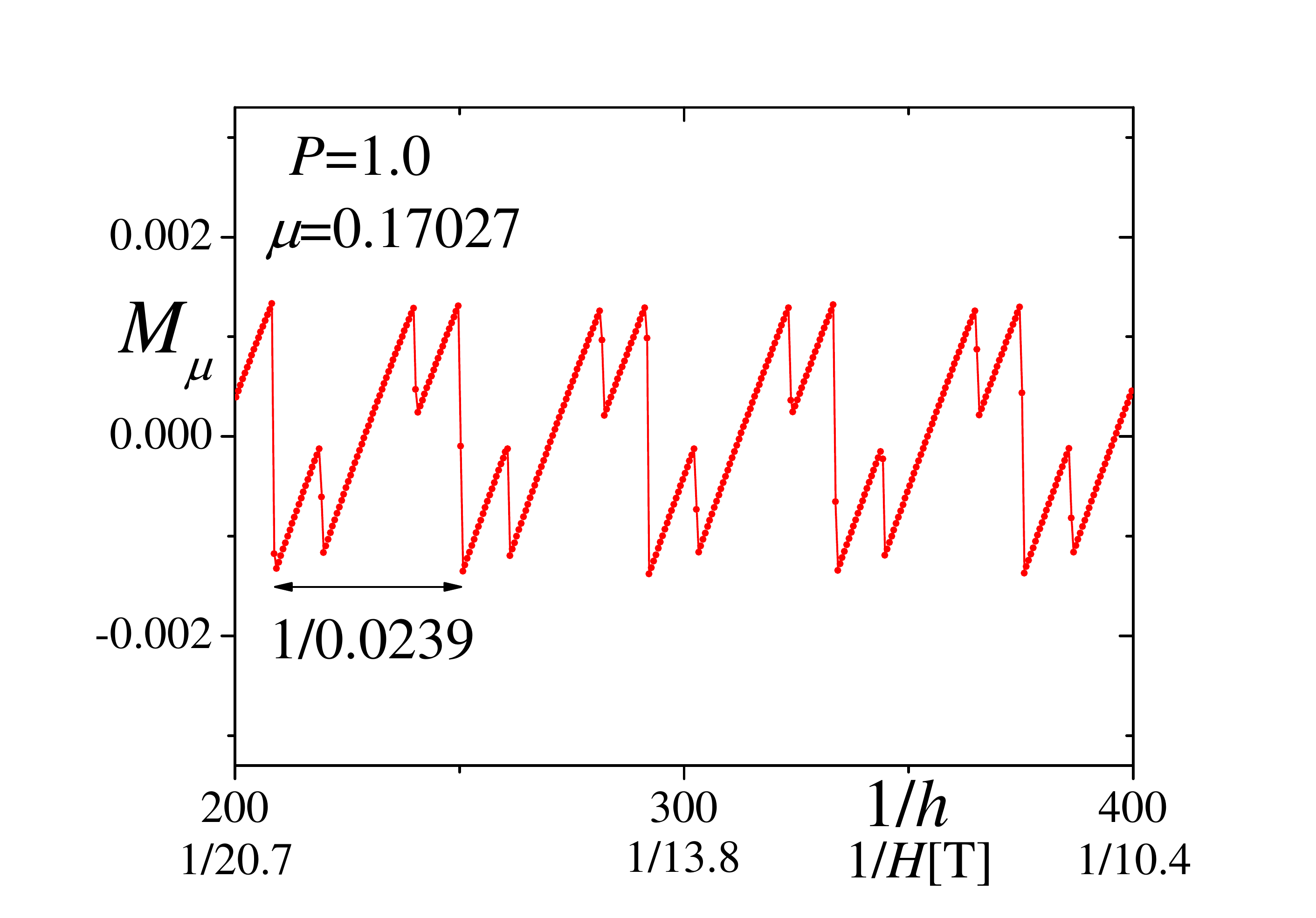}\vspace{-0.5cm}
\begin{flushleft} \hspace{0.5cm}(d) \end{flushleft}\vspace{-0.5cm}
\includegraphics[width=0.42\textwidth]{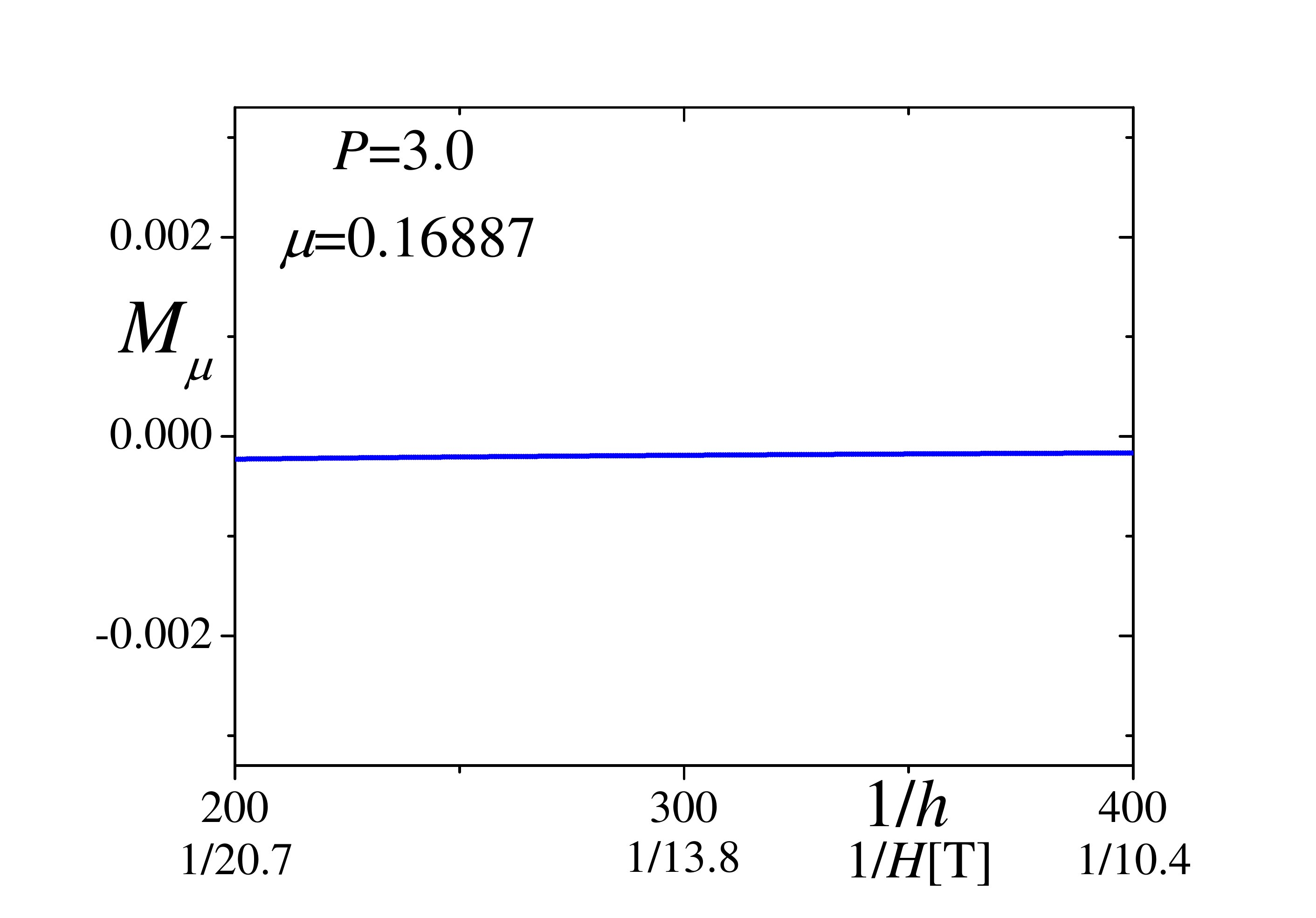}\vspace{-0.1cm}
\caption{
Magnetizations as a function of $1/h$ at $P=-1.0$ (a), $P=0$ (b), $P=1.0$ (c) and $P=3.0$ (d),
calculated by the numerical differentiation  of total energies in Fig. \ref{fig26}. 
}
\label{fig28}
\end{figure}

\section{Total energy, magnetization and de Haas van Alphen oscillation}
In this section we study the total energy and the magnetization. 
It has been known\cite{shoenberg} that the total energy
and the magnetization as a function of the magnetic field depend on 
whether we fix the chemical potential ($\mu$, i.e., grand canonical ensemble) 
or the electron number ($N$, or equivalently electron filling $\nu=N/N_s$, where $N_s$ is the site number, i.e., canonical ensemble).

In the case of fixed $\mu$, the thermodynamic potential ($\Omega$) per sites at 
the temperature $T$ is 
calculated as
\begin{equation}
\Omega=-\frac{k_BT}{4qN_k} \sum_{i=1}^{4q}\sum_{{\bf k}} \ln\left
\{\exp\left(\frac{\mu-\varepsilon(i,{\bf k})}{k_BT} \right) +1
\right\},
\end{equation}
where $k_B$ is the Boltzmann constant, $N_k$ is the number of $\mathbf{k}$ points taken in the magnetic Brillouin zone,
$4q$ is the number of bands in the presence of the magnetic field, and 
$\varepsilon(i,{\bf k})$ 
is the eigenvalues of $4q \times 4q$ matrix in Eq.~(\ref{j5}). 
The site number, $N_s$, is given by $N_s=4qN_k$. 
At $T=0$, $\Omega$ becomes the total energy for the fixed $\mu$,  
\begin{equation}
E_{\mu}=\frac{1}{4qN_k}\sum_{\varepsilon(i,{\bf k})\leq\mu}  (\varepsilon(i,{\bf k})-\mu).\label{E_mu}
\end{equation}

If the system is isolated from the reservoir of electrons, 
electron number (or electron filling $\nu$)
is conserved and the chemical potential changes depending on the magnetic field. 
Although it has been known that the magnetic-field-dependence of $\mu$ is negligibly small if we consider the effects
of the three-dimensionality, 
thermal broadening, compensated metals, electron or hole reservoirs 
 {\it etc.},\cite{fortin2008,fortin2009,kishigi_1995,nakano,harrison,fortin1998,alex1996,alex2001,champel,KH}
the magnetic-field-dependence of $\mu$ cannot be neglected in two-dimensional systems in general. 
The chemical potential, $\mu$, as a function of the magnetic field with fixed $\nu$
 should be obtained by the solution of the equation,
\begin{equation}
\nu =\frac{1}{4qN_k} \sum_{i=1}^{4q} \sum_{{\bf k}}
\frac{1}{\exp\left(\frac{\varepsilon(i,{\bf k})-\mu}{k_BT} \right) +1},\label{mu}
\end{equation}
where we take $\nu=3/4$ in this study. 
Using the magnetic-field-dependent $\mu$, the Helmholtz free energy ($F$) per sites 
at $T$ is calculated as
\begin{equation}	
F=-\frac{k_BT}{4qN_k} \sum_{i=1}^{4q}\sum_{{\bf k}} \ln\left
\{\exp\left(\frac{\mu-\varepsilon(i,{\bf k})}{k_BT} \right) +1
\right\} + \mu \nu.
\end{equation}
At $T=0$ it becomes the total energy with fixed $\nu$,
\begin{equation}
E_{\nu}=\frac{1}{4qN_k}\sum_{\varepsilon(i,{\bf k})\leq\mu}
\varepsilon(i,{\bf k}).\label{E_nu}
\end{equation}

In this paper we study the systems with fixed chemical potential and 
fixed electron number at $T=0$.
We show $E_{\nu}$ and $E_{\mu}$ at $P=-1.0, 0, 1.0$ and $3.0$ 
for the low magnetic field in Figs.~\ref{fig25} and \ref{fig26} and those for the high magnetic field at $P=-1.0, 0, 1.0, 3.0, 4.0, 5.0$ and $39.2$ 
in Figs.~\ref{fig22} and \ref{fig23}. 
We have checked 
that if $q$ is large enough as taken in the present study, 
 the wave-number dependence of the eigenvalues $\varepsilon({i,\mathbf{k}})$ 
is very small and we can take $N_k=1$.

The magnetizations are obtained for fixed $\mu$ and for fixed $\nu$ by 
\begin{eqnarray}
M_{\nu}&=& -\frac{\partial E_{\nu}}{\partial h}, \\
M_{\mu}&=& -\frac{\partial E_{\mu}}{\partial h},
\end{eqnarray}
respectively, 
where the derivative with respect to $h$ is calculated by the numerical differentiation. The magnetizations ($M_{\nu}$ and $M_{\mu}$), calculated from $E_{\nu}$ and $E_{\mu}$ in Figs.~\ref{fig25}, \ref{fig26},~\ref{fig22} 
and~\ref{fig23}, are shown in Figs.~\ref{fig27}, ~\ref{fig28} and \ref{fig24}.

\begin{figure}[bt]
\begin{flushleft} \hspace{0.0cm}(a) \end{flushleft}\vspace{-0.3cm}
\includegraphics[width=0.47\textwidth]{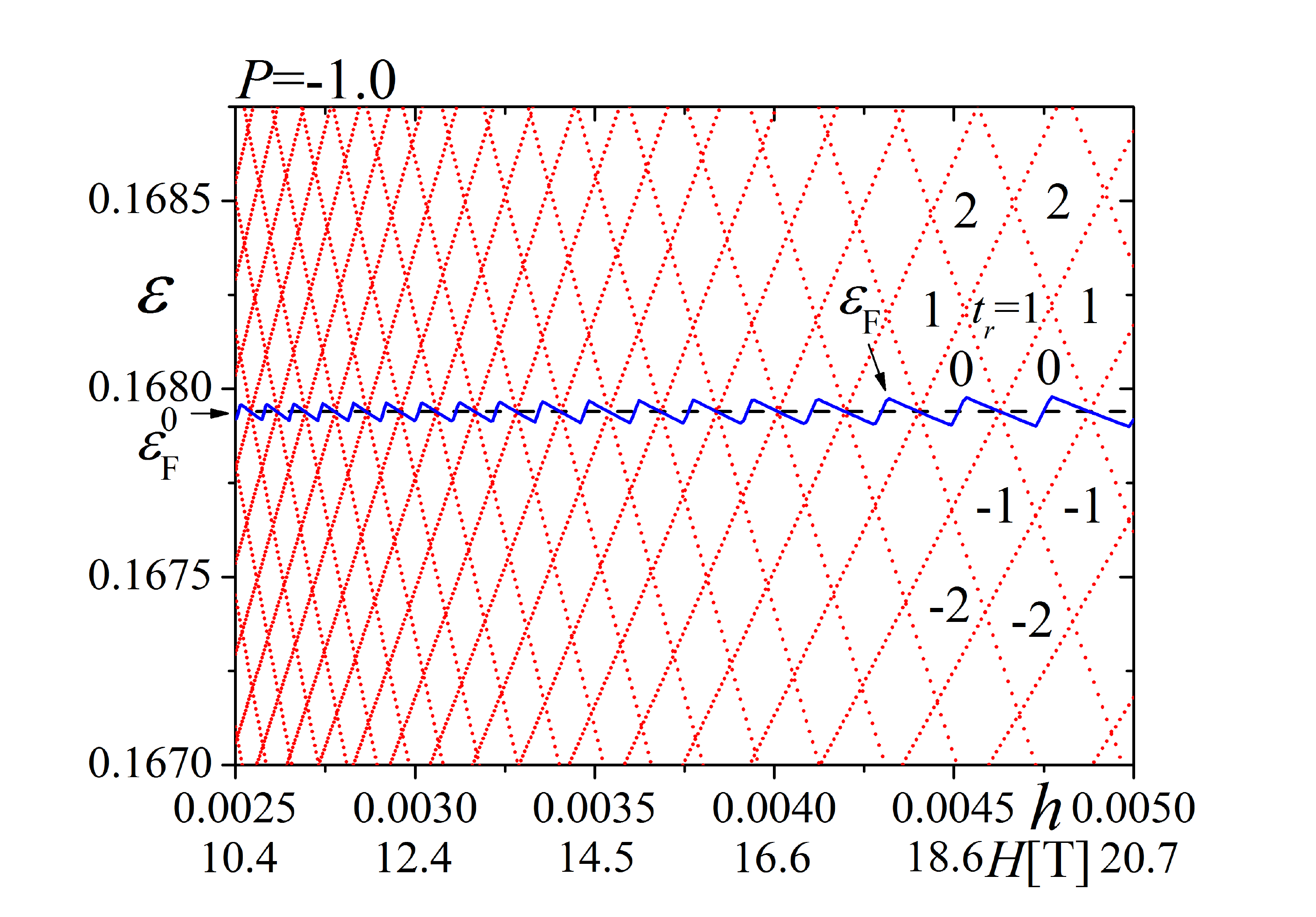}\vspace{-0.2cm}
\begin{flushleft} \hspace{0.0cm}(b) \end{flushleft}\vspace{-0.2cm}
\includegraphics[width=0.47\textwidth]{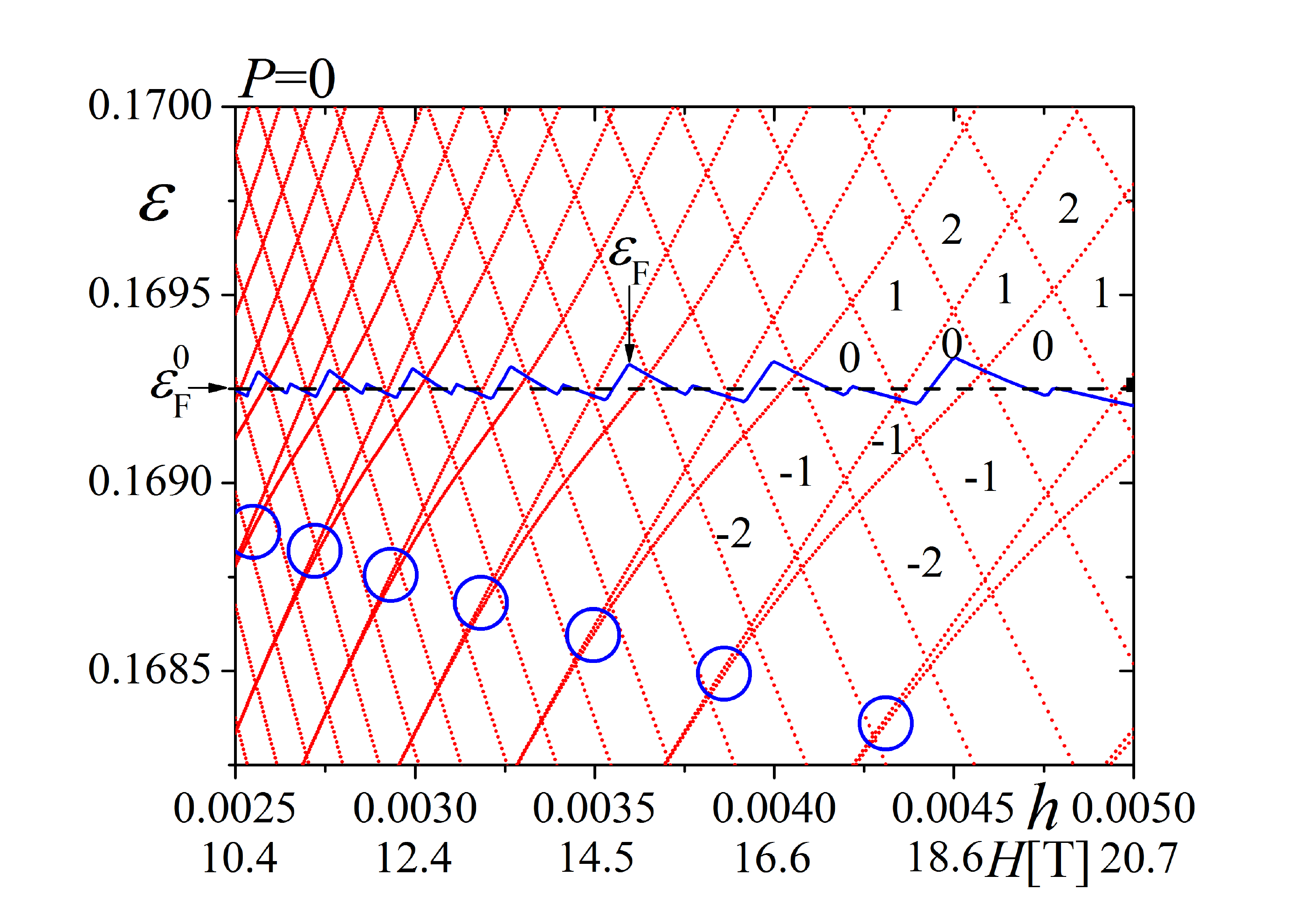}\vspace{-0.2cm}
\begin{flushleft} \hspace{0.0cm}(c) \end{flushleft}\vspace{-0.2cm}
\includegraphics[width=0.47\textwidth]{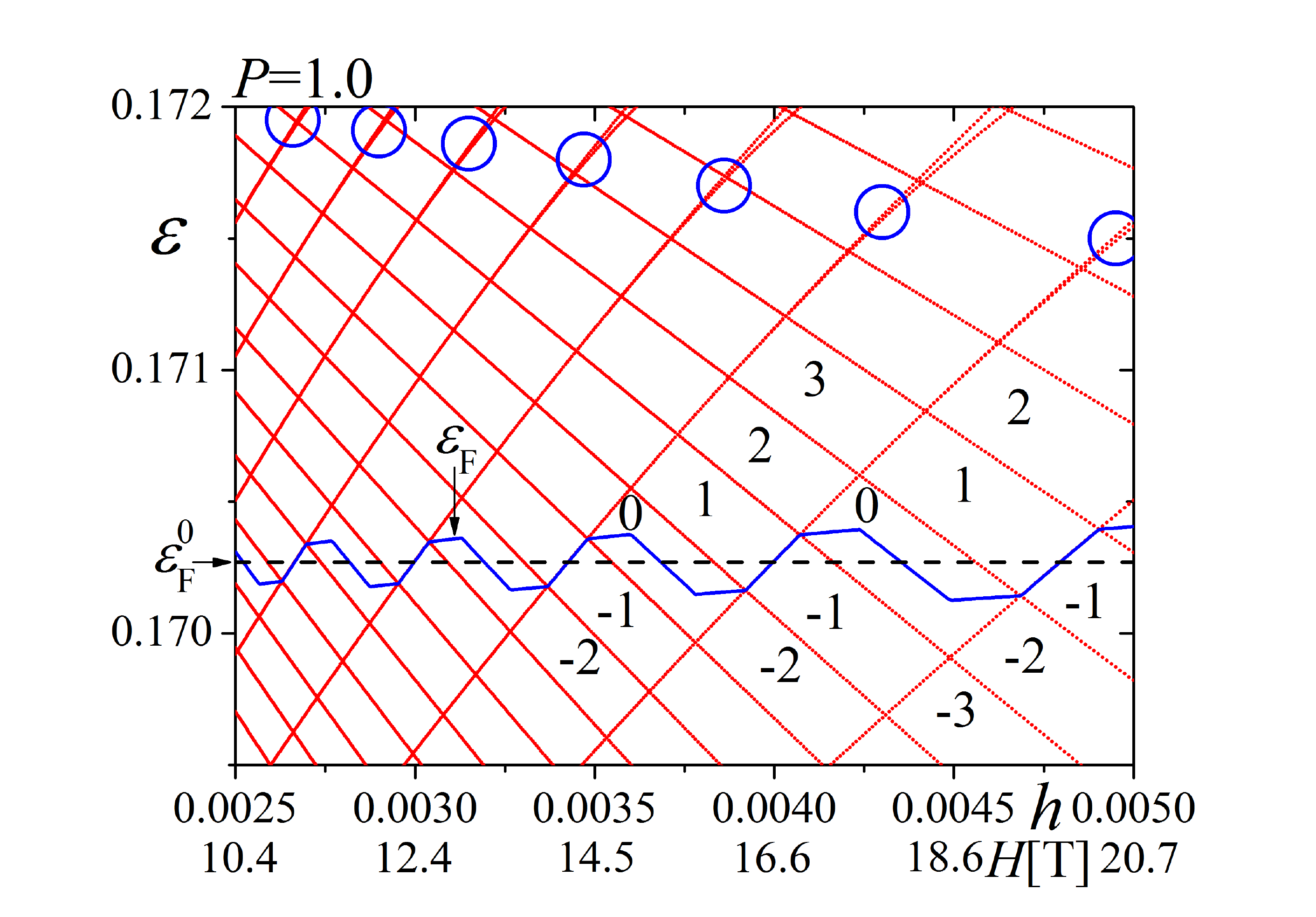}\vspace{-0.2cm}
\caption{
Energies near the Fermi energy as a function of $h$ 
at $P=-1.0$ (a), $P=0$ (b) and $P=1.0$ (c). 
We choose $p=2$ and $400\leq q \leq 800$ ($q=400, 401, \cdots, 799, 800$). 
The energy gaps are labeled by $t_r=\pm 1$, $\pm 2$ and $\pm 3$ with $s_r=3$. 
The Fermi energy $(\varepsilon_{\rm F})$ for  3/4 filling at $h\neq 0$ are 
shown by the blue thin lines. 
The Fermi energies $(\varepsilon_{\rm F}^0)$ for 3/4-filling at $h=0$ are 
 0.16794, 0.16925 and 0.17027 for $P=-1.0$, $0$ and $1.0$, respectively, which are indicated by the black broken lines.
}
\label{fig18}
\end{figure}
%

\begin{figure}[bt]
\begin{flushleft} \hspace{0.5cm}(a) \end{flushleft}\vspace{-0.5cm}
\includegraphics[width=0.5\textwidth]{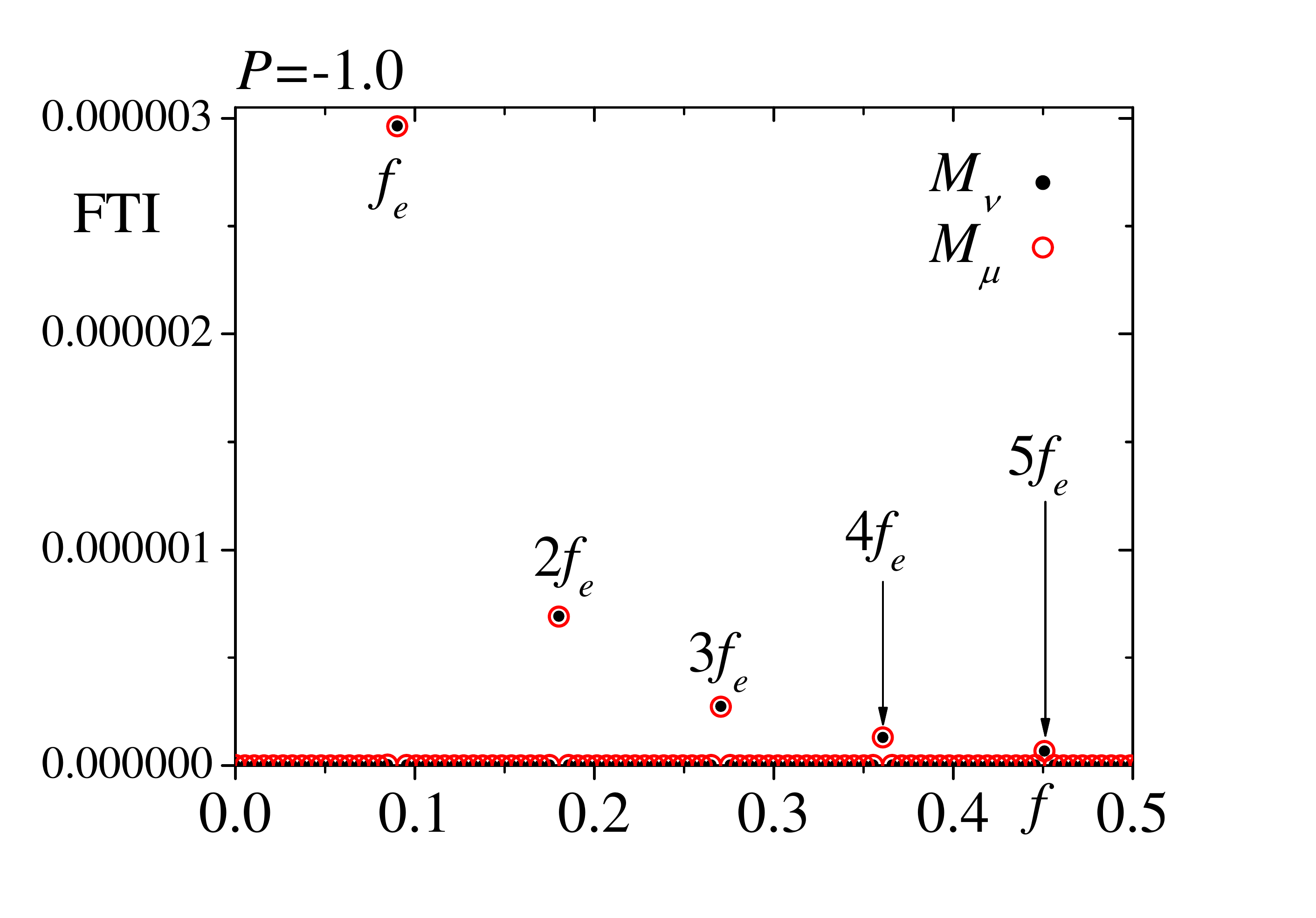}\vspace{-0.5cm}
\begin{flushleft} \hspace{0.5cm}(b) \end{flushleft}\vspace{-0.5cm}
\includegraphics[width=0.5\textwidth]{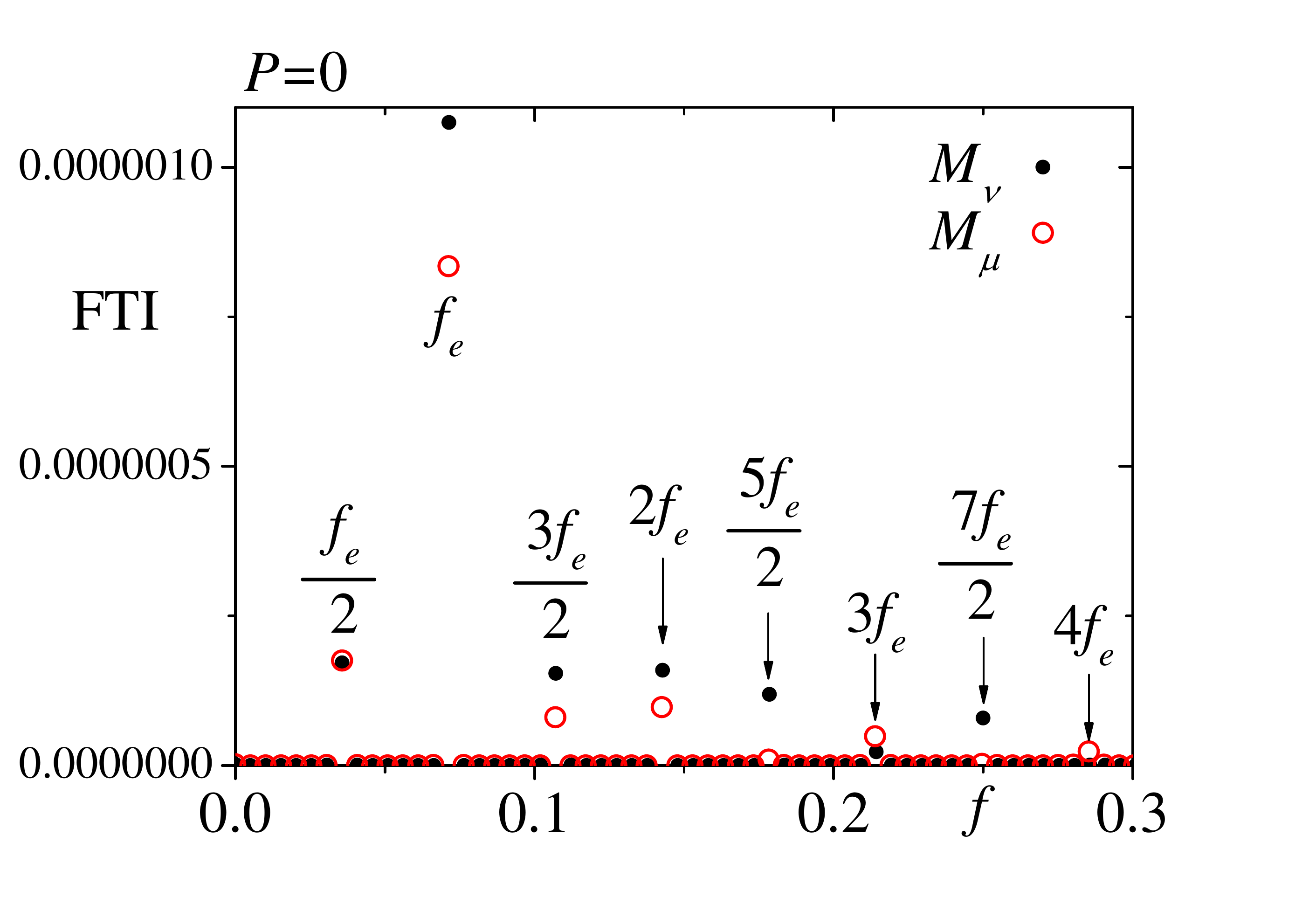}\vspace{-0.5cm}
\begin{flushleft} \hspace{0.5cm}(c) \end{flushleft}\vspace{-0.5cm}
\includegraphics[width=0.5\textwidth]{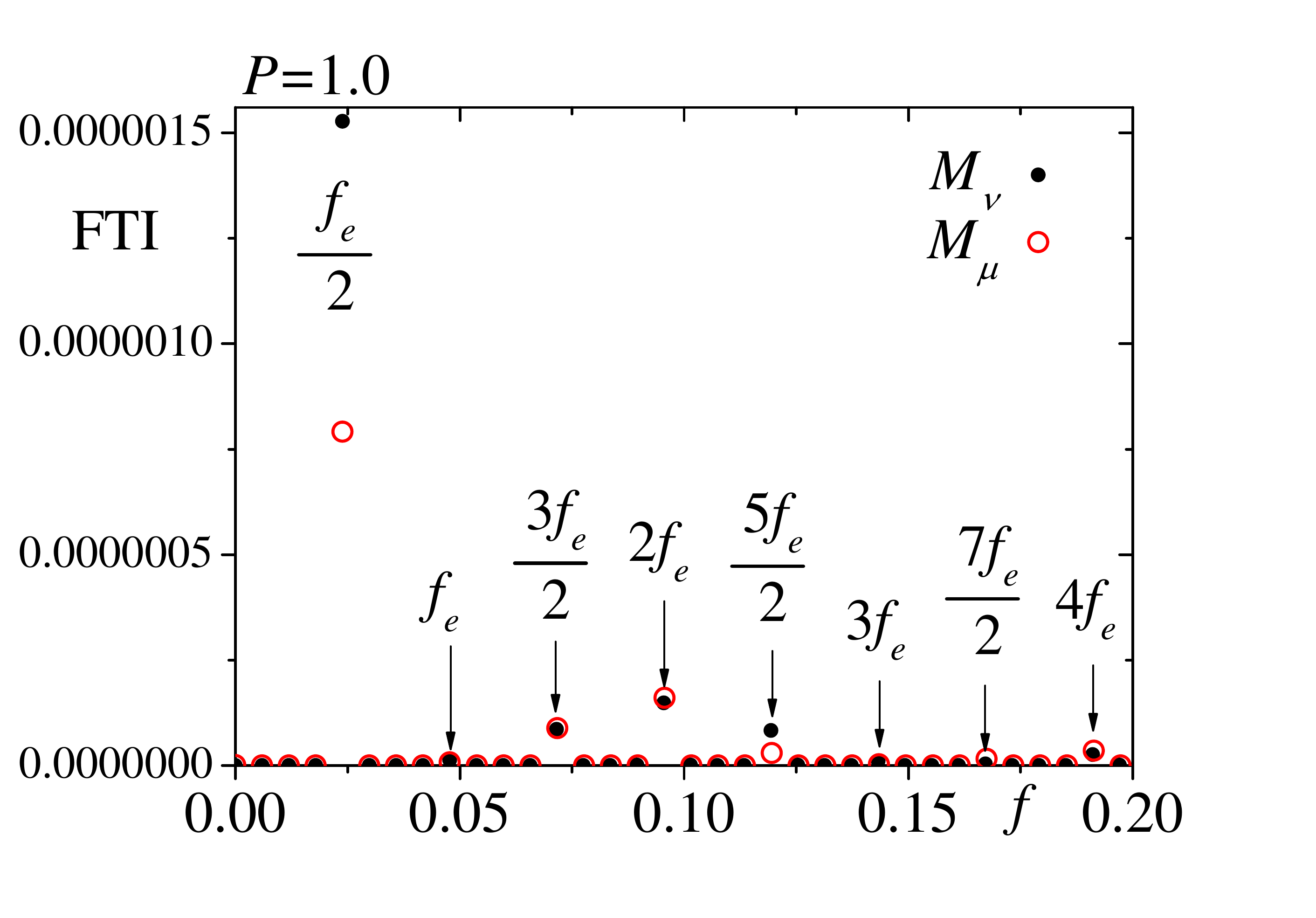}\vspace{-0.5cm}
\caption{
The FTIs of $M_{\nu}$ and $M_{\mu}$ in Fig. \ref{fig27} and Fig. \ref{fig28} for $P=-1.0$ (a), $P=0$ (b) and $P=1.0$ (c). In (a), $2L=188.5$ and $1/h_c=299.5$. In (b), $2L=196$ and $1/h_c=300.75$ for $M_{\nu}$ and $2L=196.3$ and $1/h_c=301.55$ for $M_{\mu}$. In (c), $2L=167.5$ and $1/h_c=297.25$ for $M_{\nu}$ and $2L=167.25$ and $1/h_c=292.875$ for $M_{\mu}$. 
}
\label{fig29}
\end{figure}

\subsection{Semi-metallic system at $P < 3.0$}

At $P=-1.0$ there are an electron pocket (its area is $A_e/A_{\textrm{BZ}}=0.0903$, where $A_{\textrm{BZ}}$ is the area of the first Brillouin zone) and a hole pocket ($A_h$), as shown in Fig.~\ref{fig8} (a). These areas are the same ($A_e=A_h$). These areas become $A_e/A_{\textrm{BZ}}=A_h/A_{\textrm{BZ}}=0.0715$ at $P=0$ (Fig. ~\ref{fig8} (b)). 
There is a small neck in an electron pocket around $\mathbf{k} =(\pi/a,0)$ or $\mathbf{k} =(-\pi/a,0)$, as indicated by black arrows in Figs.~\ref{fig8} (a) and (b). 
At $P\gtrsim 0.2$ an electron pocket separates around the small neck into two small electron pockets with the half area, $A_{e}/2$. At $P=1.0$, $A_e/A_{\textrm{BZ}}=A_h/A_{\textrm{BZ}}=0.0479$ and $A_e/(2A_{\textrm{BZ}})=0.0240$, as shown in Fig.~\ref{fig8}(c).

The obtained magnetizations are periodically oscillated as a function of $1/h$, as shown in Figs. \ref{fig27} 
and \ref{fig28}, where main frequencies $f$ are 0.0902 at $P=-1.0$, 0.0714 at $P=0$ and 0.0239 at $P=1.0$. 
These frequencies correspond to the areas of electron and hole pockets at $h=0$, which are considered as the dHvA oscillation. 
Actually, the Landau levels for an electron pocket (upward-sloping lines) and for a hole pocket (downward-sloping lines) 
are crossing the Fermi energy at $h=0$ (a black dotted line), as shown in Figs.~\ref{fig18} (a), (b) and (c). There is no dHvA oscillation at
$P=3.0$ in the region $1/400 \lesssim h \lesssim 1/200$, which is consistent with the fact that there are no Fermi surface at $P=3.0$ and $h=0$.

The Fourier transform intensities (FTIs) of $M_{\nu}$ and $M_{\mu}$ which are 
defined in Appendix \ref{AppendixD} are plotted in Fig.~\ref{fig29}.
The FTIs of $M_{\nu}$ and $M_{\mu}$ at $P=-1.0$ are almost the same (Fig.~\ref{fig29}(a)) because 
the oscillation of the Fermi energy as a function of $h$ (a blue thin line) is small, as shown in Fig.~\ref{fig18} (a). 
There are the peaks of the FTIs at $f=f_e=A_e/A_{\textrm{BZ}}=f_h=A_h/A_{\textrm{BZ}}$, $f=2f_e=2f_h$, $f=3f_e=3f_h$ {\it etc.} and the height of the $l$-th harmonics is smaller for larger $l$. These are the same as that of the LK formula of a closed Fermi surface.

The FTIs of $M_{\nu}$ are different from those of $M_{\mu}$ in the cases of $P=0$ and $P=1.0$ (Fig.~\ref{fig29}(b)), where the oscillations of the Fermi energy are not small, as shown in Figs.~\ref{fig18} (b) and (c).

We discuss the largest peaks at $f_e$ and the second largest peaks at $f_e/2$ in $M_{\nu}$ and $M_{\mu}$ at $P=0$, as shown in Fig.~\ref{fig29} (b). The dHvA oscillation with $f_e$ is due to the crossing of {\it not degenerated} Landau levels and $\varepsilon^0_{\textrm{F}}$ (see Fig. \ref{fig18} (b)). These Landau levels come from an electron pocket with $A_e$ in Fig.~\ref{fig8}(b). On the other hand, the frequency, $f_e/2$, corresponds to the half area of an electron pocket (the green area in Fig.~\ref{fig8}(b)). The dHvA oscillation with $f_e/2$ is explained by the magnetic breakdown in the semiclassical 
theory (i.e., a realization of an effectively closed electron's motion by the tunneling). 
In our numerical study, the effect of the magnetic breakdown 
is taken into account naturally. Therefore, 
we can understand the magnetic breakdown as the separations of the Landau levels around blue circles in Fig. \ref{fig18} (b). 
When the magnetic field and the energy are lower than blue circles, the Landau levels are almost degenerated, which are due to two small electron pockets with $A_e/2$. 
Although these degenerated Landau levels do not 
cross $\varepsilon^0_{\textrm{F}}$ (see Fig. \ref{fig18} (b)), since the separations of the Landau levels (blue circles) occur below and close to $\varepsilon^0_{\textrm{F}}$, the dHvA oscillation with $f_e/2$ becomes finite in Fig.~\ref{fig29} (b). 
At $P=-1.0$, the separations are not seen in the regions of the magnetic field and the energy (Fig. \ref{fig18} (a)). As a result, there is no peak at $f_e/2$.


There are the largest peaks at $f_{e}/2$ in $M_{\nu}$ and $M_{\mu}$ at $P=1.0$ (Fig.~\ref{fig29}(c)), which is consistent with the result expected in the LK formula because of two electron pockets with $A_e/2$ (red circles in Fig.~\ref{fig8}(c)). Since there is a hole pocket with $A_e$ (a blue circle in Fig.~\ref{fig8}(c)), the dHvA oscillations with $f_e$ and its higher harmonics are expected. 
In fact, $\varepsilon^0_{\textrm{F}}$ (a black dotted line) crosses Landau levels not only for two small electron pockets (upward-sloping lines) but also for a hole pocket (downward-sloping lines), as shown in Fig. \ref{fig18} (c).  However, peaks at $f_e$ and at $3f_e$ are very small. The anomalous smallness of these peaks is not expected in the LK formula.

\begin{figure}[bt]
\begin{flushleft} \hspace{0.5cm}(a) \end{flushleft}\vspace{-0.5cm}
\includegraphics[width=0.5\textwidth]{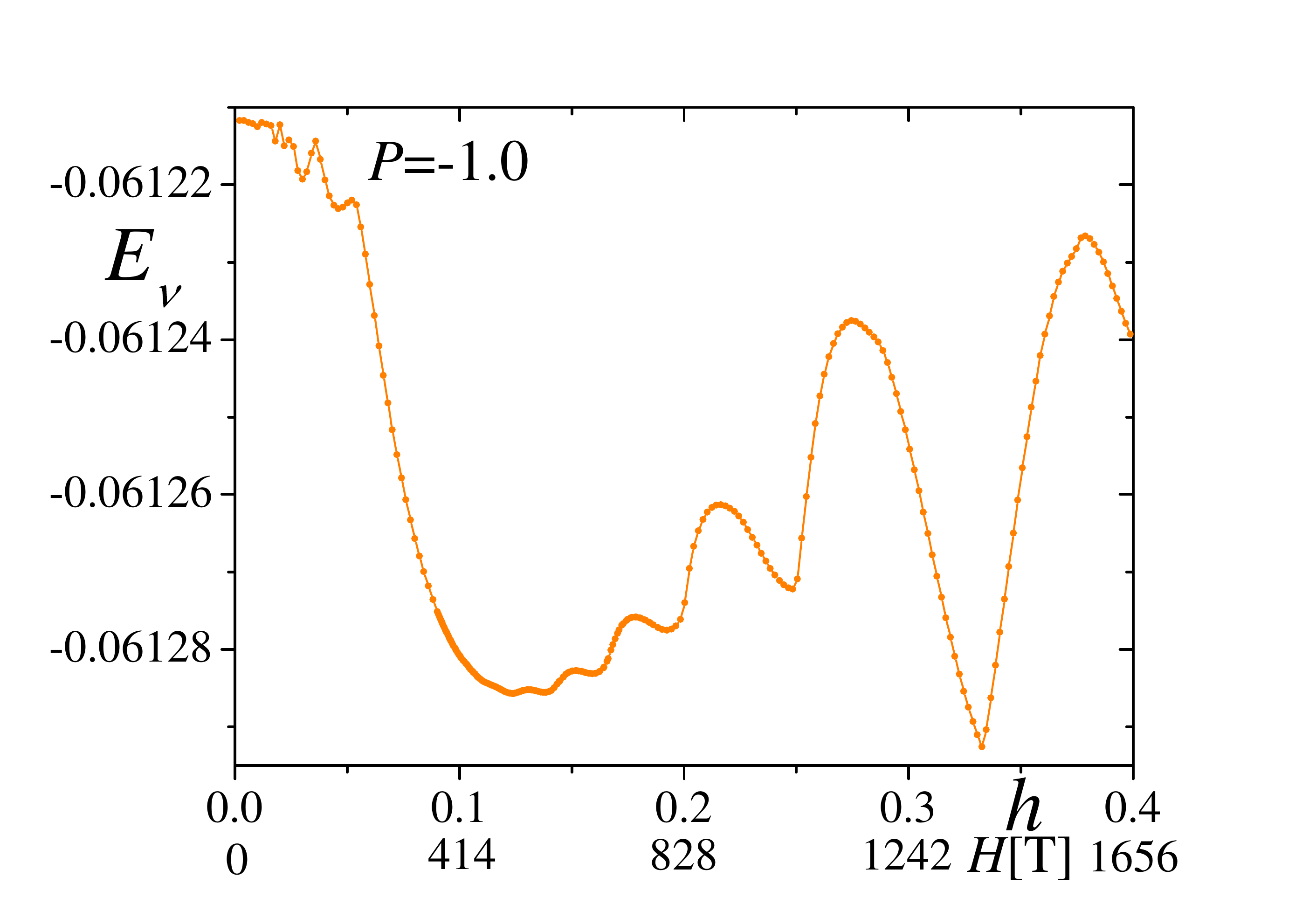}\vspace{-0.5cm}
\begin{flushleft} \hspace{0.5cm}(b) \end{flushleft}\vspace{-0.5cm}
\includegraphics[width=0.5\textwidth]{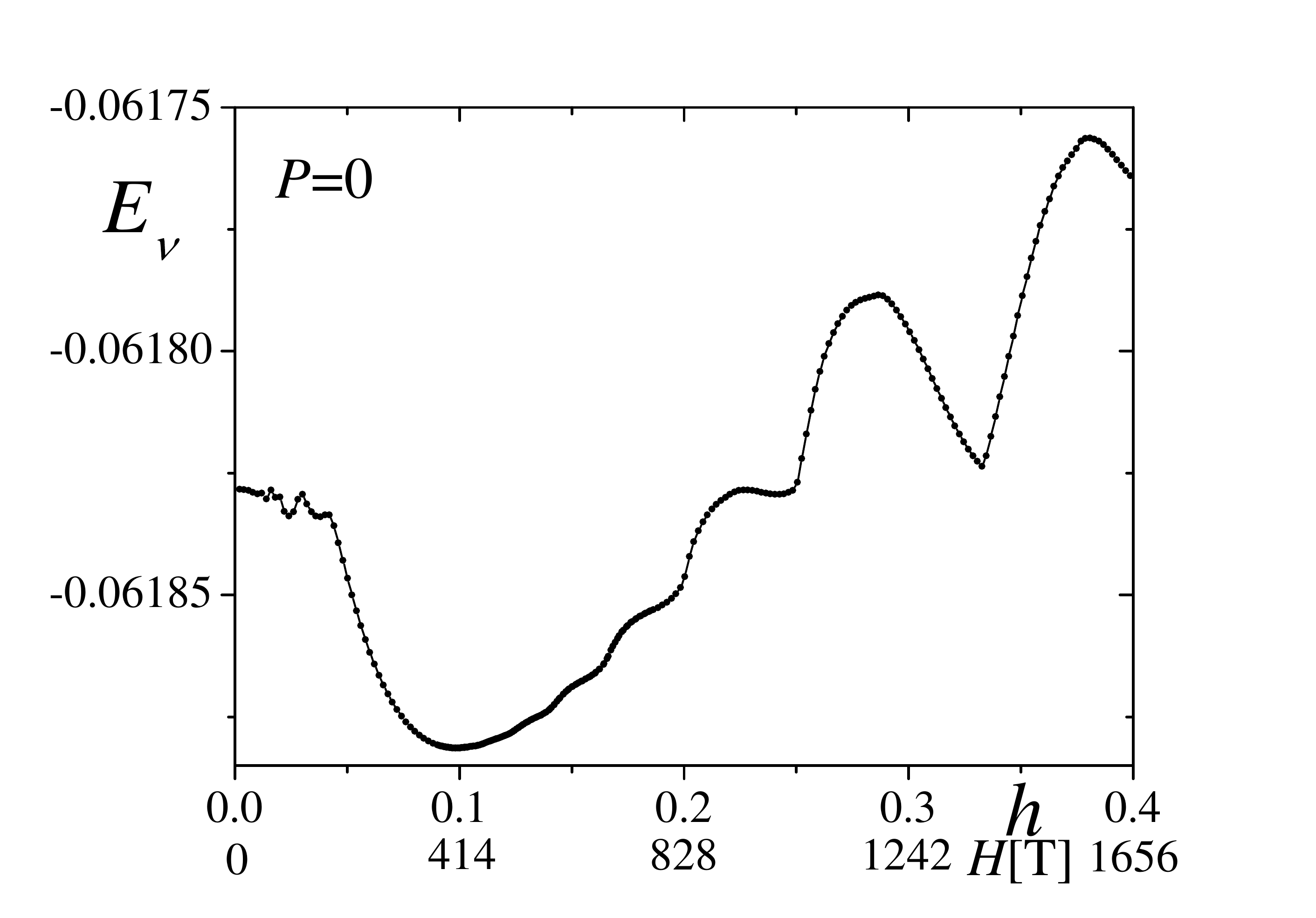}\vspace{-0.5cm}
\begin{flushleft} \hspace{0.5cm}(c) \end{flushleft}\vspace{-0.5cm}
\includegraphics[width=0.5\textwidth]{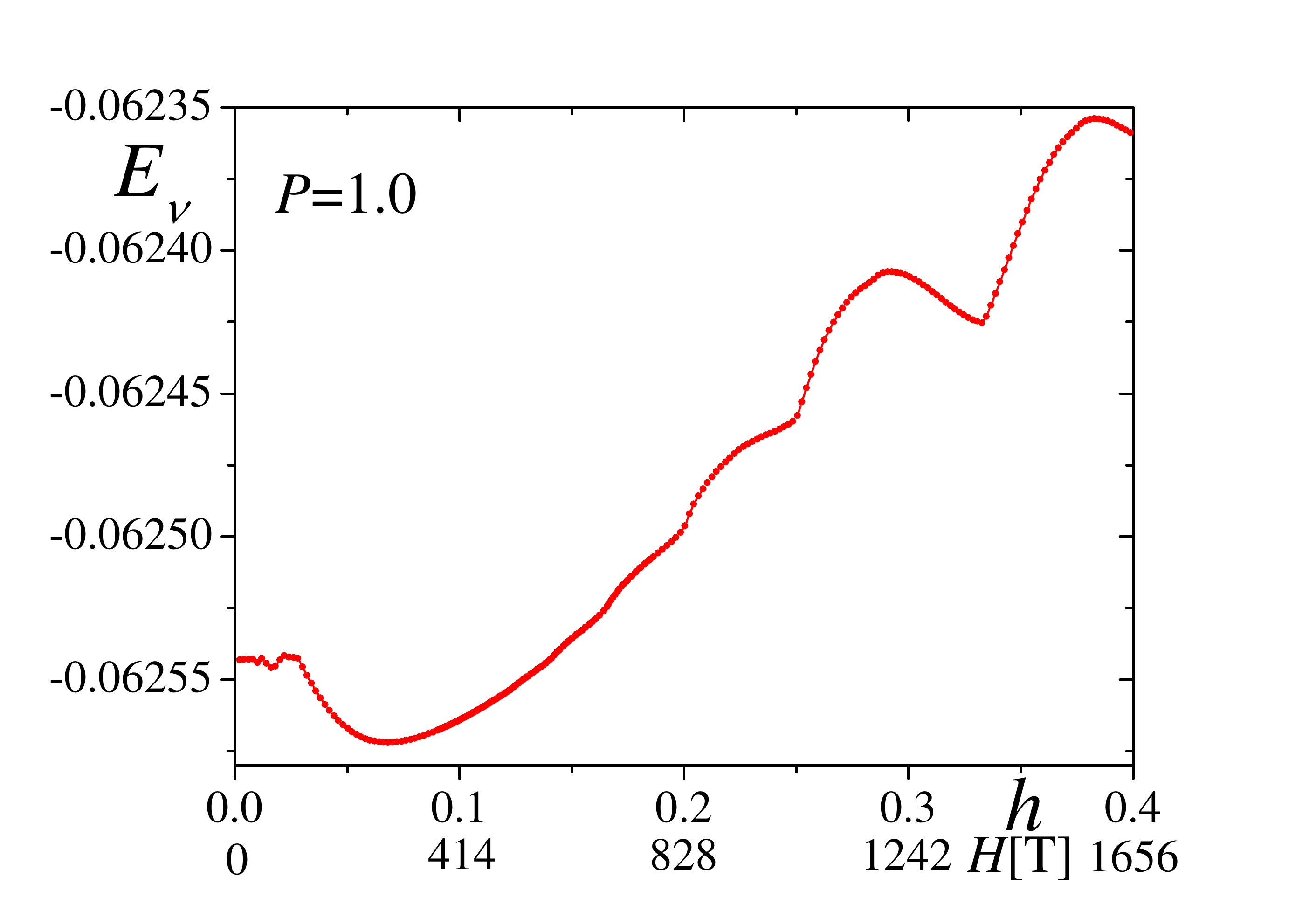}\vspace{-0.5cm}
\caption{
Total energies as a function of $h$ at $P=-1.0$ (a), $P=0$ (b) and $P=1.0$ (c). 
We take $h=\frac{2m}{499}$ ($m=1, 2, 3, \cdots, 99$), $h=\frac{2(2m-1)}{998}$  ($m=1, 2, 3, \cdots, 100$) and $h=\frac{32}{q}$  ($q=173, 175, \cdots, 355$).
}
\label{fig22}
\end{figure}
\begin{figure}[bt]
\begin{flushleft} \hspace{0.5cm}(a) \end{flushleft}\vspace{-0.5cm}
\includegraphics[width=0.5\textwidth]{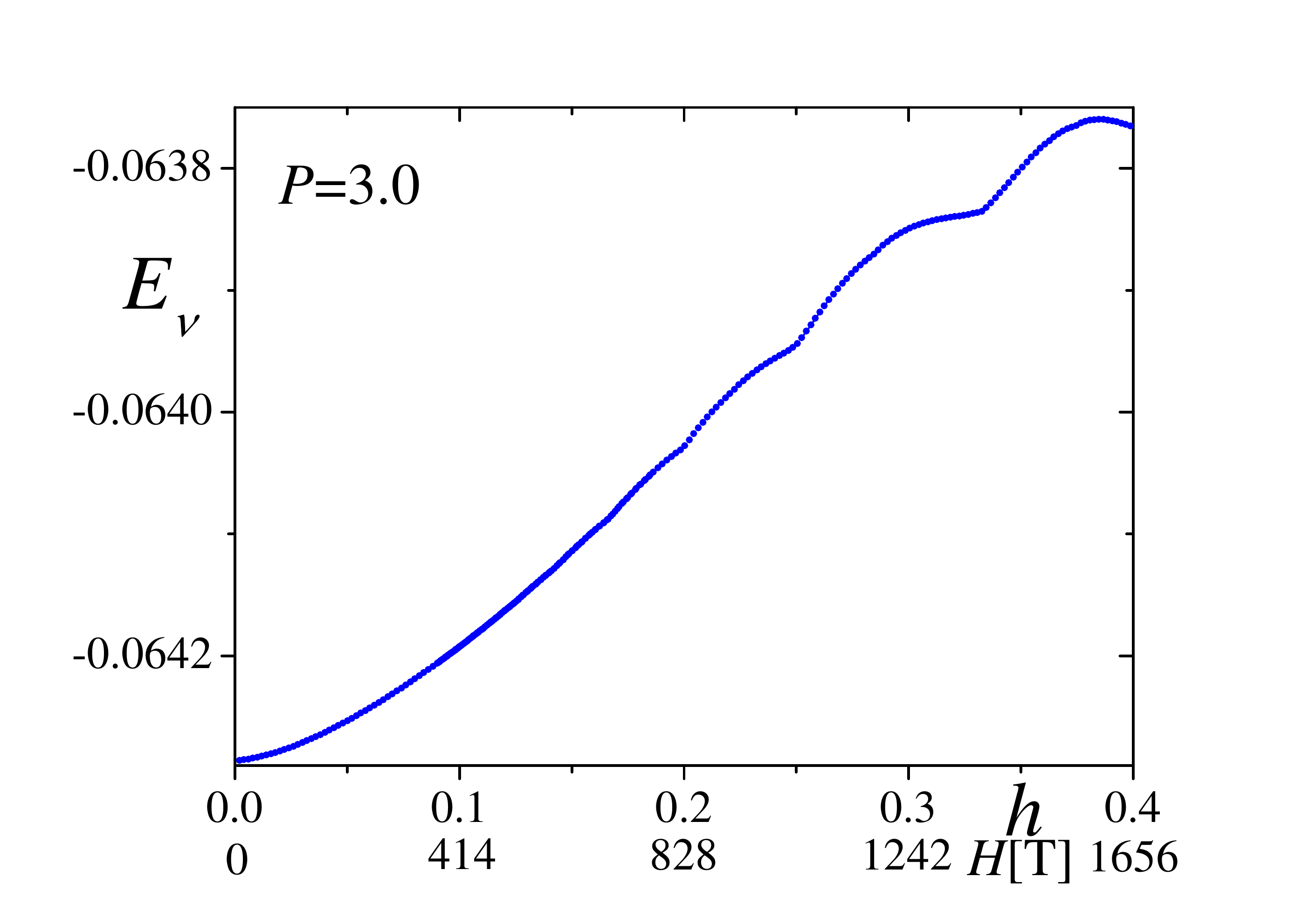}\vspace{-0.5cm}
\begin{flushleft} \hspace{0.5cm}(b) \end{flushleft}\vspace{-0.5cm}
\includegraphics[width=0.5\textwidth]{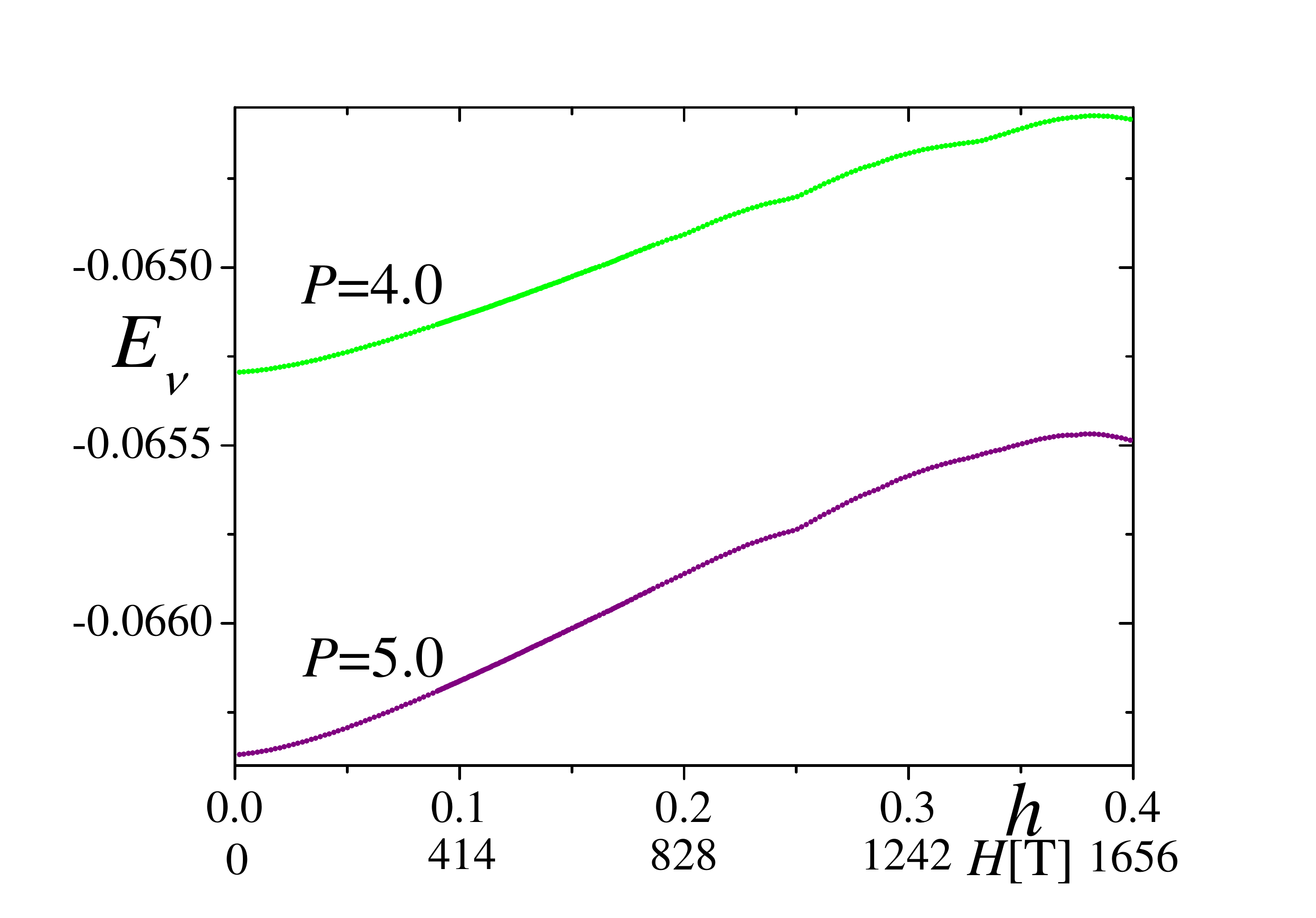}\vspace{-0.5cm}
\begin{flushleft} \hspace{0.5cm}(c) \end{flushleft}\vspace{-0.5cm}
\includegraphics[width=0.5\textwidth]{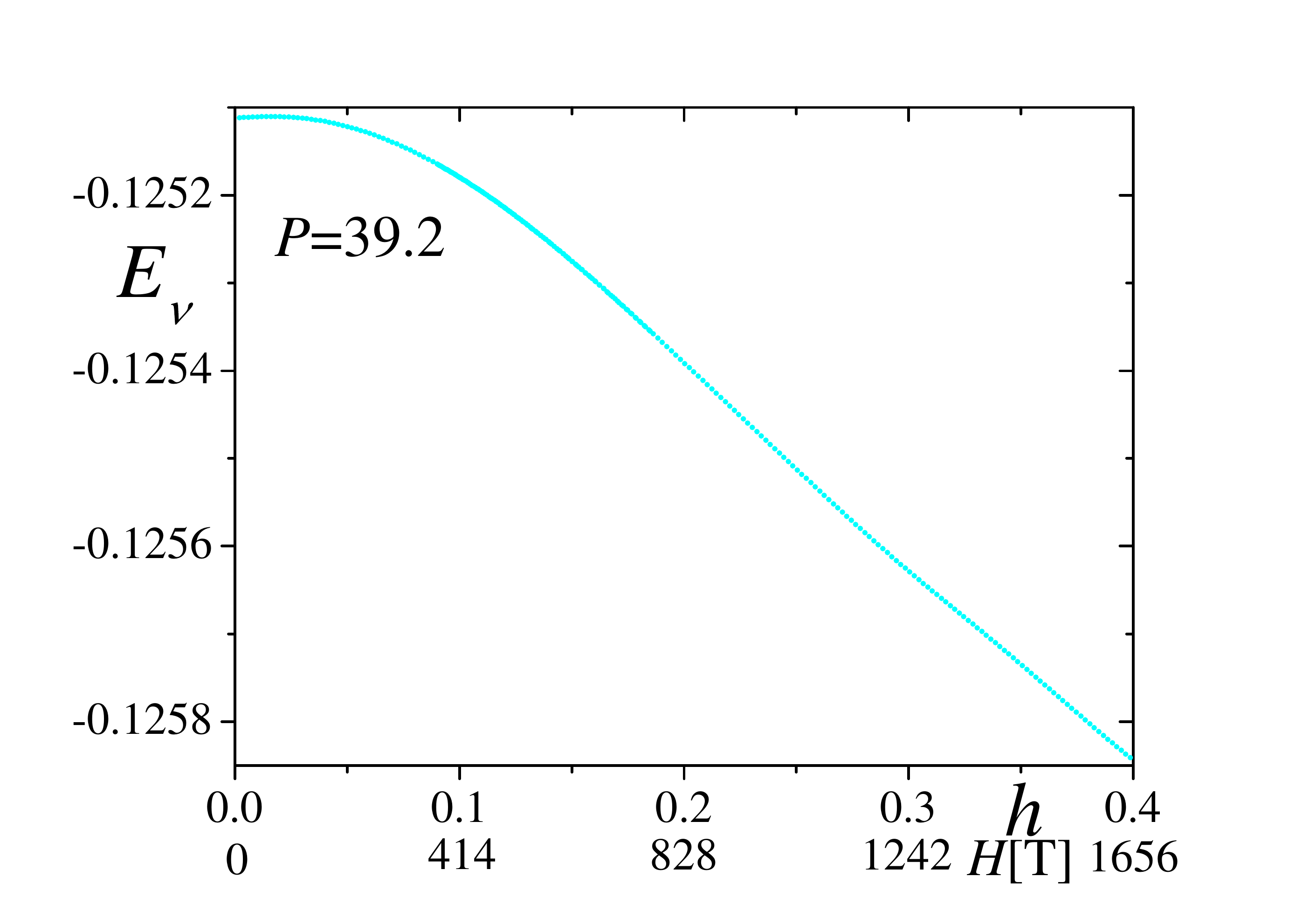}\vspace{-0.5cm}
\caption{
Total energies as a function of $h$ at $P=3.0$ (a), $4.0$ and $5.0$ (b) and $P=39.2$ (c). 
We take $h=\frac{2m}{499}$ ($m=1, 2, 3, \cdots, 99$), $h=\frac{2(2m-1)}{998}$ 
 ($m=1, 2, 3, \cdots, 100$) and $h=\frac{32}{q}$  ($q=173, 175, \cdots, 355$).
}
\label{fig23}
\end{figure}
\begin{figure}[bt]
\begin{flushleft} \hspace{0.5cm}(a) \end{flushleft}\vspace{-0.5cm}
\includegraphics[width=0.5\textwidth]{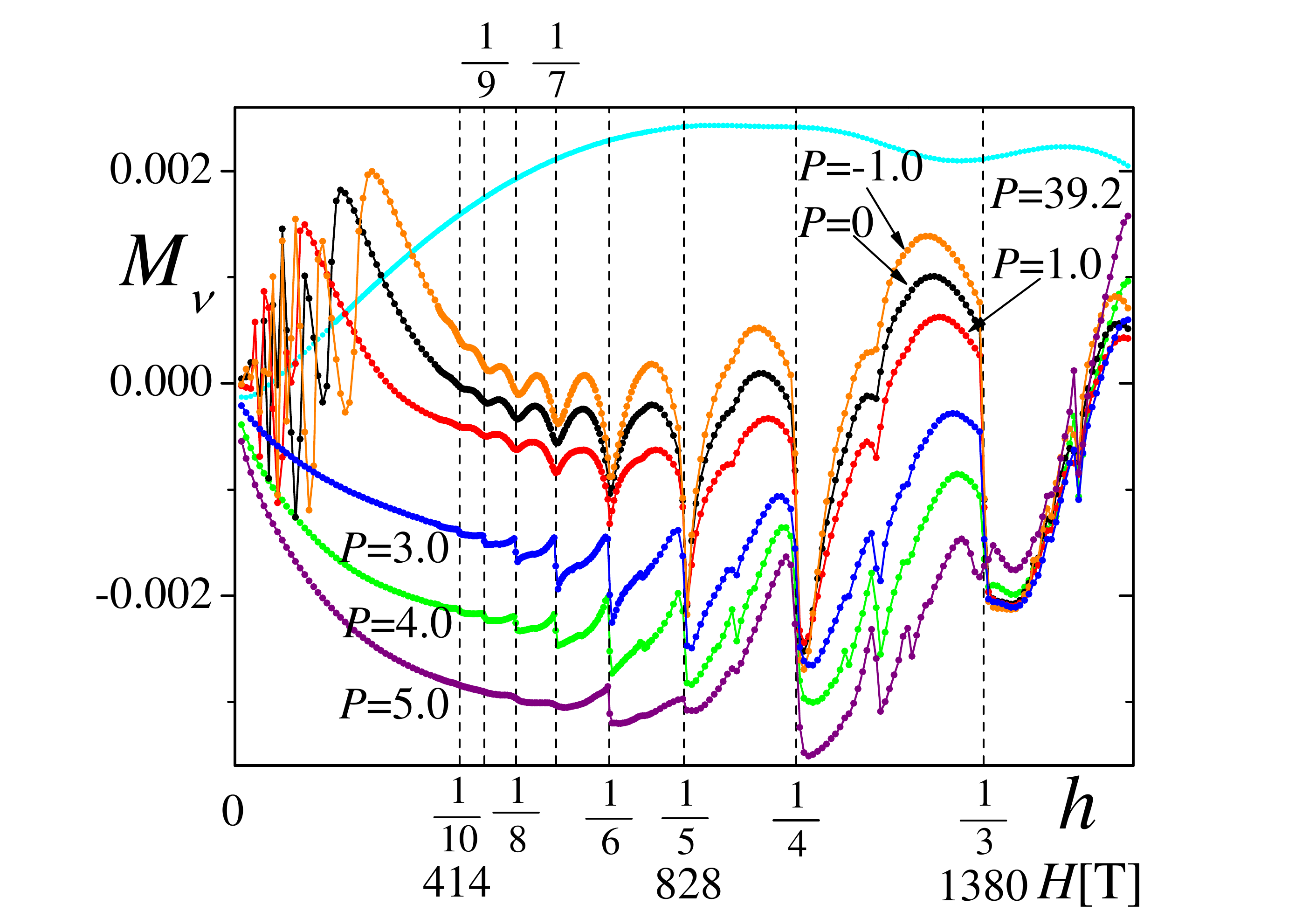}\vspace{-0.5cm}
\begin{flushleft} \hspace{0.5cm}(b)\end{flushleft}\vspace{-0.5cm}
\includegraphics[width=0.5\textwidth]{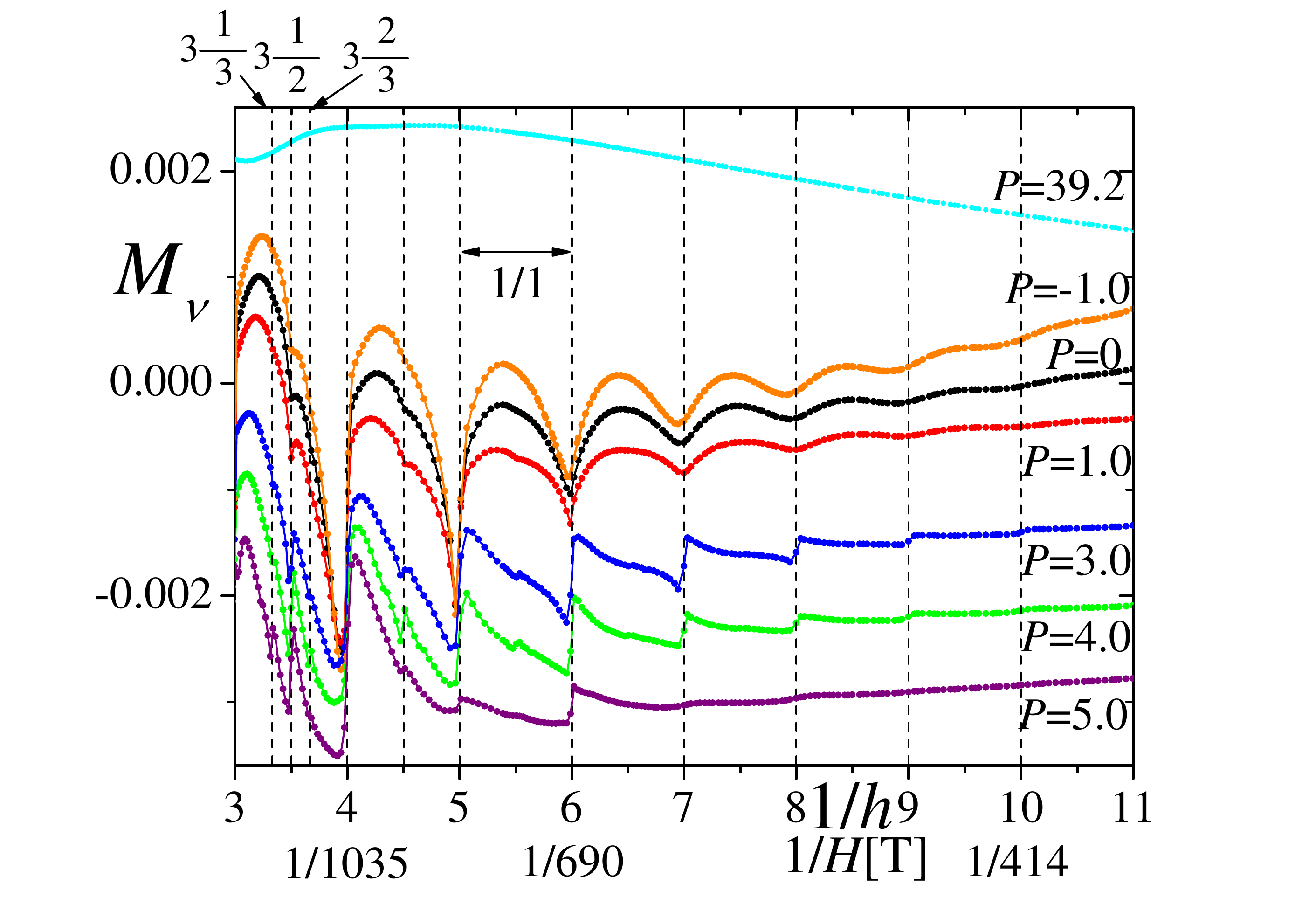}\vspace{-0.0cm}
\caption{
Magnetizations with 3/4-filling as a function of $h$ (a) and as a function of $1/h$ (b) 
at $P=-1.0$, $0$, $1.0$, $3.0, 4.0, 5.0$ and $39.2$. 
The same values of $h=p/q$ as those in Figs.~\ref{fig22} and \ref{fig23} are used.
}
\label{fig24}
\end{figure}

\subsection{Dirac fermions system at $P \geq 3.0$}\label{V_B}

The magnetization as a function of $1/h$ oscillates periodically with the frequency corresponding to the 
area of the Brillouin zone ($f=1$) at $h \gtrsim 1/10$,
corresponding to about 414 T in $\alpha$-(BEDT-TTF)$_2$I$_3$, as shown in Fig.~\ref{fig24}(b). 
This oscillation appears even in the case that there are 
no Fermi surface ($P\geq 3.0$). 
Although the usual dHvA oscillation is caused by the crossing of the chemical potential 
and the Landau levels, the obtained dHvA-like oscillation is not due to the crossing. 
The origin of the oscillation is the Harper broadening between a blue thick line and a green thin line, as shown in Fig. \ref{fig13} (c).

The magnitude of this dHvA-like oscillation becomes small as $1/h$ increases. 
The wave form of the oscillation at $P\geq 3.0$ is similar 
to the sawtooth pattern for fixed electron number rather than 
the inverse sawtooth pattern in the LK formula for the fixed chemical potential 
(see Appendix \ref{AppendixC}). 
The wave form at $P<3.0$ 
is not saw-tooth but sinusoidal-like, as shown in Fig.~\ref{fig24}.

The dHvA-like oscillation with $f=1$ has also been obtained on the honeycomb lattice before\cite{KH2014}. If we use the lattice constant ($\simeq 0.246$ nm) of graphene, the oscillation appears at a very high magnetic field 
($\sim 5000$ T)\cite{KH2014}. Since the flux through the unit cell in $\alpha$-(BEDT-TTF)$_2$I$_3$ is larger than that in graphene, 
it is expected to find the dHvA-like oscillation at lower magnetic field. 
Very recently, the dHvA-like oscillation in the system with no Fermi surface has been observed in SmB$_6$\cite{Tan2015} and studied theoretically in many models\cite{Knoll2015,Erten2016, Zhang2016,Pal2016}.

\section{Conclusions}




We find a {\it ``three-quarter''-Dirac point} in the tight-binding model of $\alpha$-(BEDT-TTF)$_2$I$_3$ 
at $P = 2.3$ kbar, although the {\it ``three-quarter''-Dirac point} is hidden by 
the metal-insulator transition at low temperature in the real system. At that pressure the Dirac cone is critically tilted and we have to take account 
of quadratic terms. Then the dispersion relation is linear in three directions and parabolic in one direction at {\it ``three-quarter''-Dirac points}.

We obtain the energy as a function of the magnetic field by taking the complex hopping integrals. We find the $H^{4/5}$-dependence due to {\it ``three-quarter''-Dirac points} at $P=2.3$~kbar. We also obtain the $H^2$-dependence at the intermediate magnetic field strength at $P=3.0$~kbar, which is caused by the laid Dirac cone.


We numerically obtain the magnetic-field-dependences of the total energy and the magnetization
(de Haas-van Alphen (dHvA) oscillation) in both cases of fixed electron number and fixed chemical potential.
At $P=0$~kbar we find the FTI at the frequency corresponding to the half of the area of an electron pocket which is not a closed orbit. This oscillation attributes to the smooth separations of the Landau levels as a function of the magnetic field. 
This is a quantum mechanical picture of the magnetic breakdown. 
At $P=1.0$~kbar the FTI at the frequency corresponding to the area of the hole pocket is shown to be quite small, which cannot be explained by the semiclassical LK formula.

When the system is considered to be massless Dirac fermions at $P > 3.0$ kbar, 
we find the unusual dHvA-like oscillation with the period corresponding to the area of the first Brillouin zone at $H \gtrsim 400$~T. This oscillation is thought to be due to the Harper broadening of the Landau levels, which is similar to the case on honeycomb lattice\cite{KH2014}.

Recently, the Landau levels in massless Dirac fermions have been directly observed from the scanning tunneling spectra\cite{Guohong}. 
The Landau levels for {\it ``three-quarter''-Dirac cones} and for almost laid Dirac cones 
are expected to be observed if the charge ordering is removed. 
Furthermore, the results for the usual dHvA oscillation and the unusual dHvA-like oscillation shown in this study will be observed. 
However, in order to suppress the charge ordering, 
the high pressures 
(the uniaxial pressure of $P\gtrsim 10$ kbar\cite{Tajima2002} and the hydrostatic 
pressure of $P\gtrsim 11-12$ kbar\cite{Dong}) 
are needed. 
Therefore, the experiments in the semi-metallic state may be difficult to be observed at low temperature. 
It will be possible to observe the obtained results, if the critically tilted Dirac cones 
or overtilted Dirac cones are realized in other systems such as ultra cold atoms\cite{Tarruell2012}
and graphene under uniaxial strain\cite{Georbig2008}.


\section*{Acknowledgement}
One of the authors (KK) thanks Naoya Tajima and Harukazu Yoshino for useful discussions and information of experiments and band calculations.

\appendix

\section{semiclassical Landau quantization of energy}
\label{Appendix0}

In the semiclassical theory, the energies of two-dimensional electrons are quantized into the Landau levels 
($\varepsilon_n$ with integer $n$)
when the area of the closed Fermi surface in the wave-number space $A(\varepsilon_n)$ at $H=0$
equals to the quantized value proportional to the magnetic field, i.e., 
\begin{equation}
A(\varepsilon_n) = (n+\gamma) \frac{2 \pi e H}{\hbar c},
 \label{eqquantization0}
\end{equation}
where 
$n$ is an integer, $e$ is the electron charge, $c$ is the speed of light, $\hbar$ is the Planck constant divided by $2 \pi$ 
and $\gamma$ is a phase factor which can be determined from the quantum mechanical calculation ($\gamma=1/2$ for massive free electrons 
and $\gamma=0$ for massless Dirac fermions).

\section{dHvA oscillation and Lifshitz and Kosevich (LK) formula}
\label{AppendixC}

The magnetization in metals oscillates periodically as 
a function of the inverse of the magnetic field at low temperatures, which is called the dHvA oscillation\cite{shoenberg}. The period of the dHvA oscillation is proportional to the extremal area of the closed Fermi surface in a plane perpendicular to the magnetic field in the semiclassical theory.

For the dHvA oscillation, the Lifshitz and Kosevich (LK) formula\cite{shoenberg,LK} based on the semiclassical theory\cite{Onsager} is derived in the case of the fixed 
chemical potential $\mu$ (the grand canonical ensemble). 
The generalized LK formula at $T=0$ for 
the two-dimensional metals with no impurity is given by
\begin{equation}
M^{\rm LK}=-\frac{e}{2\pi^2 c\hbar}
 \frac{A}{\frac{\partial A}{\partial \mu}}
\sum_{l=1}^{\infty}\frac{1}{l}\sin\left[2\pi l\left(\frac{F}{H}-\gamma\right)
\right],\label{LK_0}
\end{equation}
where its frequency ($F$) is given by 
\begin{equation}
F=\frac{c \hbar A}{2 \pi e},
\end{equation}
where 
$A$ is the area of the closed Fermi surface at $H=0$. When we use $h$ of Eq. (\ref{eq_pq}) instead of $H$ in Eq. (\ref{LK_0}), we get 
\begin{equation}
\frac{F}{H}=\frac{f^{}}{h}, 
\end{equation}
where 
\begin{equation}
f=\frac{A}{A_{\rm BZ}} 
\end{equation}
and $A_{\rm BZ}=4\pi^2/(ab)$ is the area of the Brillouin zone. 
The amplitude of the oscillation at $T=0$ is independent of $h$ in the LK formula. 

In the two-dimensional system with a closed Fermi surface at $h=0$, 
Eq. (\ref{LK_0}) becomes the saw tooth shape. If the electron number is fixed in that system, the chemical potential jumps from a Landau level to another Landau level 
as the magnetic field increases. As a result, the saw tooth pattern as a function of $1/h$ is inverted\cite{shoenberg}.

\section{energy at $H=0$}
\label{appendixA}
%
The Bravais lattice in our model (Fig. \ref{fig1} (a)) is given by 
\begin{eqnarray}
{\bf v}_1=(a, 0)  \label{eqn:1.1}
\end{eqnarray}
and 
\begin{eqnarray}
{\bf v}_2=(0, b).  \label{eqn:1.2}
\end{eqnarray}
The Hamiltonian with the hoppings between neighboring sites 
($t_{\mathrm{a1}}$, $t_{\mathrm{a2}}$, $t_{\mathrm{a3}}$,
$t_{\mathrm{b1}}$, $t_{\mathrm{b2}}$, $t_{\mathrm{b3}}$, and $t_{\mathrm{b4}}$,
see Fig.~\ref{fig1}) is given by 
\begin{align}
 \hat{\cal H}_0 &=\displaystyle 
\sum_{{\bf r}_j} \bigg\{
    t_{a1}c^\dag_{4,{\bf r}_j}c_{3,{\bf r}_j}
 + t_{a1}c^\dag_{4,{\bf r}_j}c_{3,{\bf r}_j+{\bf v}_2}
 + t_{a2}c^\dag_{1,{\bf r}_j} c_{2,{\bf r}_j}                 \nonumber \\
&+t_{a3}c^\dag_{1,{\bf r}_j}c_{2,{\bf r}_j-{\bf v}_2}
  +t_{b1}c^\dag_{1,{\bf r}_j}c_{4,{\bf r}_j}
  +t_{b1}c^\dag_{2,{\bf r}_j}c_{4,{\bf r}_j-{\bf v}_1}    \nonumber \\
 &+t_{b2}c^\dag_{1,{\bf r}_j}c_{3,{\bf r}_j}                
  +t_{b2}c^\dag_{2,{\bf r}_j}c_{3,{\bf r}_j-{\bf v}_1+{\bf v}_2} 
  +t_{b3}c^\dag_{1,{\bf r}_j}c_{3,{\bf r}_j-{\bf v}_1}   \nonumber \\
 &+t_{b3}c^\dag_{2,{\bf r}_j}c_{3,{\bf r}_j+{\bf v}_2}  
 +t_{b4}c^\dag_{1,{\bf r}_j}c_{4,{\bf r}_j-{\bf v}_1}
 +t_{b4}c^\dag_{2,{\bf r}_j}c_{4,{\bf r}_j} \nonumber \\
 &+ h.c. \bigg\},
\label{eqn:01}
\end{align}
where $c^\dag_{1,{\bf r}_j}$, $c^\dag_{2,{\bf r}_j}$, 
$c^{\dag}_{3,{\bf r}_j}$ and $c^{\dag}_{4,{\bf r}_j}$ ($c_{1,{\bf r}_j}$, $c_{2,{\bf r}_j}$, $c_{3,{\bf r}_j}$ and $c_{4,{\bf r}_j}$) are creation (annihilation) operators for 
1, 2, 3 and 4  sites in $j$-th unit cell, respectively. 
By using the following Fourier transform, 
\begin{eqnarray}
c_{1,{\bf r}_j} 
&=&\displaystyle \sum_{\bf k} e^{i\mathbf{k}\cdot{\bf r}_j}c_{1,\bf k},\label{eqn:02}\\
c_{2,{\bf r}_j}&=&\displaystyle \sum_{\bf k} e^{i\mathbf{k}
\cdot({\bf r}_j+{\bf v}_2/2)}c_{2,{\bf k}}, \label{eqn:03}\\
c_{3,{\bf r}_j} 
&=&\displaystyle \sum_{\bf k} e^{i\mathbf{k}\cdot({\bf r}_j+{\bf v}_{1}/2-{\bf v}_{2}/4)}
c_{3,{\bf k}},\label{eqn:04}\\
c_{4,{\bf r}_j}&=&\displaystyle \sum_{\bf k} e^{i\mathbf{k}
\cdot({\bf r}_j+{\bf v}_{1}/2+{\bf v}_{2}/4)}c_{4,{\bf k}}, \label{eqn:05}
\end{eqnarray}
we obtain the Hamiltonian  as 
\begin{eqnarray}
{\cal {\hat H}}_{0}=\sum_{{\bf k}}
C^{\dagger}_{{\bf k}} \hat{\varepsilon}_{\bf k}C_{{\bf k}}, 
\label{ham_0}
\end{eqnarray}
where 
\begin{eqnarray}
{C}^{\dagger}_{{\bf k}}=(c^\dag_{1,{\bf k}}, c^{\dag}_{2,{\bf k}},c^{\dag}_{3,{\bf k}}, c^{\dag}_{4,{\bf k}}),
\end{eqnarray}
\begin{eqnarray}
{C}_{{\bf k}}=\begin{pmatrix}
c_{1,{\bf k}}\\
c_{2,{\bf k}}\\
c_{3,{\bf k}}\\
c_{4,{\bf k}}
\end{pmatrix} ,
\end{eqnarray}
and $\hat{\varepsilon}_{\bf k}$ is a 4$\times 4$ matrix as follows; 
\begin{equation}
\hat{\varepsilon}_{\bf k}
= \left( \begin{array}{cccc}
0  & A_2         & B_2        & B_1   \\
A_2^*       & 0  & B_2^*      & B_1^* \\
B_2^*       & B_2         & 0 & A_1   \\
B_1^*       & B_1         & A_1        & 0
 \end{array} \right),
\label{eqbulkH}
\end{equation}
with
\begin{align}
A_1 &= 2 t_{a1} \cos \frac{k_y}{2}, \\
A_2 &= t_{a2} e^{i\frac{1}{2}k_y} + t_{a3} e^{-i\frac{1}{2}k_y}, \\
B_1 &= t_{b1} e^{i(\frac{1}{2}k_x + \frac{1}{4}k_y)} 
    +  t_{b4} e^{i(-\frac{1}{2}k_x + \frac{1}{4}k_y)},  \\
B_2 &= t_{b2} e^{i(\frac{1}{2}k_x-\frac{1}{4}k_y)} 
    +  t_{b3} e^{i(-\frac{1}{2}k_x-\frac{1}{4}k_y)}.  
\end{align}
If $t_{a1}=t_{a2}=t_{a3}=0$ (i.e. $A_1=A_2=0$), 
 the eigenvalues of the matrix in
 Eq. (\ref{eqbulkH}) have been obtained by Mori\cite{Mori2010} as
\begin{equation}
 \varepsilon^0_{\mathbf{k}} =\pm \sqrt{ |B_1|^2+|B_2|^2 
  \pm \sqrt{ (B_1^2 +  B_2^2) (B_1^{*2}+ B_2^{*2})}}.
\end{equation}
When $A_1$ and $A_2$ are not zero,
the eigenvalues are not simple, although the 
analytical solutions can be obtained because of the quartic equation. 
We studied the energy in both cases of the bulk state and edge state\cite{Hasegawa2011}.

%
\begin{figure}[bt]
\begin{center}
\begin{flushleft} \hspace{0.5cm}(a) 
\end{flushleft}\vspace{-0.0cm}
\includegraphics[width=0.5\textwidth]{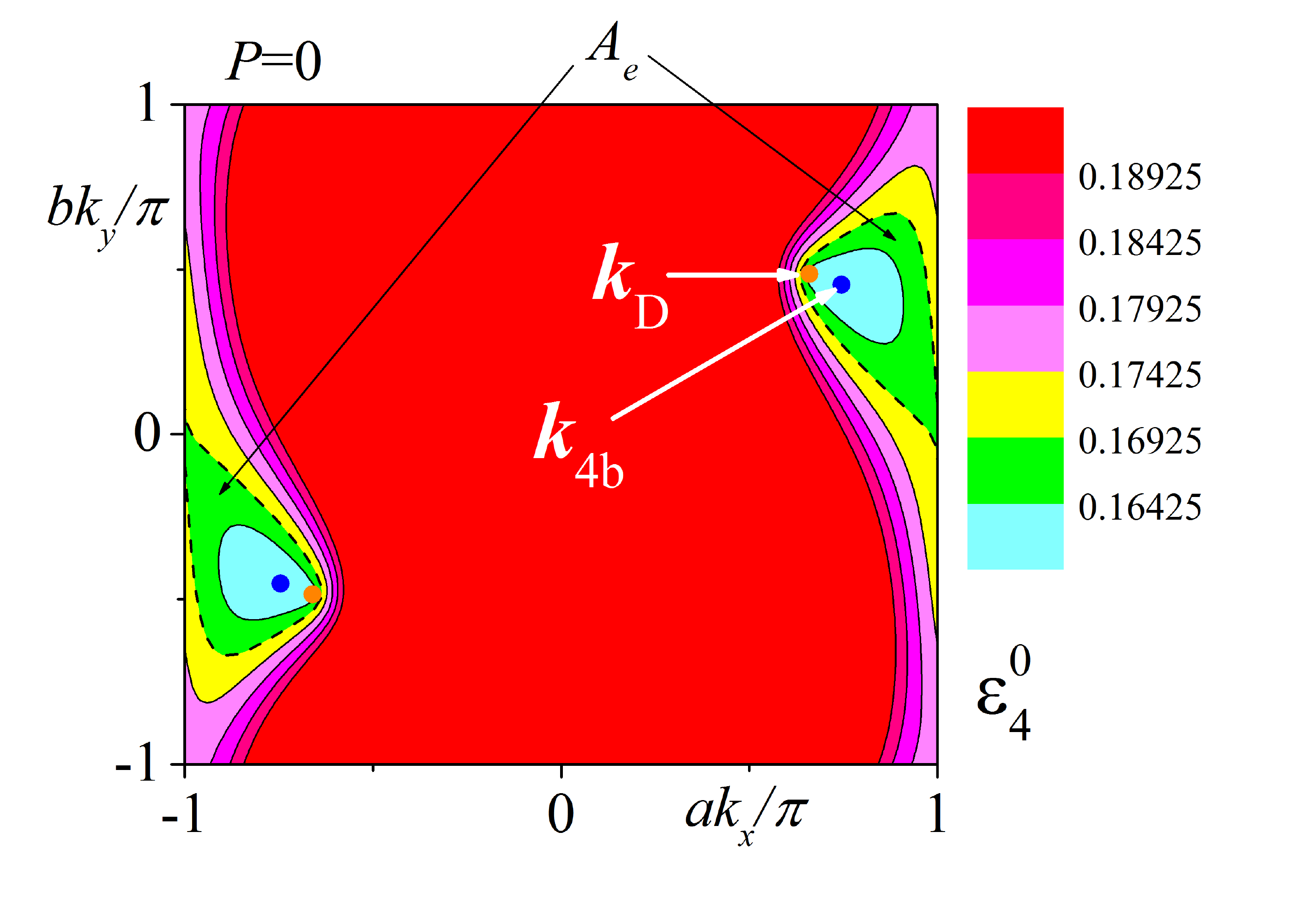}\vspace{-0.0cm}
\begin{flushleft} \hspace{0.5cm}(b) \end{flushleft}\vspace{-0.0cm}
\includegraphics[width=0.5\textwidth]{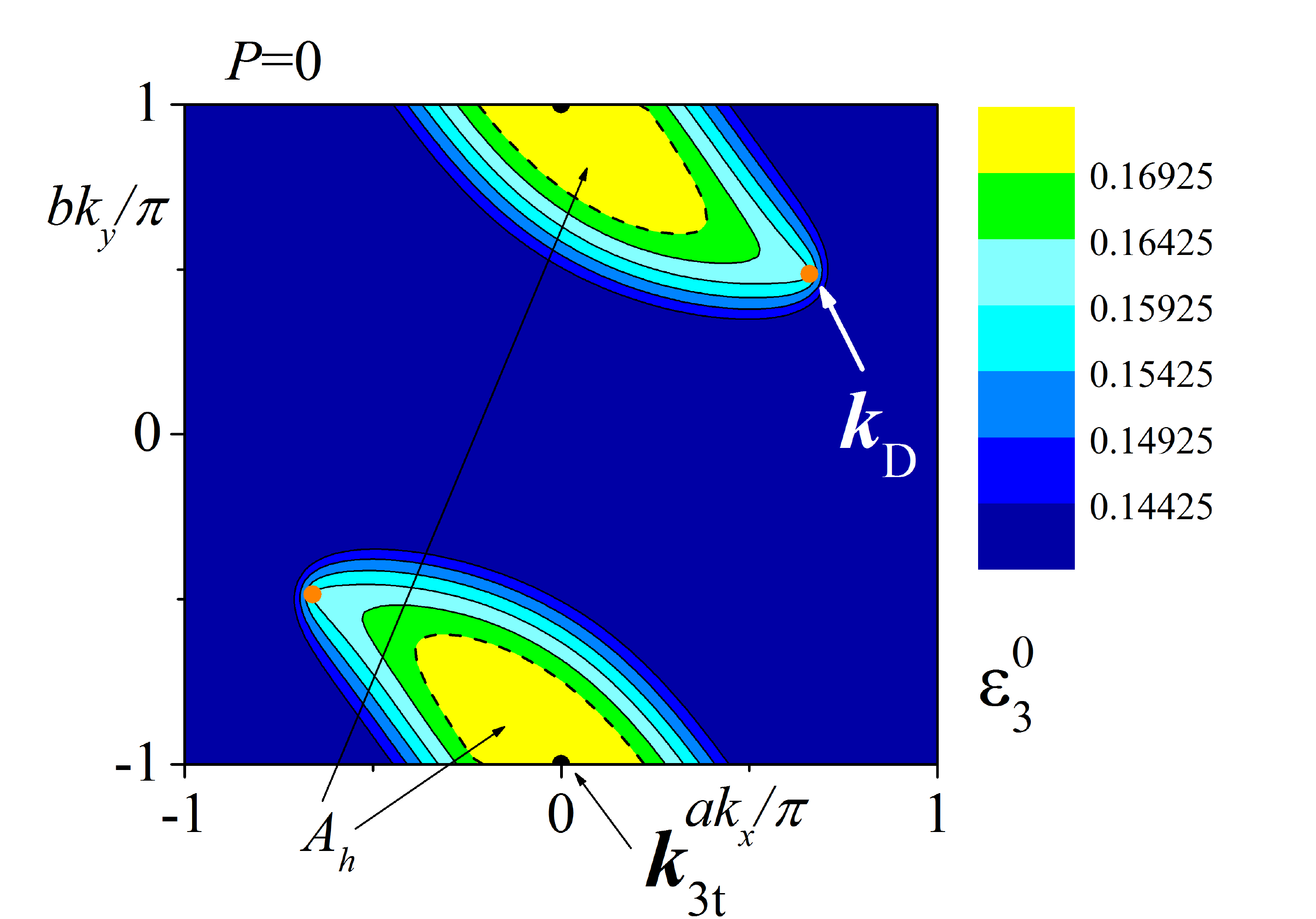}\vspace{-0.5cm}
\end{center}
\caption{
(Color online) Contour plots of the fourth 
band (a) and the third band (b) at $P = 0$.
The electron and hole pockets depicted by dotted black lines are 
the Fermi surface at $\varepsilon = \varepsilon^0_{\rm F}\simeq0.16925$. 
The areas of these electron and hole pockets ($A_e$ and $A_h$) are 
about $0.0715$ 
 of the area of the first Brillouin zone ($A_{\rm BZ}$).
The third and the fourth bands touch 
at two Dirac points (orange points), $\pm{\bf k}_{\rm D}\simeq \pm(0.6600\pi/a, 0.4854\pi/b)$. 
The third band has a top energy ($\varepsilon_{\rm 3t}^0$) at a black point,
 ${\bf k}_{\rm 3t}=(0, \pi/b)$. The fourth band has the bottom energy ($\varepsilon_{\rm 4b}^0$) at two blue points, $\pm{\bf k}_{\rm 4b}=\pm(0.7455\pi/a, 0.4530\pi/b)$.}
\label{fig3}
\end{figure}
\begin{figure}[bt]
\begin{center}
\begin{flushleft} \hspace{0.5cm}(a) \end{flushleft}\vspace{-0.0cm}
\includegraphics[width=0.5\textwidth]{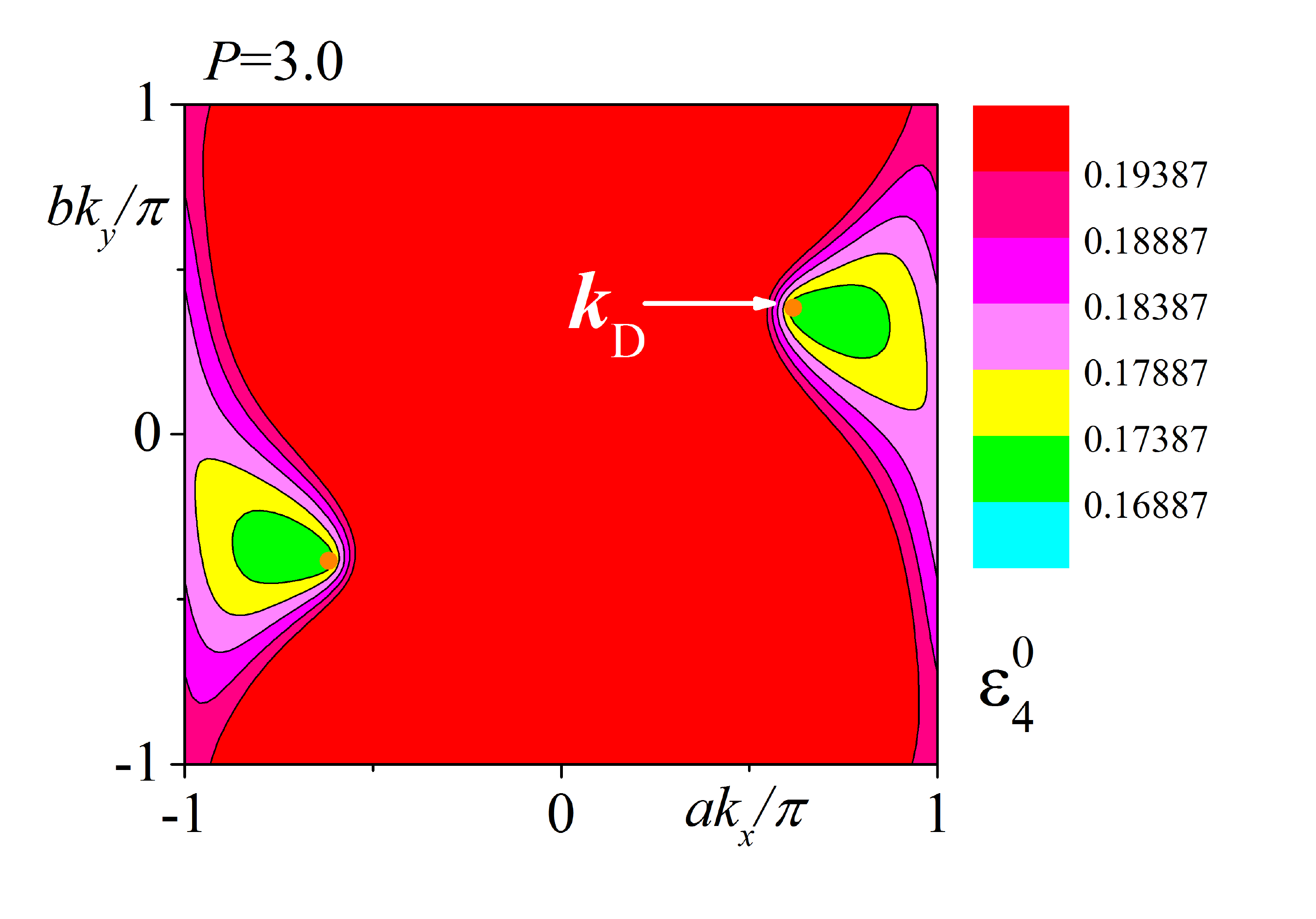}\vspace{-0.0cm}
\begin{flushleft} \hspace{0.5cm}(b) \end{flushleft}\vspace{-0.0cm}
\includegraphics[width=0.5\textwidth]{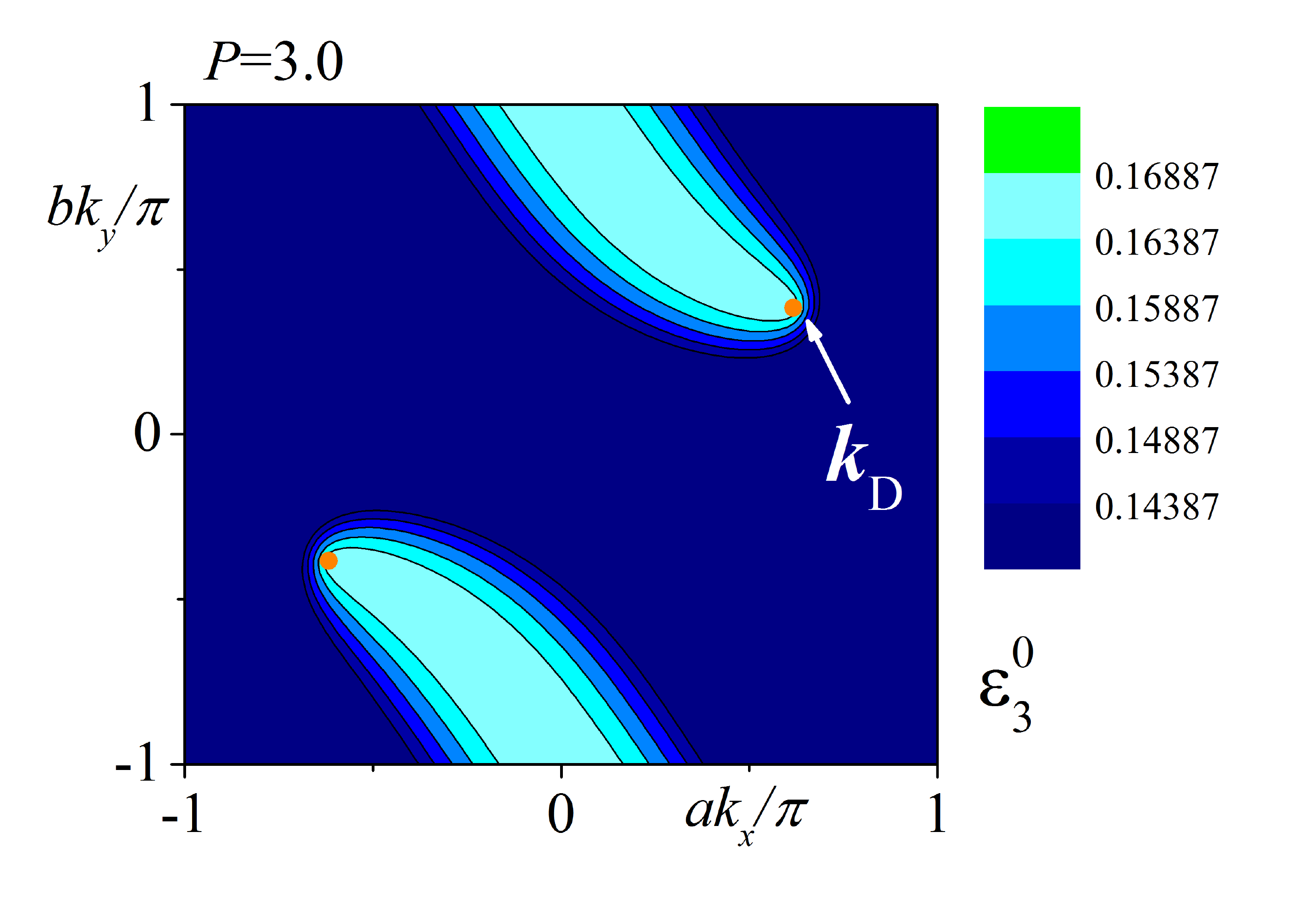}\vspace{-0.5cm}
\end{center}
\caption{
(Color online) 
Contour plots of the fourth band (a) and the third band (b) at $P=3.0$. 
The third band and four band touch 
at two orange points, $\pm{\bf k}_{\rm D}\simeq \pm(0.6169\pi/a, 0.3835\pi/b)$.
}
\label{fig5}
\end{figure}
%
\begin{figure}[bt]
\begin{center}
\begin{flushleft} \hspace{0.5cm}(a) \end{flushleft}\vspace{-0.0cm}
\includegraphics[width=0.5\textwidth]{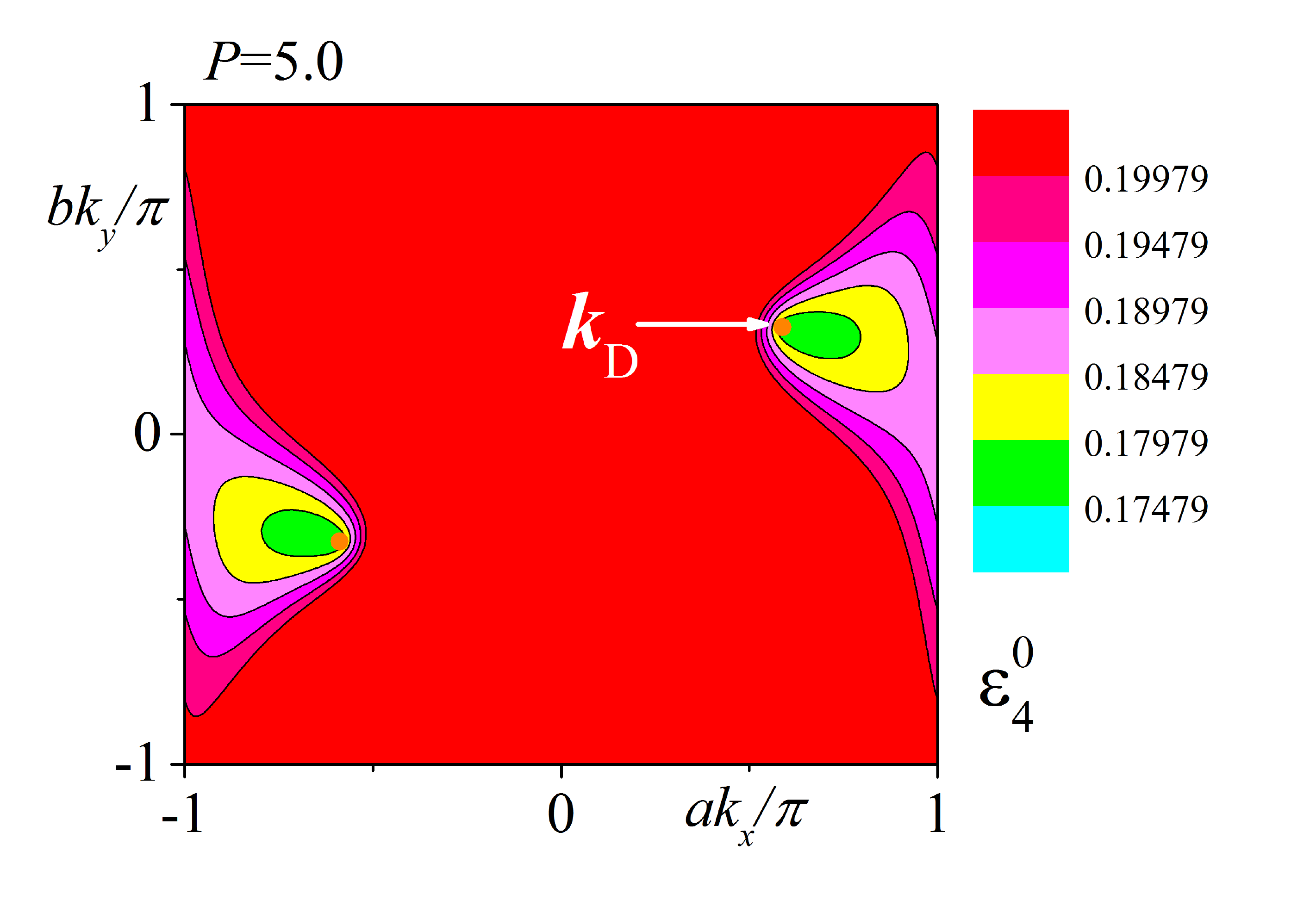}\vspace{-0.0cm}
\begin{flushleft} \hspace{0.5cm}(b) \end{flushleft}\vspace{-0.0cm}
\includegraphics[width=0.5\textwidth]{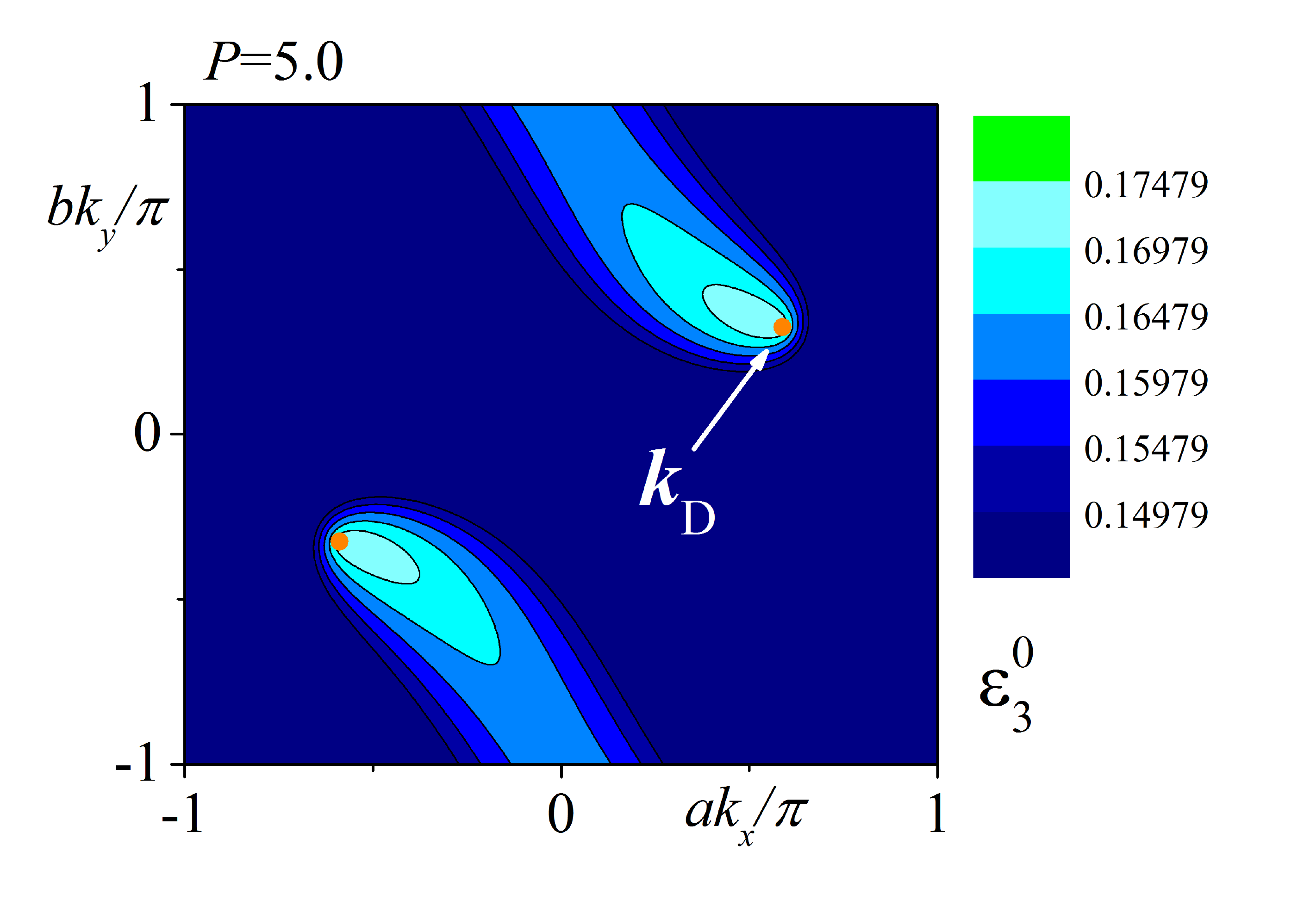}\vspace{-0.5cm}
\end{center}
\caption{
(Color online) 
Contour plots of fourth band (a) and the third band (b) at $P=5.0$. 
}
\label{fig7}
\end{figure}

\begin{figure}[bt]
\begin{center}
\begin{flushleft} \hspace{0.5cm}(a) \end{flushleft}\vspace{-0.0cm}
\includegraphics[width=0.5\textwidth]{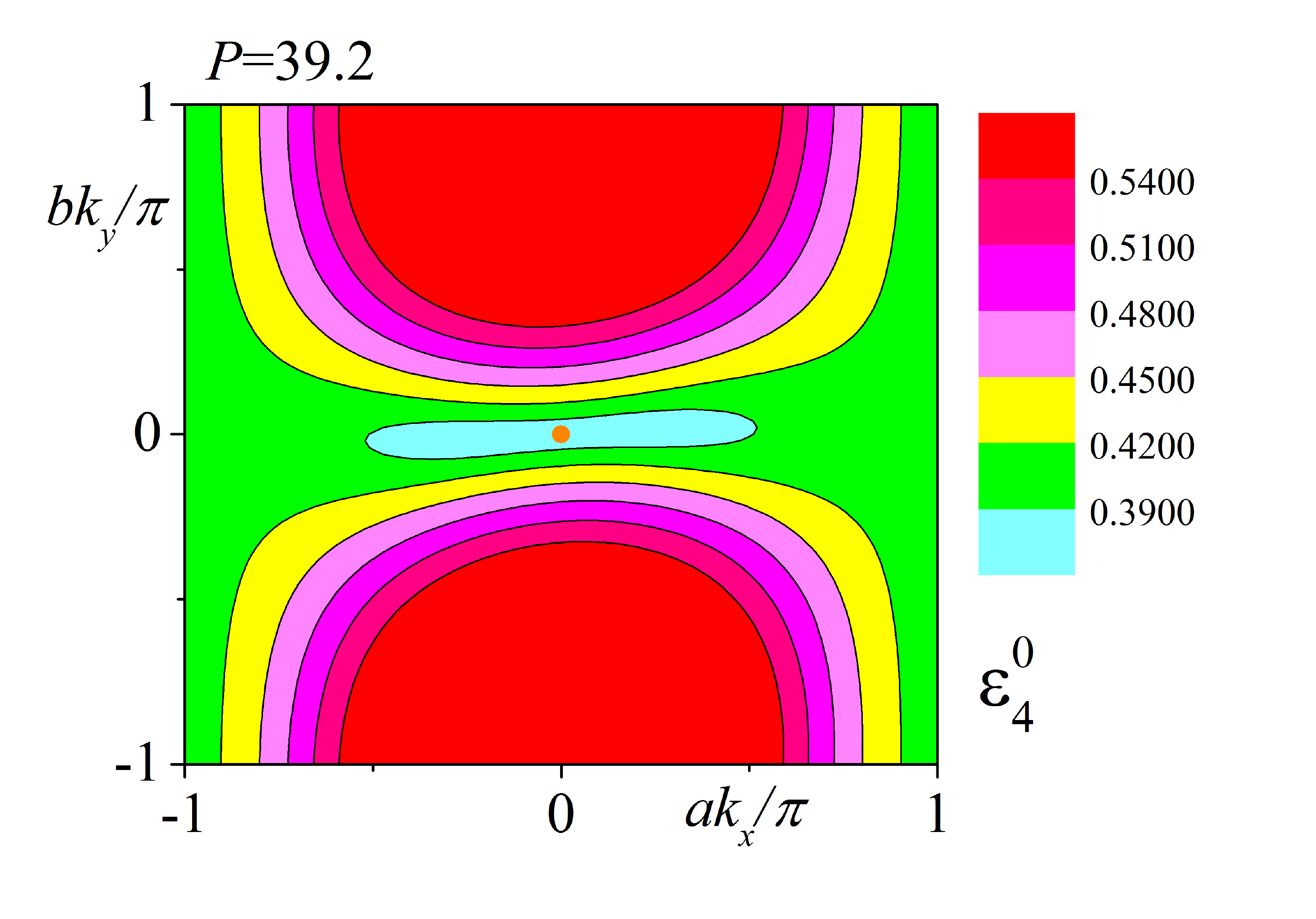}\vspace{-0.0cm}
\begin{flushleft} \hspace{0.5cm}(b) \end{flushleft}\vspace{-0.0cm}
\includegraphics[width=0.5\textwidth]{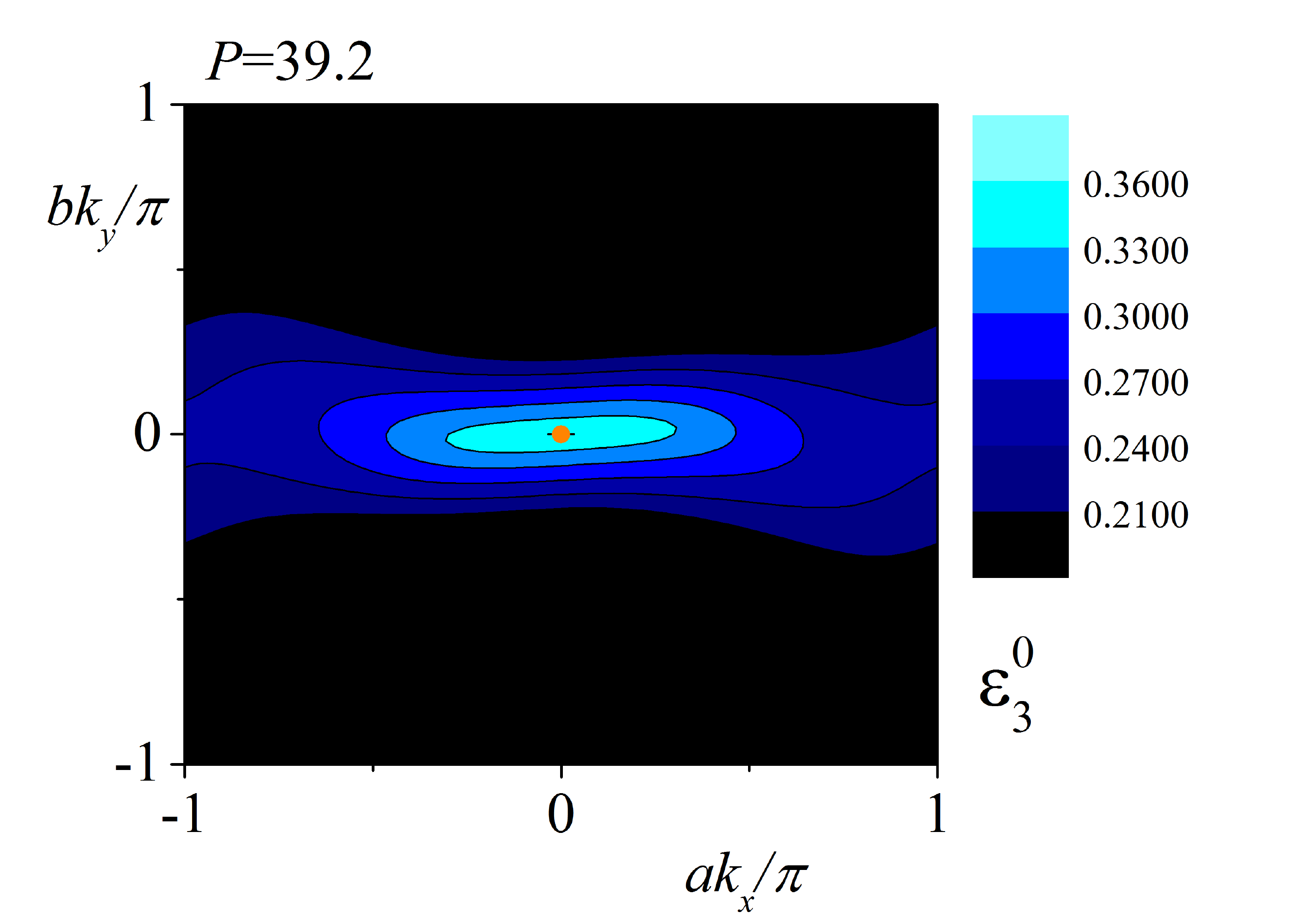}\vspace{-0.5cm}
\end{center}
\caption{
(Color online) 
Contour plots of the fourth band (a) and the third band (b) at $P=39.2$, where 
$\varepsilon^0_{\rm F}\simeq 0.36165$. Two Dirac points merge at a $\Gamma$ point (orange point,
 $(ak_x/\pi,bk_y/\pi)= (0,0)$).}
\label{fig10bc}
\end{figure}

\begin{figure}[bt]
\begin{center}
\begin{flushleft} \hspace{0.0cm}(a) \end{flushleft}\vspace{-0.1cm} 
\includegraphics[width=0.4\textwidth]{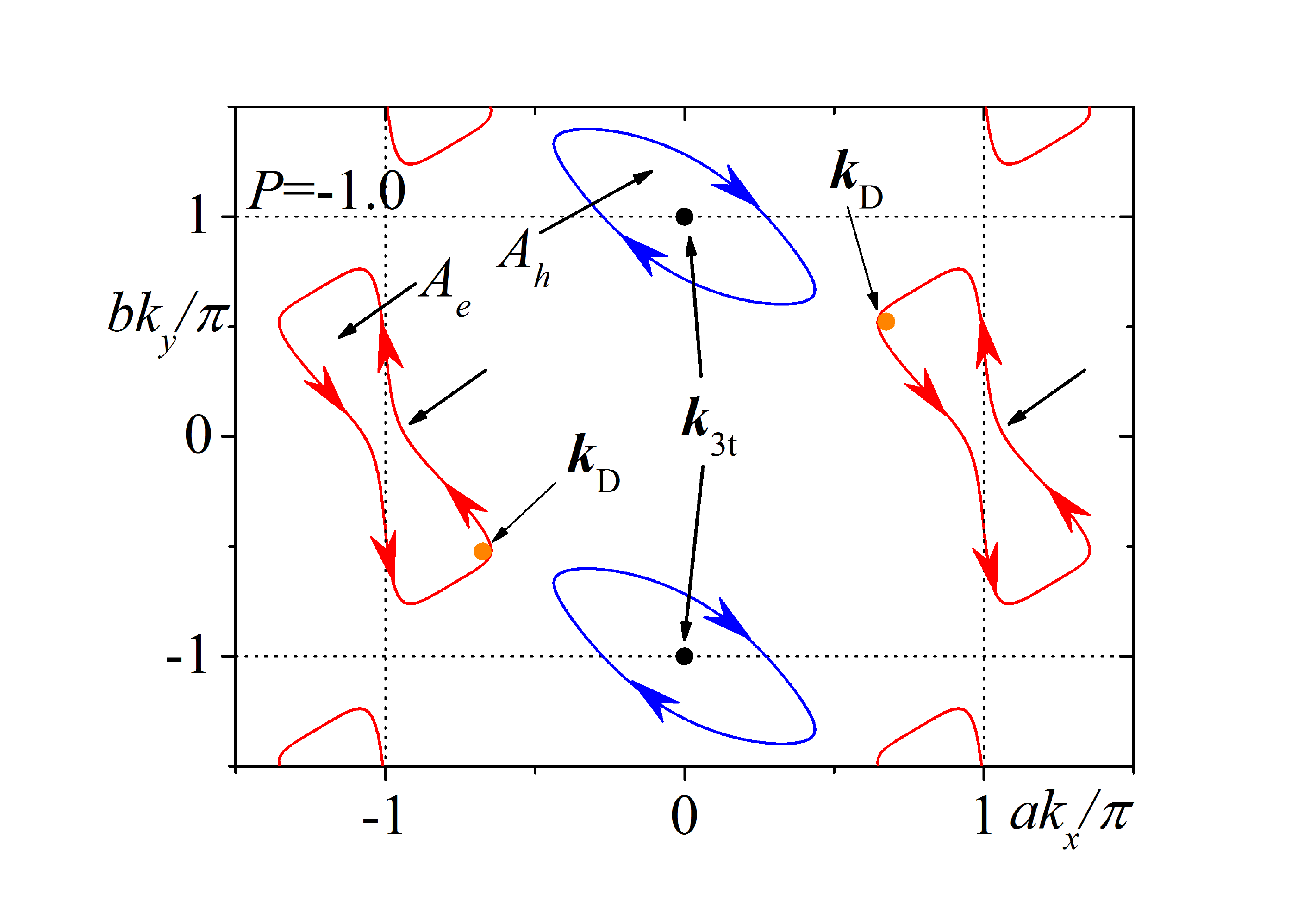}\vspace{-0.5cm}
\begin{flushleft} \hspace{0.0cm}(b) \end{flushleft}\vspace{-0.0cm}
\includegraphics[width=0.4\textwidth]{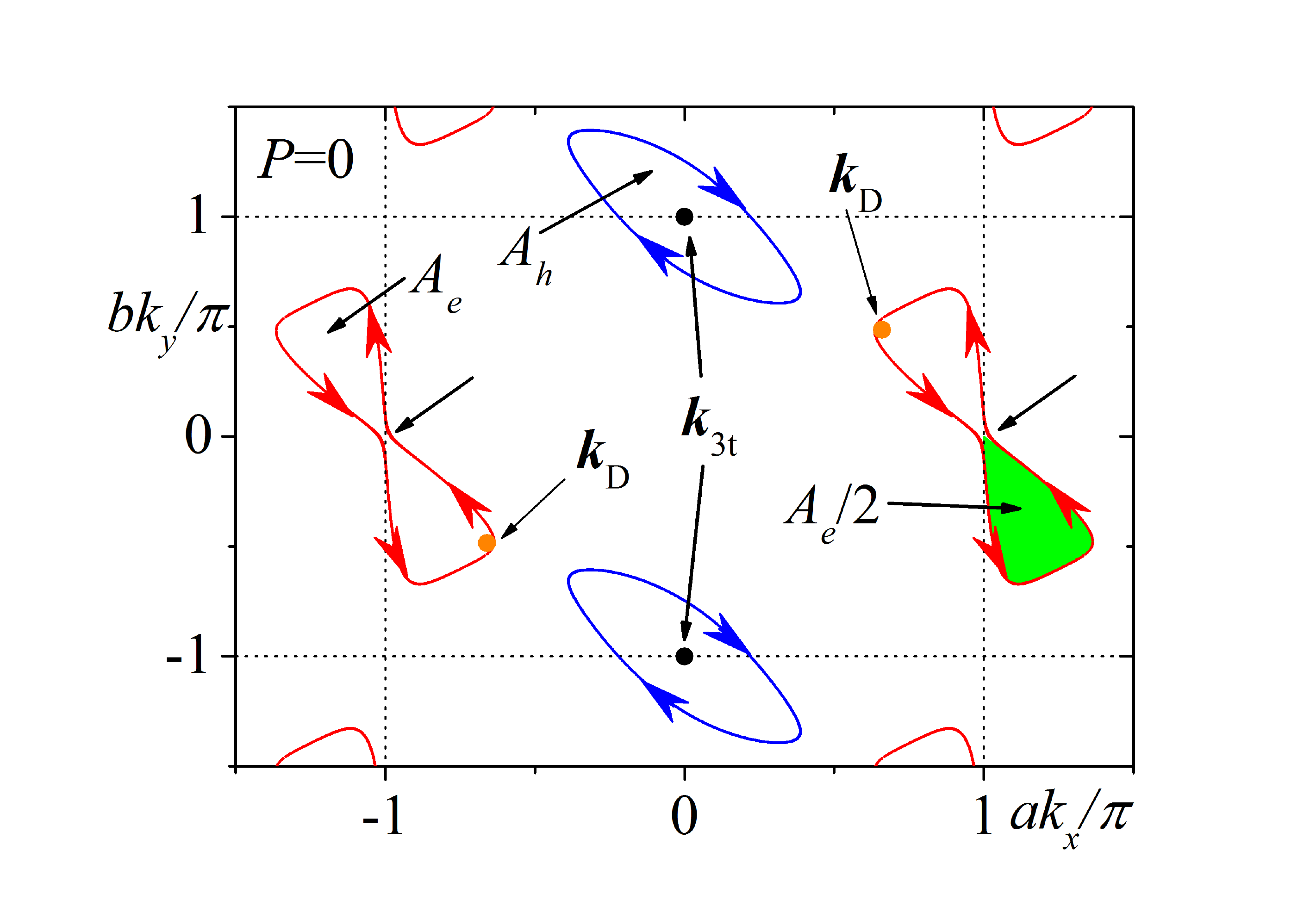}\vspace{-0.5cm}
\begin{flushleft} \hspace{0.0cm}(c) \end{flushleft}\vspace{-0.0cm}
\includegraphics[width=0.4\textwidth]{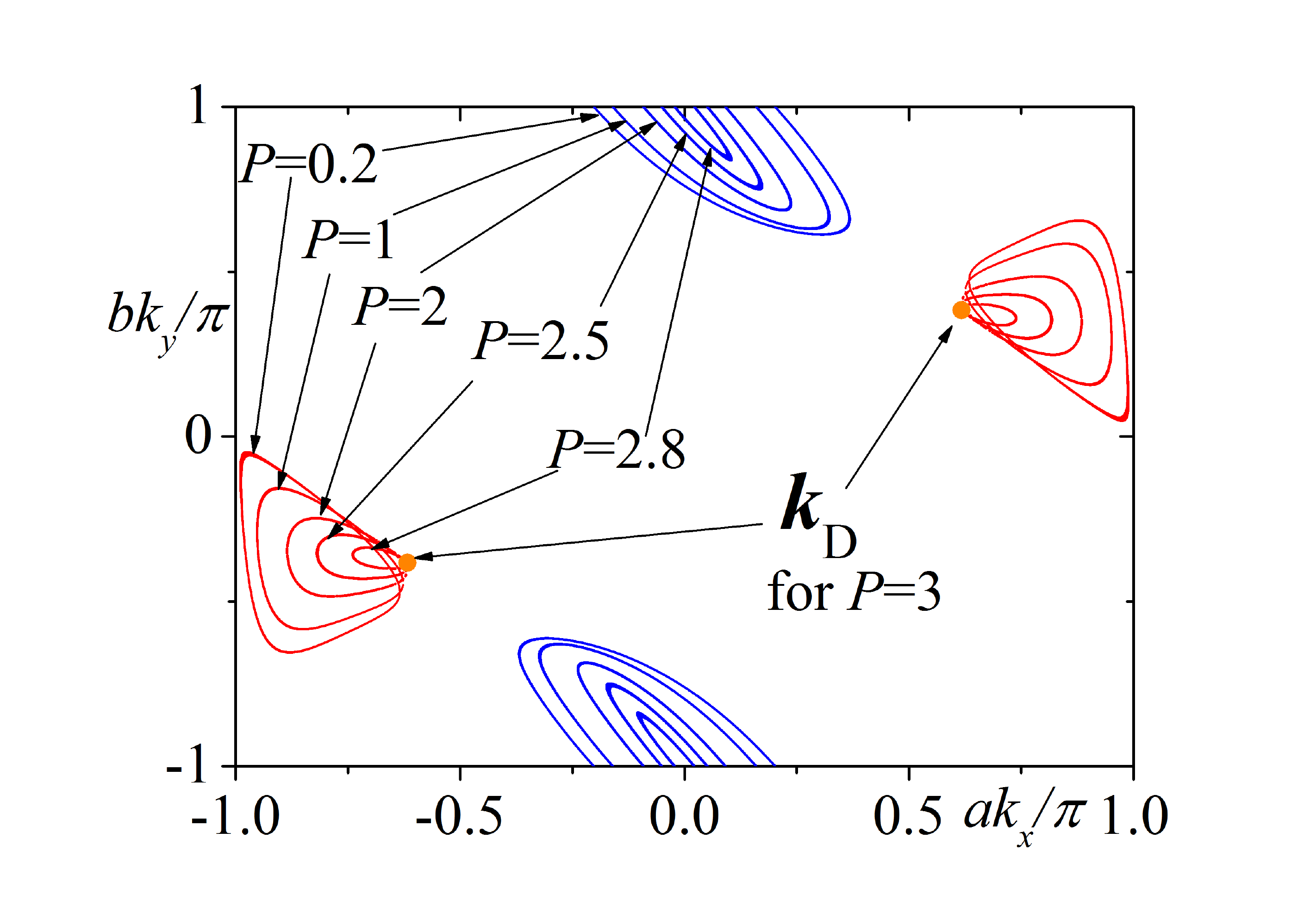}\vspace{-0.4cm}
\end{center}
\caption{(Color online) 
Fermi surfaces at $P=-1.0$ (a) and $P=0$ (b) in the extended zone, 
where arrows indicate the direction of the orbital motion for electrons in the magnetic field
(clockwise for hole pockets and counter-clockwise for electron pockets). 
In (b), a green area is a half of $A_e$. In (c), the Fermi surface for 
$P=0.2, 1.0, 2.0, 2.5$ and $2.75$ 
in the first Brillouin zone are shown, where orange points are for ${\bf k}_{\rm D}$ at $P=-1.0, 0$ and $3.0$ and black points are for ${\bf k}_{\rm 3t}$ at $P=-1.0$ and 0. The wave number, $\mathbf{k}_{\mathrm 3t}$ stays at ($0, \pi /b)$ at $P < 3.0$. 
We obtain $A_e=A_h\simeq 0.0903$ 
 at $P=-1.0$, 
 $A_e = A_h \simeq 0.0715$ at $P=0$ and $A_h \simeq 0.0479$ at $P=1.0$ 
in the unit of the area of the Brillouin zone, where 
$A_e$ and $A_h$ are the areas of an electron pocket and a hole pocket, respectively. 
The ratio of cyclotron masses for an electron pocket and a hole pocket 
($m_e$/$m_h$) are about 0.71, 0.63 and 0.61 at $P=-1.0$, 0 and 1.0.
}
\label{fig8}
\end{figure}

\section{energy at $H\neq 0$}
\label{AppendixB}
%
%
\begin{figure}[bt]
\begin{flushleft} \hspace{0.0cm}(a) \end{flushleft}\vspace{-0.0cm}
\includegraphics[width=0.49\textwidth]{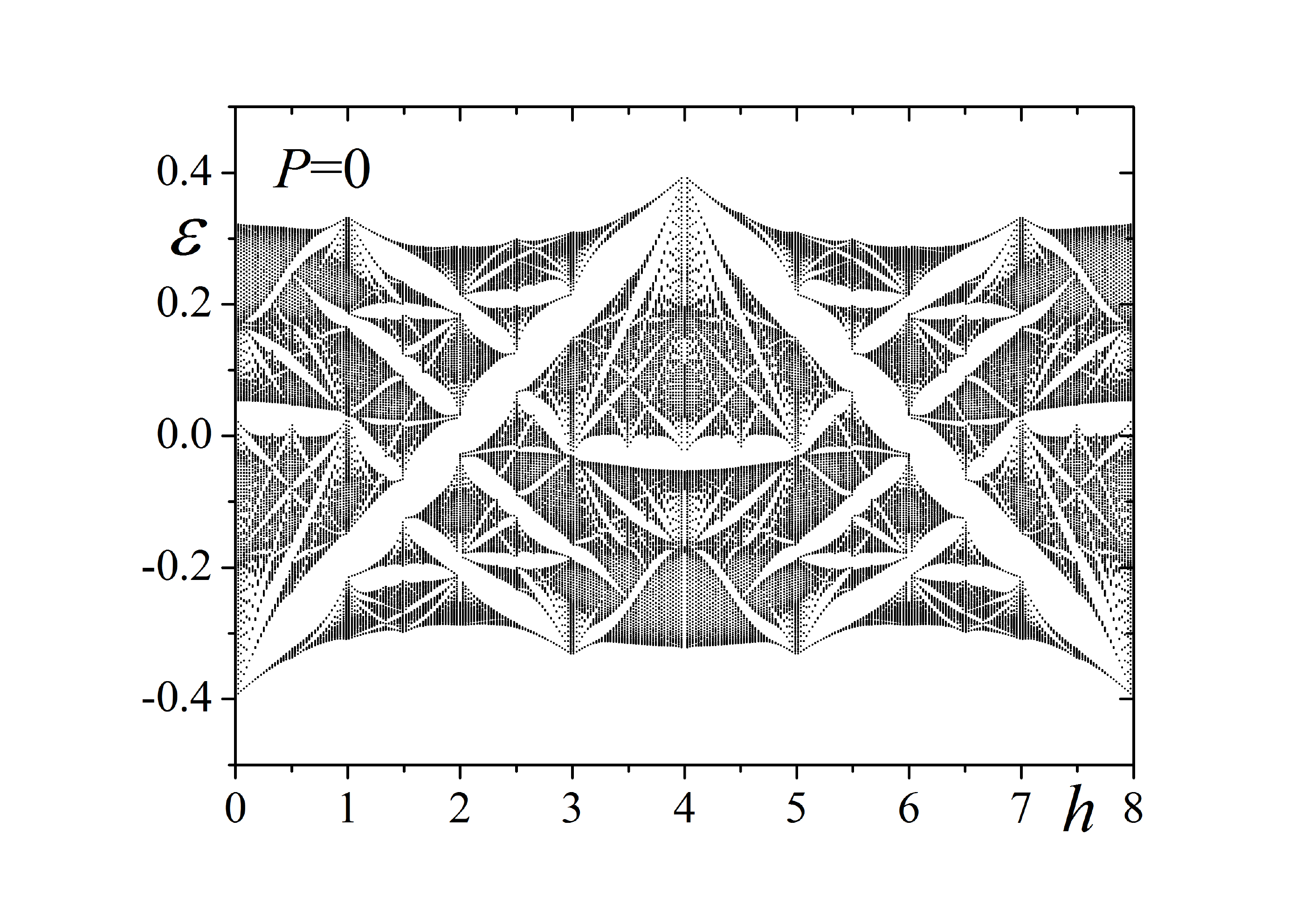}\vspace{-0.2cm}
\begin{flushleft} \hspace{0.0cm}(b) \end{flushleft}\vspace{-0.0cm}
\includegraphics[width=0.49\textwidth]{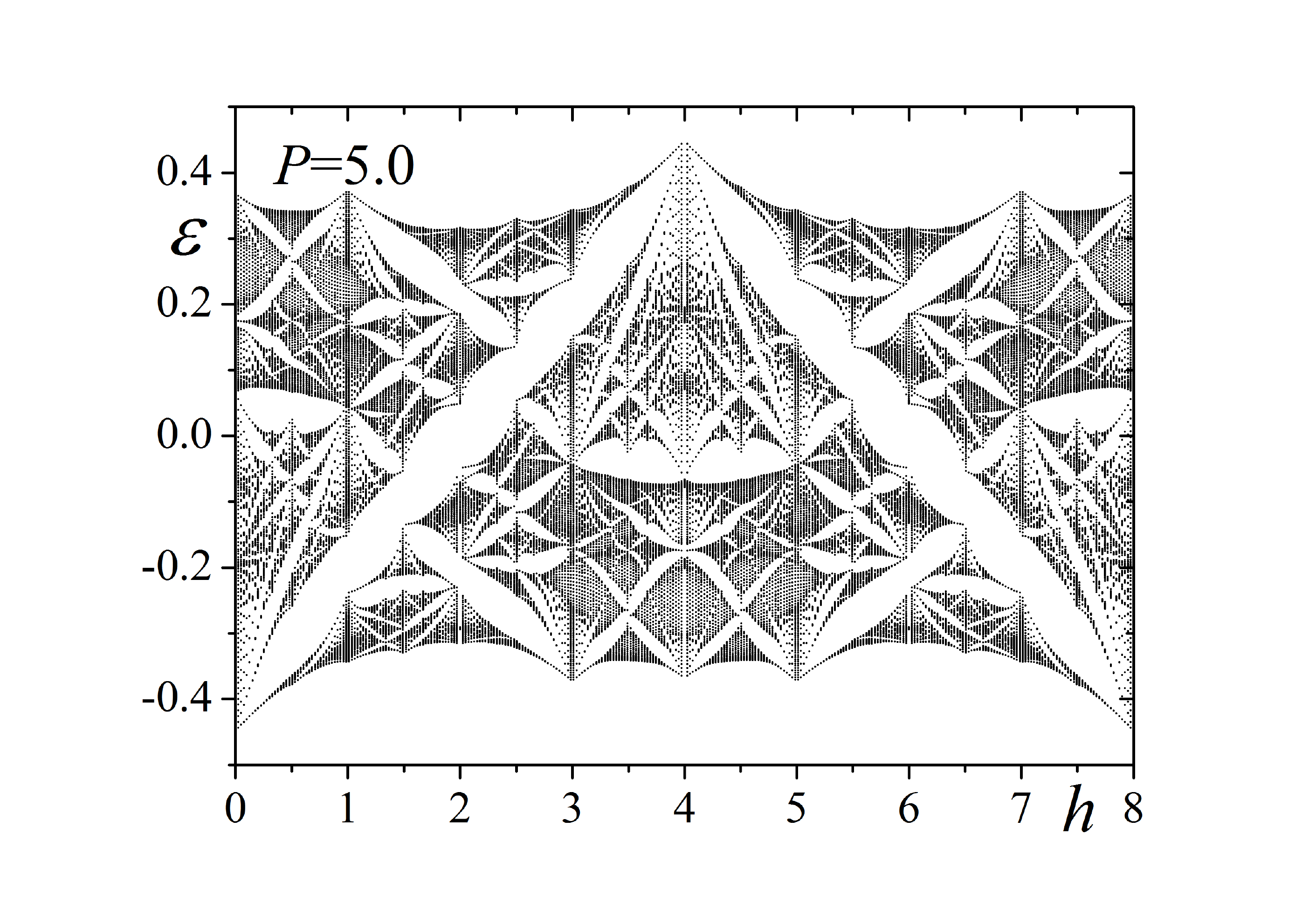}\vspace{-0.2cm}
\begin{flushleft} \hspace{0.0cm}(c) \end{flushleft}\vspace{-0.0cm}
\includegraphics[width=0.49\textwidth]{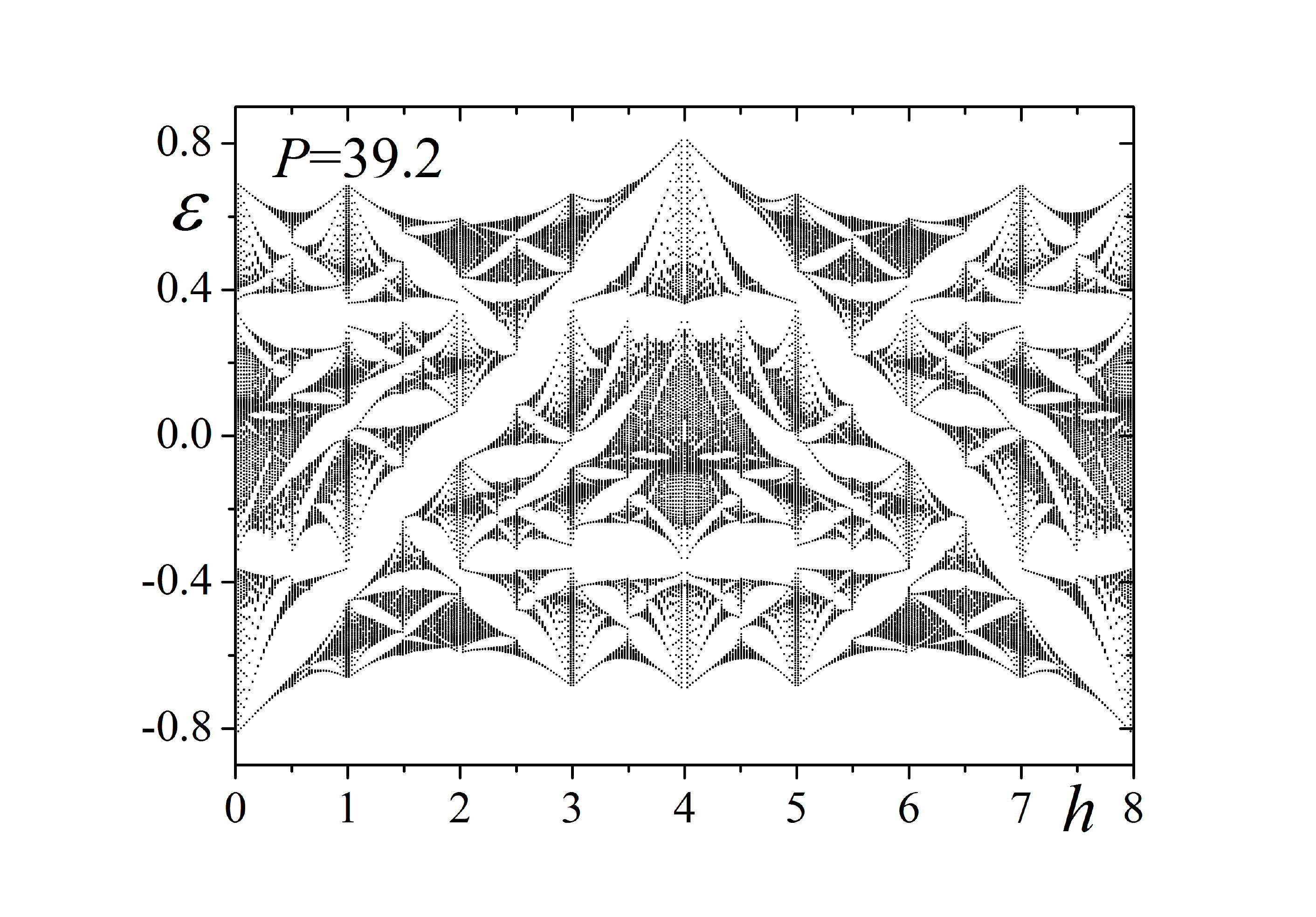}\vspace{-0.2cm}
\caption{
Energies as a function of $h$ for $P=0$ (a), $P=5.0$ (b) and $P=39.2$ (c). We 
take $h=p/q$ with $q=79$ and $p=2m$, where $m=1, 2, 3, \cdots, 4q$. 
}
\label{fig12}
\end{figure}
The Hamiltonian in the two dimensional tight-binding model in the magnetic field 
becomes 
\begin{eqnarray}
{\cal {\hat H}}=\sum_{i,j}t_{ij}  e^{i2\pi\phi_{ij}} c^{\dag}_i c_j ,
\label{ham_H}
\end{eqnarray}
where the phase factor $(\phi_{ij}$) is given by  
\begin{eqnarray}
\phi_{ij}=\frac{e}{ch}\int_j^i{\bf A}\cdot d{\bf l}.
\end{eqnarray}

In this study, we treat the magnetic field applied perpendicular to the 
$x-y$ plane by taking the ordinary Landau gauge, 
\begin{equation}
{\bf A}=(Hy,0,0).
\label{ordinaryLandau}
\end{equation} 
The flux through the unit cell 
is
\begin{eqnarray}
\Phi=abH.
\end{eqnarray}
When the magnetic field is commensurate with the lattice period, i.e.,
\begin{eqnarray}
\frac{\Phi}{\phi_0}=\frac{p}{q}, 
\label{Phi}
\end{eqnarray}
where $p$ and $q$ are integers, the magnetic unit cell is 
$a\times (qb)$, if $p$ is an even integer. Since there are two sites with half of the lattice constant in the $y$ direction in the unit cell, 
we have to take the magnetic unit cell of $a \times (2 q b)$, if we take the ordinary Landau gauge and $p$ is an odd integer. We can take a more suitable gauge (periodic Landau gauge)\cite{HK2013}, which is a powerful tool for 
the system with a large unit cell such as moire pattern in 
the twisted bilayer graphene. However, we take even integers for $p$ by using the ordinary Landau gauge in this paper, since it is possible 
to investigate magnetic-field-dependences of energies 
only by taking even integers for $p$.

The Hamiltonian in the momentum space becomes 
\begin{eqnarray}
{\cal {\hat H}}=\sum_{{\bf k}}
{\tilde C}^{\dagger}_{{\bf k}}{\tilde \varepsilon}_{\bf k}{\tilde C}_{{\bf k}}, 
\label{ham_H_k}
\end{eqnarray}
where the summation over ${\bf k}$ is taken in the magnetic Brillouin zone,
\begin{eqnarray}
-\frac{\pi}{a}&\leq &k_x<\frac{\pi}{a}, \\
-\frac{\pi}{qb}&\leq &k_y<\frac{\pi}{qb}.
\end{eqnarray}

In Eq. (\ref{ham_H_k}), 
the creation and annihilation operators have $4q$ components, 
\begin{eqnarray}
{\tilde C}^{\dagger}_{{\bf k}}=(c^{(0)\dag}_{1,{\bf k}}, c^{(0)\dag}_{2,{\bf k}}, c^{(0)\dag}_{3,{\bf k}}, c^{(0)\dag}_{4,{\bf k}},\cdots, c^{(q-1)\dag}_{3,{\bf k}}, c^{(q-1)\dag}_{4,{\bf k}}),
\end{eqnarray}
\begin{eqnarray}
{\tilde C}_{{\bf k}}=\begin{pmatrix}
c^{(0)}_{1,{\bf k}}\\c^{(0)}_{2,{\bf k}}\\
c^{(0)}_{3,{\bf k}}\\c^{(0)}_{4,{\bf k}}\\
\cdot\\\cdot\\\cdot\\
c^{(q-1)}_{3,{\bf k}}\\c^{(q-1)}_{4,{\bf k}}
\end{pmatrix}
\end{eqnarray}
and $\tilde{\varepsilon}_{\bf k}$ is the $4q\times 4q$ matrix which is given by 
\begin{align}
&\tilde{\varepsilon}_{\mathbf{k}} \nonumber \\
=& \left( \begin{array}{cccccc}
D^{(0)}_{\mathbf{k}}  & F^{(1)}_{\mathbf{k}} & 0 & \cdots & 0 &  F^{(0)\dagger}_{\mathbf{k}} \\
  F^{(1) \dagger }_{\mathbf{k}} & D^{(1)}_{\mathbf{k}}  & F^{(2)}_{\mathbf{k}} & \ddots & \ddots & 0 \\
  0  &   F^{(2) \dagger}_{\mathbf{k}}   &  D^{(2)}_{\mathbf{k}} &    F^{(3)}_{\mathbf{k}}    & \ddots & \vdots \\
  \vdots & \ddots & \ddots & \ddots & \ddots & 0 \\
  0 & \ddots & \ddots &  F^{(q-2) \dagger}_{\mathbf{k}}  & D^{(q-2)}_{\mathbf{k}} &  F^{(q-1)}_{\mathbf{k}} \\
   F^{(0)}_{\mathbf{k}} & 0 & \dots & 0 &  F^{(q-1) \dagger}_{\mathbf{k}} & D^{(q-1)}_{\mathbf{k}}\\
   \end{array} \right),
\label{j5} 
\end{align}
where
\begin{equation}
D_{\mathbf{k}}^{(n)} = \left( \begin{array}{cccc}
0& \epsilon_{\mathbf{k} 12}^{(n)} & \epsilon_{\mathbf{k} 13}^{(n)} & \epsilon_{\mathbf{k} 14}^{(n)} \\
\epsilon_{\mathbf{k} 12}^{(n)*} &0 &   0 & \epsilon_{\mathbf{k} 24}^{(n)} \\
\epsilon_{\mathbf{k} 13}^{(n)*} & 0 &   
0& \epsilon_{\mathbf{k} 34}^{(n)} \\
\epsilon_{\mathbf{k}14}^{(n)*}                        & \epsilon_{\mathbf{k}24}^{(n)*} &   \epsilon_{\mathbf{k}34}^{(n)*} & 0 \\
 \end{array} \right),
\end{equation}
\begin{equation}
F_{\mathbf{k}}^{(n)} = \left( \begin{array}{cccc}
   0                      &   0                         &   0                         & 0 \\
\epsilon_{\mathbf{k}21}^{\prime(n)}  &   0                         &     \epsilon_{\mathbf{k}23}^{\prime(n)}                          &    0                        \\
   0                        &   0                         &     0                         &    0                         \\
0   & 0 & \epsilon_{\mathbf{k}43}^{\prime(n)}& 0\\
 \end{array} \right),
\end{equation}
and
\begin{equation}
F_{\mathbf{k}}^{(n)\dag} = \left( \begin{array}{cccc}
   0                      &   \epsilon_{\mathbf{k}21}^{\prime(n)*}                         &   0                         & 0 \\
0  &   0                         &     0                          &    0                        \\
   0                        &   \epsilon_{\mathbf{k}23}^{\prime(n)*}                         &     0                         &    \epsilon_{\mathbf{k}43}^{\prime(n)*}                         \\
0   & 0 & 0 & 0\\
 \end{array} \right). 
\end{equation}
The matrix elements, $\varepsilon^{(n)}_{\mathbf{k}\alpha\beta}$, are the hoppings from the $\beta$ site ($j$) in the $n$th unit cell to the $\alpha$ site ($i$) 
in the $n$th unit cell in the magnetic field ($nb \leq y_j <(n+1)b$ and 
$nb \leq y_i <(n+1)b)$ given by
\begin{align}
\epsilon^{(n)}_{\mathbf{k} 12}
&=t_{a{2}}e^{i\frac{1}{2}bk_y},\\
\epsilon^{(n)}_{\mathbf{k} 13}
&=t_{b2}\exp\bigg[i\left(\frac{1}{2}ak_x- \frac{1}{4}bk_y+2\pi  \phi_{b2,13}^{(n)}\right)\bigg]\nonumber \\
&+t_{b3} \exp\bigg[i\left(-\frac{1}{2}ak_x- \frac{1}{4}bk_y+2\pi  \phi_{b3,13}^{(n)}\right)\bigg],\\
\epsilon^{(n)}_{\mathbf{k} 14}
&=t_{b1}\exp\bigg[i\left(\frac{1}{2}ak_x +\frac{1}{4}bk_y+2\pi  \phi_{b1,14}^{(n)}\right)\bigg]\nonumber \\
&+t_{b4}\exp\bigg[i\left(-\frac{1}{2}ak_x+\frac{1}{4}bk_y+2\pi  \phi_{b4,14}^{(n)}\right)\bigg],\\
\epsilon^{(n)}_{\mathbf{k} 24}
 &=t_{b1}\exp\bigg[i\left(-\frac{1}{2}ak_x -\frac{1}{4}bk_y+2\pi  \phi_{b1,24}^{(n)}\right)\bigg]\nonumber \\
&+t_{b4}\exp\bigg[i\left(\frac{1}{2}ak_x-\frac{1}{4}bk_y+2\pi  \phi_{b4,24}^{(n)}\right)\bigg], \\
\epsilon^{(n)}_{\mathbf{k} 34}
&=t_{a1}e^{i\frac{1}{2}bk_y},
\end{align}
where 
\begin{eqnarray}
\phi^{(n)}_{b2,13}&=&\frac{\Phi}{\phi_0}(\frac{n}{2}-\frac{1}{16}), \\
\phi^{(n)}_{b3,13}&=&-\frac{\Phi}{\phi_0}(\frac{n}{2}-\frac{1}{16}), \\
\phi^{(n)}_{b1,14}&=&\frac{\Phi}{\phi_0}(\frac{n}{2}+\frac{1}{16}), \\
\phi^{(n)}_{b4,14}&=&-\frac{\Phi}{\phi_0}(\frac{n}{2}+\frac{1}{16}), \\
\phi^{(n)}_{b1,24}&=&-\frac{\Phi}{\phi_0}(\frac{n}{2}+\frac{3}{16}), \\
\phi^{(n)}_{b4,24}&=&\frac{\Phi}{\phi_0}(\frac{n}{2}+\frac{3}{16}). 
\end{eqnarray}
The matrix elements, $\varepsilon^{\prime(n)}_{\mathbf{k}\alpha\beta}$, are 
the hoppings from 
the $\beta$ site ($j$) in the $(n+1)$th unit cell to the $\alpha$ site ($i$) 
in the $n$th unit cell in the magnetic field 
($(n+1)b \leq y_j < (n+2)b$ and $nb \leq y_i <(n+1)b$) given by 
\begin{align} 
\epsilon^{\prime(n)}_{\mathbf{k} 21}&= t_{a3}e^{ i\frac{1}{2}bk_y},\\ 
\epsilon^{\prime(n)}_{\mathbf{k} 23} &= 
t_{b2}\exp\bigg[i\left(-\frac{1}{2}ak_x +\frac{1}{4}bk_y+2\pi 
\phi^{\prime(n)}_{b2,23}\right)\bigg]\nonumber \\ 
&+t_{b3}\exp\bigg[i\left( \frac{1}{2}ak_x + \frac{1}{4}bk_y+2\pi 
\phi_{b3,23}^{\prime(n)}\right)\bigg],\\ 
\epsilon^{\prime(n)}_{\mathbf{k} 43}&= t_{a1}e^{ i\frac{1}{2}bk_y}, 
\end{align} 
where 
\begin{align} 
\phi^{\prime(n)}_{b2,23} &= 
-\frac{\Phi}{\phi_0}(\frac{n}{2}-\frac{3}{16}), \\ 
\phi^{\prime(n)}_{b3,23} &= \frac{\Phi}{\phi_0}(\frac{n}{2}-\frac{3}{16}). 
\end{align}

\section{Derivation of Eq. (\ref{P_c})
}
\label{AppendixB2}

\begin{figure}[bt]
\centering 
\includegraphics[width=0.37\textwidth]{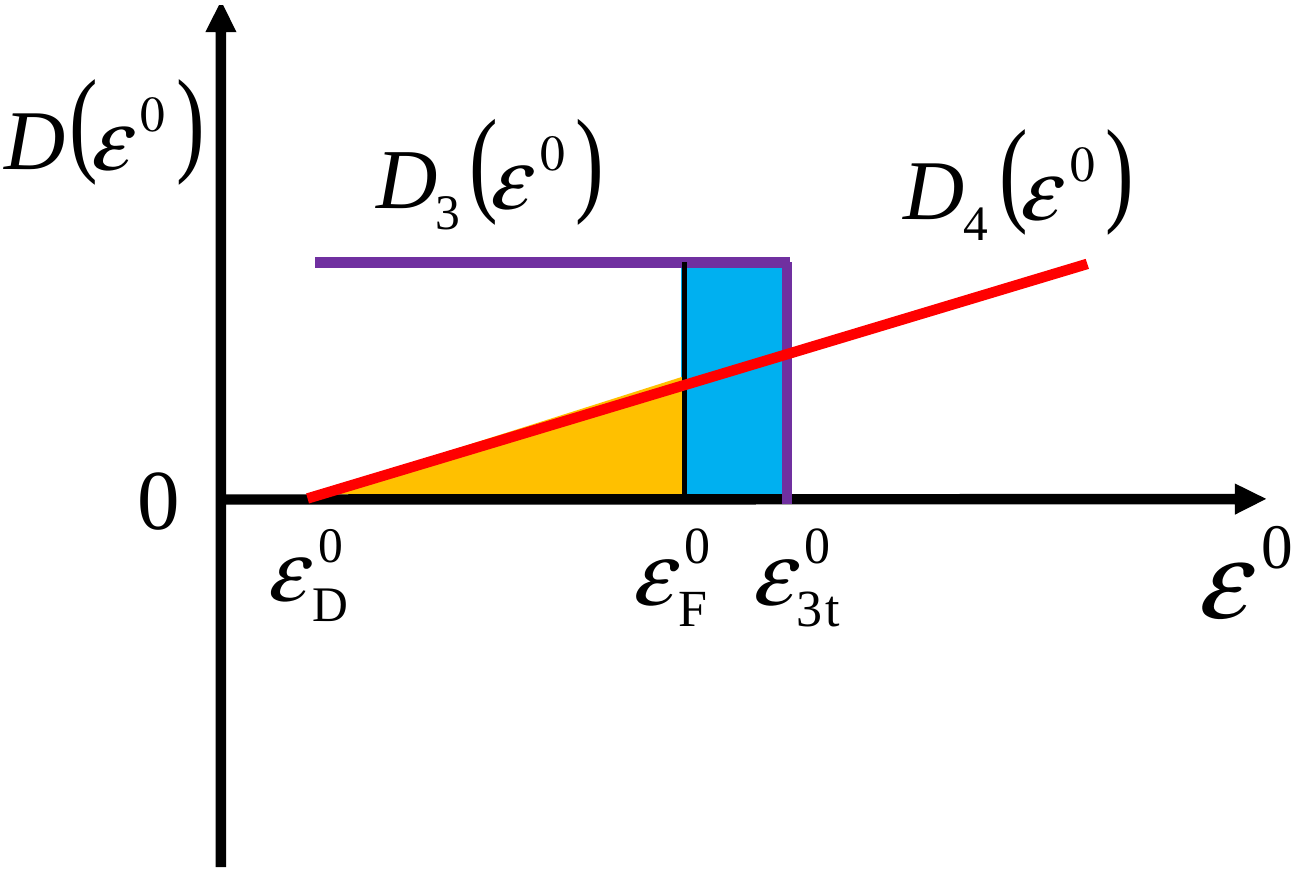} 
\caption{
(Color online)
A schematic plot of the density of states at $P < P_c$.
}
\label{figdos}
\end{figure}

The critical pressure, $P_c = 3.0$, is defined by the pressure at which
the global maximum energy of the third band,
 $\varepsilon^0_{\textrm{3t}}$, becomes the same as 
the energy at the Dirac points $\varepsilon^0_{\textrm{D}}$. 
At $2.3< P \lesssim P_c$, $\varepsilon^0_{\textrm{3t}}$ and $\varepsilon^0_{\textrm{D}}$ depend on pressure as
\begin{align}
\varepsilon^0_{\textrm{3t}} &= a_{\textrm{3t}}(P_c-P) + \varepsilon^{00}_{\rm D},  \\
\varepsilon^0_{\textrm{D}} &= -a_{\textrm{D}}(P_c-P) + \varepsilon^{00}_{\rm D}, 
\end{align}
respectively, 
where $a_{\textrm{3t}}$ and $a_{\textrm{D}}$ are pressure-independent constant and
$\varepsilon^{00}_{\rm D}$ is the energy at the Dirac points at $P=P_c$.
The density of states at the third band and the fourth band are given by
\begin{align}
D_3(\varepsilon^0) &= d_3 \theta (\varepsilon^0_{\textrm{3t}} -\varepsilon^0), \label{D_3}\\
D_4(\varepsilon^0) &= c_{\textrm{D}} (\varepsilon^0-\varepsilon^0_{\textrm{D}}),\label{D_4}
\end{align}
where $d_3$ and $c_{\textrm{D}}$ are constants and 
$\theta (\varepsilon^0_{\textrm{3t}} -\varepsilon^0)$ is the step function, 
as shown in Fig. \ref{figdos}.

The Fermi energy, $\varepsilon^0_{\textrm{F}}$, at $P \lesssim P_c$ is obtained
by the condition that the area of a hole pocket equals to that of an electron pocket, i.e., 
\begin{equation}
\int_{\varepsilon^0_{\textrm{F}}}^{\varepsilon^0_{\textrm{3t}}} 
D_3(\varepsilon^0) d \varepsilon^0
=\int_{\varepsilon^0_{\textrm{D}}}^{\varepsilon^0_{\textrm{F}}}
D_4(\varepsilon^0) d \varepsilon^0.
\label{eqfermi}
\end{equation}
By putting Eqs. (\ref{D_3}) and (\ref{D_4}) into Eq. (\ref{eqfermi}), 
we obtain
\begin{equation}
d_3 (\varepsilon^0_{\textrm{3t}}-\varepsilon^0_{\textrm{F}}) 
=\frac{1}{2} c_{\textrm{D}}
(\varepsilon^0_{\textrm{F}}-\varepsilon^0_{\textrm{D}})^2.
\label{eqequation0}
\end{equation}
We study the cases of 
\begin{equation}
 P_c-P \ll P_c
\end{equation} 
and
\begin{equation}
\varepsilon^0_{\textrm{D}} < \varepsilon^0_{\textrm{F}} < 
\varepsilon^0_{\textrm{3t}}.
\end{equation}
 Since
\begin{equation}
\varepsilon^0_{\textrm{3t}}- \varepsilon^0_{\textrm{D}} = (a_{\textrm{3t}} 
+a_{\textrm{D}} ) (P_c-P)
\end{equation}
goes to zero when $P_c-P \to 0$, we obtain both 
$\varepsilon^0_{\textrm{3t}}-\varepsilon^0_{\textrm{F}}$ and 
$\varepsilon^0_{\textrm{F}}-\varepsilon^0_{\textrm{3t}}$
go to zero when $P_c-P \to 0$. 
By using 
\begin{equation}
\varepsilon^0_{\textrm{F}}-\varepsilon^0_{\textrm{D}}= (a_{\textrm{3t}} + 
a_{\textrm{D}} ) (P_c-P)-
(\varepsilon^0_{\textrm{3t}}-\varepsilon^0_{\textrm{F}}   )
\end{equation}
and Eq. (\ref{eqequation0}), 
we obtain 
\begin{align}
\varepsilon^0_{\textrm{F}}-\varepsilon^0_{\textrm{D}} &\simeq (a_{\textrm{3t}} + a_{\textrm{D}})(P_c-P) +O 
((P_c-P)^2), \\
\varepsilon^0_{\textrm{3t}} - \varepsilon^0_{\textrm{F}} &\simeq \frac{c_{\rm D}}{2 d_3} 
(a_{\textrm{3t}} + a_{\textrm{D}})^2 (P_c-P)^2. \label{E12}
\end{align}

\section{``three-quarter''-Dirac and derivation of Eq. (\ref{area})
}
\label{Appendix_area}

In this Appendix we derive the area as a function of energy around 
{\it ``three-quarter''}-Dirac point, Eq. (\ref{area}). 
The minimal Weyl Hamiltonian studied by Goerbig, Fuchs and Montambaux\cite{Georbig2008} is given by 
\begin{equation}
\mathcal{H}_{\textrm{Weyl}}^0 = \mathbf{w}_0\cdot \mathbf{q} \sigma^0 + w_x q_x \sigma^x + w_y q_y \sigma^y, 
\label{eqGoerbig}
\end{equation}
where $\sigma^0$ is a $2\times 2$ unit matrix, $\sigma^x$ and $\sigma^y$ are 
the Pauli matrices and $\mathbf{w}_0=(w_{0x}, w_{0y})$, $w_x$ and $w_y$ are constants. The energy dispersion is given by 
\begin{equation}
\varepsilon^0_{\pm} (\mathbf{q})=
\mathbf{w}_0 \cdot \mathbf{q} \pm \sqrt{w_x^2 q_x^2 + w_y^2 q_y^2}.
\end{equation}
The anisotropy and the tilting of the Dirac cone are described by $w_x$ and $w_y$ and
by $w_{0x}$ and $w_{0y}$, respectively. For simplicity we take $w_x >0$ and $w_y >0$.
The energy of the tilted Dirac cone has also been studied in the linearized form 
in the context of type II Weyl semimetals\cite{Soluyanov2015}.
%

When we take 
\begin{align}
 w_{0y} &= 0 \\
 w_{0x} &= - w_x <0,
\end{align}
the Dirac cone is critically tilted. 
In this case, we have to introduce quadratic terms along the $q_x$ axis as
\begin{equation}
\mathcal{H}_{\textrm{tqD}}^0 = (-w_x q_x + \alpha_{2}' q_x^2) \sigma^0 
+ (w_x q_x + \alpha_{2}'' q_x^2) \sigma^x + w_y q_y \sigma^y 
\label{eqHtqD}
\end{equation}
and obtain
\begin{align}
\varepsilon^0_{\textrm{tqD}\pm} (\mathbf{q}) &= -w_x q_x + \alpha_{2}' q_x^2 \nonumber \\
&\pm \sqrt{(w_x q_x+{\alpha_{2}^{\prime\prime}} q_x^2)^2 + (w_y q_y)^2}.\label{tq-DE}
\end{align}
%
The energy dispersions of the upper band ($\varepsilon^{0}_{\textrm{tqD}+}(\mathbf{q}))$ and the lower band ($\varepsilon^{0}_{\textrm{tqD}-}(\mathbf{q}))$ near $\mathbf{q} = (0, 0)$ are given by
\begin{align}
\varepsilon^0_{\textrm{tqD}+}(q_x, q_y=0)
&= 
 \left\{ \begin{array}{ll}
 \alpha_2 q_x^2              & \mbox{if $q_x>0$} \label{E7}\\
 2 w_x |q_x| + \tilde{\alpha}_2 q_x^2  & \mbox{if $q_x < 0$} 
\end{array} \right. \\
 \varepsilon^0_{\textrm{tqD}+} (q_x=0, q_y) &= w_y |q_y| \\
 \varepsilon^0_{\textrm{tqD}-} (q_x, q_y=0) &= 
 \left\{ \begin{array}{ll}
-2w_x q_x + \tilde{\alpha}_2 q_x^2  & \mbox{if $q_x > 0$} \label{E9}\\
  \alpha_2 q_x^2              & \mbox{if $q_x < 0$}
\end{array} \right.  \\
 \varepsilon^0_{\textrm{tqD}-} (q_x=0, q_y) &= - w_y |q_y|,
\end{align}
where 
\begin{equation}
  \alpha_2 = \alpha_2' + {\alpha_2''}
\end{equation}
and
\begin{equation}
  \tilde{\alpha}_2 = \alpha_2' -{\alpha_2''}.
\end{equation}
We take $\alpha_2>0$ for simplicity. Then $\mathbf{q}=0$ is a local minimum of 
$\varepsilon_{\textrm{tqD}+}(\mathbf{q})$. 
From Eqs. (\ref{E7}) and (\ref{E9}), it is found that the dispersions of $\varepsilon_{\textrm{tqD}+}(\mathbf{q})$ and 
$\varepsilon_{\textrm{tqD}-}(\mathbf{q})$ near $\mathbf{q}=0$ are 
linear in three directions and quadratic in one direction. 
This can reproduce the dispersion near the Fermi energy in $\alpha$-(BEDT-TTF)$_2$I$_3$ at $P=2.3$. 
Therefore, we consider Eq. (\ref{eqHtqD}) at $\alpha_2>0$ as a model of 
{\it ``three-quarter''-Dirac cone}. The point of $\mathbf{q}=0$ is 
{\it ``three-quarter''-Dirac point}.

Next, we calculate the area of the closed constant energy line of the forth band by using Eq. (\ref{tq-DE}). 
We set 
$\varepsilon_{\textrm{tqD}+}(\mathbf{q})=\varepsilon$ and $\varepsilon>0$. 
The constant energy line is described by 
\begin{align}
& w_y q_{y\mathrm{F}} (q_x) \nonumber \\
 &= \pm \sqrt{ (\varepsilon + w_x q_x - \alpha_2' q_x^2)^2 
- (w_x q_x + {\alpha^{\prime\prime}_2} q_x^2)^2 } \nonumber \\
  &= \sqrt{\alpha_2 \tilde{\alpha}_2 (q_x-q_{x0}) (q_x-q_{x1}) (q_x-q_{x2}) (q_x-q_{x3})}, \label{E13}
\end{align}   
where
\begin{align}
 q_{x0} &=  -\sqrt{\frac{\varepsilon}{\alpha_2}}, \\
 q_{x1} &= \frac{w_x -\sqrt{w_x^2 + \tilde{\alpha}_2 \varepsilon}}{\tilde{\alpha}_2} \nonumber \\
          & \simeq - \frac{\varepsilon}{2 w_x}, \\
 q_{x2} &= \sqrt{\frac{\varepsilon}{\alpha_2}},\\
 q_{x3} &=  \frac{w_x +\sqrt{w_x^2 + \tilde{\alpha}_2 \varepsilon}}{\tilde{\alpha}_2} \nonumber \\
          & \simeq \frac{2 w_x}{\tilde{\alpha}_2}. 
\end{align}
Note
\begin{equation}
 q_{x0} \ll q_{x1} <0 < q_{x2} \ll | q_{x3} |. 
 \label{eqapprox}
\end{equation}
The area is calculated by  
\begin{align}
A(\varepsilon) = 2 \int_{q_{x1}}^{q_{x2}} q_{y\mathrm{F}} (q_x) d q_x. \label{Ae}
\end{align}
By taking an approximation that an electron pocket is elliptic, we obtain 
from Eq. (\ref{E13}) and Eq. (\ref{Ae}) 
\begin{align}
 A(\varepsilon)  &\simeq \frac{2}{w_y} \int_{q_{x1}}^{q_{x2}}
    \sqrt{\alpha_2 \tilde{\alpha}_2 q_{x_0} q_{x_3} (q_x-q_{x1}) (q_x-q_{x2})} dq_x \nonumber \\
   &\simeq \frac{2}{w_y} \sqrt{\alpha_2 \tilde{\alpha}_2 \sqrt{\frac{\varepsilon}{\alpha_2}}
\frac{2 w_x}{\tilde{\alpha}_2} }   \frac{\pi}{8} 
\left( \sqrt{\frac{\epsilon}{\alpha_2}}+\frac{\epsilon}{2 w_x} \right)^2 \nonumber \\
&\simeq \frac{\sqrt{2 w_x}\pi}{4 w_y} \alpha_2^{-\frac{3}{4}} \varepsilon^{\frac{5}{4}}.
\end{align}

\section{Fourier transform intensities}
\label{AppendixD}
In order to analyze the oscillations in the magnetization, 
we calculate the Fourier transform intensities numerically  as follows. 
By choosing the center $h_c$ and a finite range $2L$, we calculate 
\begin{equation}
  \mathrm{FTI}^{(1/h)}(f, \frac{1}{h_c},L) =\left|\frac{1}{2L} 
  \int_{\frac{1}{h_c}-L}^{\frac{1}{h_c}+L} M (h) e^{2\pi i \frac{f}{h}} d \left(\frac{1}{h}\right)\right|^2, \label{FTI}
\end{equation}
where we take $f=j/(2L)$ with integer $j$ ($j=512$ is used in this study).


\end{document}